%% file: sifin_main.tex
\PassOptionsToPackage{dvipsnames}{xcolor}
\PassOptionsToPackage{colorlinks=true,linkcolor=blue,citecolor=blue,urlcolor=blue,hypertexnames=false}{hyperref}
\PassOptionsToPackage{algo2e,ruled,lined,linesnumbered,commentsnumbered,longend}{algorithm2e}
\documentclass[final,onefignum,onetabnum]{siamonline250211}

\usepackage{amsfonts,amssymb}
\usepackage{algorithm2e}
\usepackage{array}
\DeclareMathOperator*{\argmax}{arg\,max}
\DeclareMathOperator*{\argmin}{arg\,min}

\let\sifinfootnote\footnote
\input{sifin/macros}
\let\footnote\sifinfootnote

\newsiamremark{remark}{Remark}
\newsiamthm{assumption}{Assumption}
\crefname{assumption}{Assumption}{Assumptions}

\bibpunct{[}{]}{,}{n}{}{,}

\allowdisplaybreaks

\AtBeginDocument{%
  \setlength{\abovedisplayskip}{2pt}%
  \setlength{\belowdisplayskip}{2pt}%
  \setlength{\abovedisplayshortskip}{0pt}%
  \setlength{\belowdisplayshortskip}{1pt}%
  \setlength{\jot}{0.5pt}\setlength{\textfloatsep}{4pt plus 1pt minus 1pt}\setlength{\floatsep}{3pt plus 1pt minus 1pt}\setlength{\intextsep}{3pt plus 1pt minus 1pt}\setlength{\abovecaptionskip}{2pt}\setlength{\belowcaptionskip}{0pt}%
}

\makeatletter\AtBeginDocument{\setlength{\parskip}{0pt}\setlength{\topsep}{3pt}\setlength{\partopsep}{0pt}\setlength{\itemsep}{1pt}\setlength{\parsep}{0pt}\def\thm@space@setup{\thm@preskip=3pt\relax\thm@postskip=3pt\relax}}\makeatother\headers{Multi-Task Contextual Dynamic Pricing}{A. Javanmard, J. Ji, and R. Xu}

\title{Multi-Task Dynamic Pricing in Credit Market with Contextual Information}

\author{Adel Javanmard\thanks{Marshall School of Business, University of
Southern California (\email{ajavanma@usc.edu}).}
\and Jingwei Ji\thanks{Management Science and Engineering, Stanford University
(\email{jingwei.ji@stanford.edu}, \email{renyuanu@stanford.edu}).}
\and Renyuan Xu\footnotemark[2]}

\ifpdf
\hypersetup{
  pdftitle={Multi-Task Dynamic Pricing in Credit Market with Contextual Information},
  pdfauthor={Adel Javanmard, Jingwei Ji, and Renyuan Xu}
}
\fi

\begin{document}

\maketitle

\begin{abstract}
We study the dynamic pricing problem faced by a broker seeking to learn prices for a large number of credit market securities, such as corporate bonds, government bonds, loans, and other credit-related securities. A major challenge in pricing these securities stems from their infrequent trading and the lack of transparency in over-the-counter (OTC) markets, which leads to insufficient data for individual pricing. Nevertheless, many securities share structural similarities that can be exploited. 
Moreover, brokers often place small ``probing'' orders to infer competitors’ pricing behavior. Leveraging these insights, we propose a multi-task dynamic pricing framework that leverages the shared structure across securities to enhance pricing accuracy.
In the OTC market, a broker wins a quote by offering a more competitive price than rivals. The broker's goal is to learn winning prices while minimizing expected regret against a clairvoyant benchmark. We model each security using a $d$-dimensional feature vector and assume a linear contextual model for the competitor's pricing of the yield, with parameters unknown a priori. We propose the Two-Stage Multi-Task (TSMT) algorithm: first, an unregularized MLE over pooled data to obtain a coarse parameter estimate; second, a regularized MLE on individual securities to refine the parameters. We show that the TSMT achieves a regret bounded by $\widetilde{\mathcal{O}} \paren{ \deltamax \sqrt{T M d} + M d } $, outperforming both fully individual and fully pooled baselines, where $M$ is the number of securities and $\deltamax$ quantifies their heterogeneity. 
Finally, our empirical experiment on U.S. corporate bonds shows promising performance of our TSMT algorithm and supports the theoretical findings.

\end{abstract}



\section{Introduction}

As of 2022, the average daily turnover of corporate bonds in the U.S. is around \$36 billion \citep{coalitionGreenwichReport}, making it one of the largest security markets in the world.  
In most credit markets such as the corporate bond market, there is no central limit order book (CLOB) to provide common prices to trade on, and instead, dealers quote prices in response to a client who sends a request for trading. 
Subsequently, the client selects the most favorable one to trade with. Hence from the dealers' perspective, they need to learn and predict the best competitor level (BCL), i.e., the quote provided by the best competitor given the current market contexts, meanwhile proposing a compelling price that is profitable. In addition, dealers in such markets are motivated to respond to requests across a vast array of securities, as it is important for large players to preserve their market share and enhance client loyalty by offering proactive responses.

This is a rather challenging problem in practice because of the coexistence of the {\it scarcity of historical transaction data} and {\it a vast number of different bonds}. As of 2023, there are currently about 66,000 U.S. corporate bonds available to trade, which compares to about 4,500 U.S. listed stocks.
Meanwhile, even the most liquid bonds (such as investment-grade bonds in the financial sector) trade only 300 times per day \citep{finra_trace}. 
Meanwhile, we usually see this many transactions in minutes in equity market \citep{nasdaq_data}.

Furthermore, the information disclosure regulation in the Europe, Middle East, and Africa (EMEA) market even makes the problem more challenging \citep{fermanian2016behavior, gueant2019deep}.
In EMEA, only the dealer who wins the quote has access to the second-best price for the request for quote, and other lost dealers only have the information that they do not win the transaction. This leads to a {\it one-sided censored feedback} in the information structure. 
These challenges leave the dynamic pricing of financial securities in the credit market largely unexplored in the literature, despite its critical importance.

{ 
A further distinguishing feature is the need for {\it real-time} pricing. Dealers must respond immediately whenever RFQs arrive, which contrasts with most existing corporate bond asset-pricing studies, where pricing and return predictions are typically made at low (e.g., monthly) frequency \citep{gebhardt2005cross, chen2007corporate}. Table \ref{tab:literature_position} summarizes differences across asset classes and sampling frequencies. 
To the best of our knowledge, real-time pricing is primarily studied in markets organized around a predominant limit order book (LOB).
In contrast, no existing work considers real-time data-driven pricing in credit markets, where such infrastructure is absent.
}

\begin{figure}[htbp]
    \centering
    \includegraphics[width=0.75\linewidth]{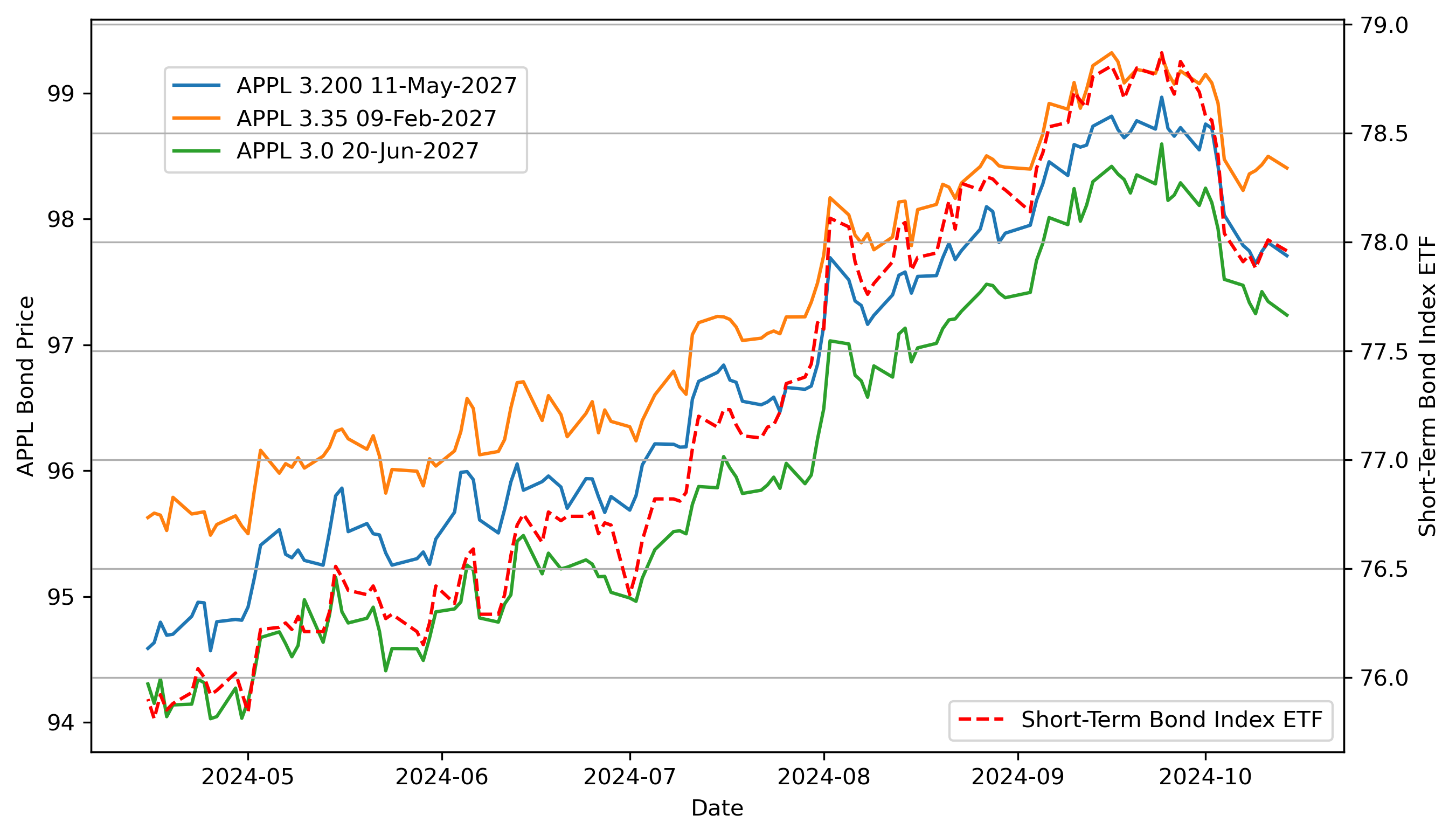}
    \caption{ Three outstanding bonds (as of October 2024) issued by Apple and the Vanguard Short-Term Bond Index ETF.  
        Apple bonds track each other and respond to the ETF in a similar fashion. }
    \label{fig:apple}
\end{figure}

Amidst the complexities, there remains a silver lining. Bonds, particularly those issued by the same company or within the same sector, often exhibit similarities. For instance, their prices may be affected by some macroeconomic indicators in the same direction, but with different magnitudes. See Figure  \ref{fig:apple}. 
This empirical observation leads us to the idea of developing a multi-task learning framework to price a vast array of securities, effectively overcoming the challenges associated with data scarcity and censored feedback by leveraging structural similarities. 

{ 

Our problem naturally relates to the broader asset-pricing literature, which aims to explain and predict expected excess returns across assets. While earlier work predominantly employed linear factor models \citep{fama1970efficient, black1972capital, fama1993common}, more recent research has increasingly used machine learning to enhance predictive accuracy given the explosion of candidate predictors \citep{gu2020empirical,chen2024deep,bianchi2021bond,weigand2019machine,kelly2023financial}. However, these contributions almost exclusively adopt offline learning on extensive historical datasets, typically monthly observations over decades. 
In contrast, our setting requires adaptive, real-time decision-making using limited and rapidly evolving data. This motivates an online learning framework that both incorporates the structural similarities described above and updates continually with new observations.

Finally, while inventory control, market making, and pricing are all central questions in OTC markets \citep{bergault2021size,cohen2024inventory,cont2024dynamics,atkins2024reinforcement}; they arise at different decision layers. 
Pricing involves determining fair quotes under informational frictions; market making governs the intermediation of client flow; and inventory control manages positions under capital and risk constraints. 
Because these layers involve different decision variables, objectives, and timescales, it is common in practice to treat them separately. 
In this paper, we focus on the pricing layer, developing adaptive quoting rules under informational frictions that can serve as a foundation for inventory-aware market-making policies.
}

\subsection{The research question, our result and contributions}
We propose a multi-task learning framework {to price a broad class of securities in credit market, which} leverages the potential similar structure shared by the securities {\it without} prior knowledge of the similarity. 
We measure the performance of a dynamic pricing policy via regret, which is the expected revenue loss compared to a clairvoyant that knows the model parameters a priori. {For ease of exposition, we focus on (client) buying.
In view of \cite{cesa2024market}, our framework can be generalized to a market-making problem where the dealer simultaneously buys and sells.}

Specifically, at each timestamp, a client who wants to buy a security $j \in [M]$  asks for quotes from multiple firms (including ``us", the decision maker). 
The client will buy the security from the firm which gives the best (lowest) quote. 
We assume a linear contextual model for the yield of the quote of the best competitor, which has a parameter $\bftheta_\star^j \in \mathbb{R}^d$. 

To capture the similarity among different securities, we decompose the individual security model parameter $\bftheta_\star^j$ into a common part $\bftheta_\star$ shared by all the securities and an idiosyncratic deviation $\bfdelta_\star^j$ for security $j$. 
Precisely, we assume that for each $j \in [M]$:
\begin{equation}  
   \bftheta^j_\star  = \bftheta_{\star} + \bfdelta_{\star}^j  .
\end{equation}
To measure the {\it degree of similarity} among different securities, we define 
\begin{equation}
    \deltamax   = \max_{j \in [M]}~~  \| \bfdelta_{\star}^j \|_2 .
\end{equation}

We study the dynamic pricing problem of a firm when the number of securities $M$ is {\it large}. 
At each timestamp $t$, a buyer comes with a request to purchase security $Z_t \in [M]$. 
If the firm's quote $p_t$ is better than or equal to the best competitor's quote, then the security is sold, and the firm collects a revenue of $p_t$.
In particular, we answer the following question:

\begin{adjustwidth}{1cm}{1cm}
    {\textit{
    How can we design a good pricing policy which utilizes the similarities between securities without knowing $\|\bfdelta_\star^j\|$'s a priori, and whose regret scales gracefully in both $T$ and $M$? 
    }}
\end{adjustwidth}


\noindent In Theorem~\ref{thm:regret_simple_version},  we give an affirmative answer to this important setting which applies to many modern and complex data-driven decision-making problems. Intuitively, when securities are extremely similar to each other (i.e., when $\deltamax$ is close to zero), it is beneficial to pool the data together and use a single model for all securities, which we call the \textit{pooling strategy}. Conversely, when securities differ significantly (i.e., when $\deltamax$ is very large), it is advantageous to train separate models for each security, which we call the \textit{individual learning strategy}, avoiding using data from other securities. Ideally, an effective pricing policy should {\it automatically} adapt to the actual similarity structure of the securities, outperforming both the pooling and individual learning strategies without prior knowledge of the similarities.
To achieve this, we introduce the Two-Stage Multi-Task (TSMT) pricing algorithm. 
The policy runs in an episodic fashion, and updates the estimates of the model at the beginning of every episode.  
When updating, the estimation consists of two stages. 
In the first stage, all data from different bonds are aggregated to estimate a common part of all tasks. 
In the second stage, data points of individual bonds are used to refine the estimate in the first stage. 


    The analysis of our algorithm presents challenges from two distinct perspectives. 
    {\bf (1)} From a modeling and optimization standpoint, we assume that the yield follows a linear contextual structure which is supported by empirical evidence, and embed the yield-to-price function to address the pricing problem. 
    While this modeling choice captures the essential structure of pricing credit market securities, it departs from the standard dynamic pricing framework, which typically assumes a linear relationship between price and context (see, e.g., \cite{javanmard2019dynamic, cohen2020feature, bu2022context}). 
    As a result, our novel framework introduces a more complex optimization landscape, making it difficult to control the regret by the statistical error (i.e., Lemma~\ref{lemma:perround_regret_to_estimation_error}). {\bf (2)} From the statistical perspective, we need to show that the TSMT {\it automatically outperforms} both the pooling strategy and the individual learning strategy in terms of the statistical error of the coefficients (i.e., Lemma~\ref{lemma:expectation_bound_stage_2}), even without knowing the similarity structure. Although practically important, this automatic guarantee has not been established in the literature.
  In addition, the online nature of the problem introduces randomness in security arrivals,  making estimation errors harder to control and creating unique challenges that have been largely overlooked in the primarily offline settings.
    By addressing all the above-mentioned difficulties, we show that the TSMT, without knowing how tasks are related in the first place, achieves a regret of order $ \widetilde{\mathcal{O}} \paren{ \deltamax \sqrt{T M d} + {M} d } $.

    Finally, in Section~\ref{sec:experiments}, we showcase our algorithm on a dataset of the U.S. corporate bonds. 
    The experiments demonstrate how our method outperforms the benchmarks, highlighting its capacity to effectively utilize the available information among different securities while facing the challenge of data scarcity.


\subsection{Related literature}
Our work contributes to the asset pricing literature from a topical perspective, and to dynamic pricing and multi-task learning from a technical standpoint. In the following sections, we provide a brief overview of the studies most closely related to ours in these areas and highlight the novel aspects of our framework.

\paragraph{Asset pricing across different data frequencies}

{

Table \ref{tab:literature_position} summarizes representative asset-pricing studies across different asset classes and data frequencies, highlighting where our work fits within the broader literature.
Commodities, foreign exchange (FX), and equities are predominantly traded on centralized exchanges or electronic communication networks (ECNs), where LOBs provide rich intra-day or even real-time information. As a result, asset-pricing research in these markets spans monthly, intra-day, and real-time horizons.
In contrast, corporate bond trading is decentralized and occurs primarily OTC. The absence of consolidated LOBs and the limited availability of high-frequency quote data have led existing corporate bond studies to focus almost exclusively on monthly pricing. 
Consequently, real-time pricing in credit markets remains largely unexplored, which is precisely the gap our paper addresses.

 \begin{table}[H]
  \centering
  \caption{Pricing by asset class and data frequency (selected work)}
    \begin{tabular}{c>{\centering\arraybackslash}p{10.715em}>{\centering\arraybackslash}p{10.715em}c}
    \toprule
    \toprule
       & Monthly & Intra-Day  & Real-Time  \\
    \midrule
    Commodities & \cite{gorton2013fundamentals, gorton2006facts, szymanowska2014anatomy} & \cite{ready2022order} & LOB  \\
    \hline
    FX & \cite{lustig2011common} & \cite{andersen1998deutsche, evans2002order} & LOB \\
    \hline
    Equities  & \cite{fama1970efficient, fama1993common} & \cite{aleti2025news, ait2025continuous} & LOB \\
    \hline
    Corporate bonds & \cite{gebhardt2005cross, chen2007corporate} &  None identified  & \textbf{Our paper} \\
    \bottomrule
    \end{tabular}%
  \label{tab:literature_position}%
\end{table}%

}

\paragraph{Machine learning for asset pricing} 
Recent technology advancements have sparked increasing interests in applying machine learning techniques to enhance the predictive accuracy of asset pricing models, particularly as the dimensionality of the feature space grows. Machine learning methods are well-suited to handle the complexity and high dimensionality of modern datasets, offering improved performance where traditional linear models may face limitations. 
In particular, \cite{gu2020empirical} conduct an empirical comparative study exploring how various popular machine learning techniques can enhance the forecasting performance of stock excess returns compared to traditional linear regression models. 
{  \cite{bryzgalova2019forest} show how to use a tree-based approach to estimate the stochastic discount factor based on firm characteristics, which empirically perform better than the traditional 25 size-and-value \cite{fama1993common} portfolios. 
}
\cite{gu2021autoencoder} propose an autoencoder latent factor model which subsumes the linear latent structure in \cite{kelly2019characteristics}. 
\cite{chen2024deep} explore how deep learning techniques, such as GANs and LSTMs, can be integrated within the fundamental framework of no-arbitrage pricing (see also \cite{dai2024data}).
In the study of corporate bond returns, alongside the literature that directly follows the Fama and French framework for equities, \cite{kelly2023modeling} diverge by proposing an instrumented principal components analysis. 
In addition, \cite{bianchi2021bond} show that tree-based methods and neural networks leveraging macroeconomic and yield information data offer strong statistical evidence supporting the bond return predictability. 
We refer readers to \cite{weigand2019machine, giglio2022factor, kelly2023financial} for comprehensive reviews of the potential and limitations of machine learning techniques across different problems in empirical asset pricing.

It is worth noting that the aforementioned studies adopt offline frameworks, whereas we focus on an online learning approach--motivated by data challenges, the trading mechanism  on credit markets, and the brokers' strategic use of small ``probing'' orders to infer competitors' pricing behavior. 
See also \cite{jia2022policy,huang2025sublinear,huang2024mean} for related developments in continuous-time reinforcement learning.

\paragraph{Dynamic pricing}
Dynamic pricing (or posted-price auction) garners attention from the fields of computer science, economics and operations management. 
The online nature of this literature makes it particularly well-suited for applications where data is scarce, such as pricing illiquid assets in finance.
Early works on dynamic pricing focus exclusively on a single type of items (products) \citep{myerson1981optimal, kleinberg2003value, besbes2009dynamic, broder2012dynamic, chen2023robust}. 
In particular, \cite{kleinberg2003value} tackle the problem using a multi-armed bandit approach, by allowing the firm to use discrete grid of prices within the continuum of feasible prices. 
More recent papers on dynamic pricing consider models with features/covariates to differentiate products, motivated by data-driven decision-making approaches. 
For example, \cite{qiang2016dynamic} study a linear contextual model where the firm observes the demand entirely. 
 \cite{javanmard2019dynamic} employ techniques from high-dimensional statistics to exploit the sparse structure in the model parameter.
\cite{cohen2020feature} consider a model where the contexts are adversarially chosen and the valuations are deterministic, without random noise. 
Other works that apply dynamic pricing framework to financial market includes \cite{xu2024dynamic}.

Extending dynamic pricing frameworks to accommodate multiple products is a natural progression, broadening the scope of potential applications. While several works delve into this direction, each of which has a different focus than ours. 
\cite{keskin2014dynamic} design a myopic policy that learns the demand of multiple products at the same time. 
\cite{javanmard2017perishability} considers a setting where there is a large number of products. 
Their primary objective is to devise an algorithm capable of adeptly adjusting to rapidly changing model parameters of different products, while our goal is to deal with constant model parameters from a large number of products which arrive in sequence.
\cite{den2024pricing} consider a setting with differentiated products with customers purchasing according to a locational choice model. 
Both \cite{bastani2022meta} and \cite{kveton2021meta} study the setting where there is a large number of related products, modeled via a Bayesian structure. 
In their setups, each product has a selling horizon of $T$ rounds. 
A new product does not arrive \textit{until} the complete selling horizon of the old product has passed. 
In contrast, in our setting, any of the $M$ securities might arrive randomly during any of the $T$ rounds.

In addition, all the above mentioned works consider a linear relationship between context and price, which cannot be directly applicable to our setting. 
It is more commonly believed that yield is linear in market and security-specific contexts, while the relationship between yield and price is nonlinear and nontrivial, which is a key challenge to address. Complementary work by Cohen and Treetanthiploet \cite{cohen2021generalised} develops a Bayesian bandit approach to dynamic pricing that exploits dependence across price choices.

\paragraph{Multi-task learning}

The concepts of multi-task learning \citep{caruana1997multitask, breiman1997predicting, romera2013multilinear, yu2020gradient}, transfer learning \citep{taylor2009transfer, zhuang2020comprehensive}, and meta-learning \citep{finn2017model, finn2019online, hospedales2021meta} exhibit inherent connections, often with blurred boundaries between them. 
Many formulations have been proposed by researchers across different communities,  especially after the empirical success of \cite{finn2017model}. 
However, the overarching objective remains consistent: to devise algorithms capable of swiftly adapting to similar (new) tasks based on past experience, whether for classification \citep{cavallanti2010linear}, quantile regression \citep{fan2016multitask}, or other applications.
Rather than attempting to classify the overwhelmingly extensive literature (we refer interested readers to the review paper \cite{zhang2018overview}), we focus on summarizing works closely aligned with our objective: leveraging data across different tasks to  learn faster. 
There is a stream of literature in the statistics community, which uses the very natural idea of $\ell_2$-distance and $\ell_1$-distance  between model parameters to measure task similarity \citep{xu2021learning, li2022transfer, duan2022adaptive, tian2023learning}. However, these existing developments cannot be applied to address our challenges due to differences in the problem settings. Specifically, \cite{tian2023learning} focuses on designing multi-task learning and transfer learning algorithms within a linear representation framework for an offline setting, robust to outlier tasks.
Additionally, \cite{xu2021learning} investigates multi-task contextual bandit problems, emphasizing high-dimensional and sparse structures.  The recent work by \cite{duan2022adaptive} is closely related to ours.  However, their focus is exclusively on the adaptivity and robustness issue in the offline setting.   Hence, their algorithm and analysis do not automatically overcome the unique challenges that arise in the online setting that we consider. In addition, directly applying their result does not yield a satisfactory dependence on the number of securities $M$ in the online setting.  We defer a more detailed discussion on the technical perspective to Remark~\ref{rmk:technical_diff}.

Transfer learning \citep{gu2022robust, bastani2021predicting,li2022transfer}, though relevant, focuses on a different training process where data from source tasks are used to learn a new task. In contrast, we need to learn multiple tasks that arrive randomly.



\subsection{Organization and notations}

The remainder of this paper is organized as follows. In Section~\ref{sec:setup}, we formulate the problem and introduce the notion of regret.  
In Section~\ref{sec:algorithm}, we propose the Two-Stage Multi-Task (TSMT) pricing algorithm and analyze the regret of the TSMT algorithm.  
In Section~\ref{sec:experiments}, we support our theoretical assertions with numerical experiments conducted on real datasets. 
In Section~\ref{sec:main_proof}, we lay out the proof for our main result, Theorem~\ref{thm:regret_simple_version}. Finally, the proofs of several technical lemmas are deferred to the appendices.

\paragraph{Notations} 
We reserve $M$ for the number of securities, $T$ for the number of rounds. 
Bold lowercase letters refer to column vectors. 
Bold uppercase letters denote matrices. 
The function $\lambda_{\max} \paren{\cdot}$ and $\lambda_{\min} \paren{\cdot}$ map a matrix to its maximum and minimum eigenvalues, respectively. 
We write $\normtwo{\cdot}$ for both the $\ell_2$ vector norm and the associated operator norm. 
Inner product in Euclidean space is denoted by $\inner{\cdot, \cdot}$.
The symbol $x \lesssim y$ means that there exists some absolute constant $C$ such that $x \leq C y$. 
Given an event $A$ and a random variable $X$, $\E{X ; A}$ is a shorthand for $\E{X \one{A}}$.
We use $\mathcal{B}(r)$ to represent the $L_2$ ball centered at the origin with radius $r$ in $\mathbb{R}^d$. For an integer $m$, we use the shorthand $[m] = \{1,2,\dotsc, m\}$.

\section{The Problem Setup} \label{sec:setup}

In this section, we introduce the problem set-up and justify our modeling assumptions.

\paragraph{Context and competitors offer} 
We consider $M$ distinct securities. 
Let $T$ be the length of the overall rounds. 
In round $t$, the following events happen in sequence: 
\begin{enumerate}
    \item A buyer sends a request for quote (RFQ) to multiple dealers (including us) to buy one unit of security $Z_t \in [M]$. 
    \item Every firm that receives the RFQ observes $Z_t$ and the contextual feature $\bfx_t \in \mathbb{R}^d$, which is sampled from a security specific distribution. 
    \item Each firm offers a quote (i.e., price to sell) to the buyer, among which we call the best price the Best Competitor Level (BCL). 
    \item If our quote $p_t$ is {better than (lower than)} or equal to the BCL, then the buyer purchases it from us. In this case, we can further observe the value of BCL. 
    Otherwise, our security is not sold and we can only observe the event that our quote $p_t$ is worse (larger) than BCL.
\end{enumerate}
 The above formulation and information structure in particular corresponds to regulation rules in the EMEA market \citep{fermanian2016behavior}, which we will adopt throughout the paper.

\smallskip

\begin{remark}[Formulation and information structure]
    A few remarks in place:
\begin{itemize}
\item This set-up also applies to other applications such as double auctions \citep{friedman2018double}.
\item In addition, our framework can be applied to scenarios where a firm receives requests from both the buy and sell sides. For simplicity, we assume the firm only receives buy requests, offering a price to sell one unit of security to each potential client, or ``buyer".
\end{itemize}

\end{remark}

\smallskip

We assume that $Z_t$ $i.i.d.$ follows a categorical distribution $\operatorname{CG} (\bfpi )$ with $\bfpi = [\pi_1, \cdots, \pi_M]^\top$, which we call the \textit{arrival distribution}.  
At the beginning of each round, a context vector $\bfx_t \in \mathbb{R}^d$ associated with the security is observed by us, the competing firms, and the buyer. 
Our algorithm can handle very generic arrival distributions and, in fact, the algorithm performance highly depends on the arrival distribution (see Theorem \ref{thm:regret_simple_version} and Corollary \ref{cor:distribution_decay}).

We make the assumption that conditioned on $Z_t = j$, the context $\bfx_t$ is $i.i.d.$ sampled from a fixed but unknown distribution $\mathcal{P}^{j}$ with a bounded support. Namely, 
\begin{equation} \label{eq:x_iid}
    (\bfx_t\,|\,Z_t=j) \overset{\text{i.i.d.}}{\sim} \mathcal{P}^{ j } , 
\end{equation}
for which there exists some constant $\xmax$ such that $ \mathcal{P}^{ j } \paren{ \normtwo{\bfx_t} \leq \xmax } = 1$.
We denote $ \bfSigma^j \defeq \E{  \bfx_t \bfx_t^\top| Z_t =j} $ and $\bfSigma  \defeq \E{\bfx_t \bfx_t^\top} $.
We assume that the contexts are $i.i.d.$ as stated in \eqref{eq:x_iid} primarily for simplicity.
We remark that our analysis can be extended to accommodate contexts generated by, e.g., a Markov chain, under a minorization condition with respect to a coupling measure, whose support spans $\mathbb{R}^d$. 

\paragraph{BCL yield and reward}
Bonds are priced and hedged based on yield spreads, making it easier to compare them with Treasury yields, which serve as a risk-free benchmark for different maturities.
Following empirical studies in the literature \citep{bianchi2021bond, huang2023bond}, we assume a linear structure for the yield. More specifically,
provided that security $Z_t = j$ arrives, the yield-to-maturity of the BCL $y_t$ is assumed to take the following form,
\begin{equation} \label{eq:linear_bcl}
    y_t = \inner{ \bftheta^j_\star , \,\bfx_t } \,+\, \epsilon_t ,  
\end{equation}
where $\epsilon_t$ is the idiosyncratic noise. 
We stress that $\bfx_t $ contains security-specific contextual information. 
We assume that $\braces{\epsilon_t}_{t \geq 1}$ are drawn $i.i.d.$ from a distribution with zero mean and density function $f(x) = F'(x)$. 
Let $\bar{F}(x) = 1 - F(x)$. 
Furthermore, we assume that the firm has the knowledge of $F$. 

\smallskip
\begin{remark}[Linear form of $y_t$]
The linear  model, though simple, is a prevailing practice in the finance literature to model  yield \citep{bianchi2021bond} as well as
excess return \citep{gabbi2005factors, bongaerts2017asset}. 

From a practical perspective, the linear model remains popular in practice due to its explainability and tractability, which also aligns with the risk management mandates.
\end{remark}

\smallskip

We consider bonds with fixed coupon payments.  Given bond $Z_t$, let $\braces{ \tau_1^{(t)}, \tau_2^{(t)}, \cdots, \tau_{n^{(t)}}^{(t)}}$ be the set of \textit{future} payment dates (in years) associated with the bond, where each $\tau_i^{(t)}$ represents the time to a scheduled cash flow, including both coupon payments and the par value. 
Let $c^{(t)} \in [0,1]$ be the coupon rate, $P^{(t)} \in \mathbb{R}^+ $ the par value. 
We refer to the tuple $\paren{ \braces{ \tau_1^{(t)}, \tau_2^{(t)}, \cdots, \tau_{n^{(t)}}^{(t)}} ,  c^{(t)} , P^{(t)}}$ as the \textit{bond primitives}, and assume that these primitives take values in a common compact set $\bfXi$ uniformly over $t$. 
 The yield-to-price mapping is defined as
\begin{equation}  \label{eq:bond_yield-to-price}
    V_{t} (y) \defeq \sum_{ i=1}^{ n^{(t)} }  \frac{ c^{(t)} P^{(t)} }{ (1+y)^{\tau_i^{(t)}} }  + \frac{P^{(t)}}{(1+y)^{\tau_{n^{(t)}}^{(t)}}}  . 
\end{equation}
Clearly, the yield-to-price function is twice differentiable. It is strictly decreasing and convex in yield $y$, namely, $V_t' < 0, V_t'' > 0$, and $V_t'''<0$.  The function is injective, and hence its inverse exists.

Let $\bar{y}_t$ be a random variable defined as:
$
\bar{y}_t =
\begin{cases}
y_t, & \text{if } V_t(y_t) \geq p_t, \\
\emptyset, & \text{otherwise}, 
\end{cases}
$
where $\emptyset$ represents null information, indicating the absence of an observable value. 
This corresponds to the information structure in EMEA market: the BCL is revealed only when the dealer wins the bid. 
Define the filtration $\{\mathcal{F}_t\}_{t \geq 1}$ as the sequence of $\sigma$-algebras generated by the available information for the dealer to quote at time $t$:
$
\mathcal{F}_t = \sigma\braces{\bfx_1, Z_1, \bar{y}_1, \cdots, \bfx_{t-1}, Z_{t-1}, \bar{y}_{t-1},  \bfx_t, Z_t }
$. 

The instantaneous reward of the decision maker at round $t$ is 
\begin{equation} \label{eq:per_round_reward}
     \operatorname{reward}_t \paren{ p_t } = (p_t - \gamma_t) \one{p_t  \leq V_t(y_t) }  , 
\end{equation}
where in the context of market making, $\gamma_t \geq 0$ can be seen as a monetary ``cost" of the bond, or a hyperparameter to control the aggressiveness of the quote.
A smaller value $\gamma_t$ encourages more aggressive quotes, as acquiring information about the best competitor's quote is among the top priorities of many dealers.  On the other hand, a larger $\gamma_t$ sometimes is suitable as quoting too cheap can be worse than losing the competition. Any underpriced quote will soon be taken advantage of in a competitive market.

Conditional on observing $Z_t$ and context $\bfx_t$, the expected reward from quoting $p$ is 
\begin{equation}  \label{eq:per_round_reward_expected}
  R_t(p) \defeq  \E{  \operatorname{reward}_t (p)  \mid \mathcal{F}_t  } = 
    \paren{ p - \gamma_t } \pr{ p \leq V_t( y_t ) \mid \mathcal{F}_t  }  . 
\end{equation}

\smallskip
\begin{remark}[Pricing vs inventory control] Market makers care about inventory risk; however, we did not model it in the reward function \eqref{eq:per_round_reward}. This is because it is more practical to view pricing and inventory control (or optimal trading with inventory consideration) as two separate problems in the over-the-counter market.  Here, we focus on pricing in response to small orders from clients, aiming to infer competitors’ pricing behavior and determine the optimal quote to outperform competitors at desired probability levels.  Optimal trading with inventory control, which adjusts prices to manage levels of financial instruments effectively, can be treated as a distinct problem; see, e.g., \cite{atkins2024reinforcement}.
\end{remark}
\smallskip

\paragraph{Structure similarity} Next, we impose a structure on how securities are similar to each other. 
We decompose each  $\bftheta_\star^j$ to a common part $\bftheta_\star$ and an idiosyncratic deviation $\bfdelta_\star^j$ with respect to it, i.e. 
\begin{equation}  \label{eq:coefficient_structure}
   \bftheta^j_\star  = \bftheta_{\star} + \bfdelta_{\star}^j  .  
\end{equation}
We assume that $\normtwo{\bftheta^j_\star}  \leq W$ for some absolute constant $W>0$.
The quantity $ \deltamax   := \max_{j \in [M]}  \| \bfdelta_{\star}^j \|_2 $ is to measure the similarity across the securities. 
We assume that the firm has the knowledge of $W$ but {\it neither} of $\deltamax$ \textit{nor} of $\normtwo{\bfdelta_\star^j}$. 
This assumption is not restrictive since 
$W$ can be set as a sufficiently large constant that exceeds any reasonable model parameter. The information of  $\normtwo{\bfdelta_\star^j}$ is critical in the context of multi-task learning. 
Since an undesirable design, such as sharing knowledge among unrelated tasks, can negatively impact multi-task learning (Yu et al. 2020), a good algorithm should {\it automatically} adapt to the structure of task relationships, even without prior knowledge of their similarity.

\paragraph{Regret}

Let $\bar{p}$ be an upper bound on the price we quote. 
Given $\hat{\bftheta}_t$, an estimation of $\bftheta_\star$, there is a natural pricing rule given by $ p_{t} = p_t^\star(\langle \bfx_t, \hat{\bftheta}_t\rangle) $, where 
\begin{equation} \label{eq:p_star_as_b}
    p^\star_t (b) = \argmax_{ p \in ( - \infty, \bar{p} ] } ~ ( p - \gamma_t ) F ( V^{-1}_{t} (p) - b  )  .
\end{equation}
The optimal price, which is also the benchmark we compare to, is
$
p_{t}^{\star} \defeq p^\star_t (\inner{ \bfx_t, \bftheta^{Z_t}_{\star}})  . 
$

Compared with the benchmark policy which quotes the optimal prices $\braces{p_t^\star}_{t \geq 1}$,
the expected regret of a policy which quotes prices $\braces{p_t}_{t \geq 1}$ for $T$ rounds is defined to be 
\begin{equation}  \label{eq:regret}
   \regret \paren{T}  =
    \E{ \sum_{t=1}^{T}  \paren{ \operatorname{reward}_t \paren{p_t^\star} - \operatorname{reward}_t \paren{p_t} } }  . 
\end{equation}

\smallskip

\begin{remark}[Exogeneity of competitors quotes] When we use the notion of regret (defined in \eqref{eq:regret}) as the criterion for evaluation, we assume that the best competitor’s level $y_t$ is given exogenously. This means there are no strategic interactions between us and other competitors. Although this assumption might seem restrictive initially, our setup and results serve as an essential foundation for understanding more complex models, such as those involving equilibrium analysis \citep{aqsha2024strategic,wu2024broker,boyce2024market}.
We revisit this exogeneity assumption in Appendix~\ref{app:strategic_foundation}, where we provide an equilibrium interpretation of the linear BCL model based on strategic competition among dealers.
\end{remark}

\section{The Two-Stage Multi-Task Pricing Policy}  \label{sec:algorithm} 
In this section, we present the  pricing algorithm and provide theoretical results on its regret.

First, we introduce some notations. 
We denote by $\ell_t \paren{\bftheta; p_t, y_t, \bfx_t}$ the likelihood function of the observation at round $t$. 
Given our model, $\ell_t$ is given by 
\begin{equation}
\begin{aligned}
\ell_t \paren{\bftheta; p_t, y_t, \bfx_t} \defeq{}&
\log \paren{\bar{F}\paren{V_t^{-1}(p_t)-\inner{\bftheta,\bfx_t}}}\one{V_t(y_t)<p_t} \\
&+\log \paren{f\paren{y_t-\inner{\bftheta,\bfx_t}}}\one{V_t(y_t)\geq p_t}.
\end{aligned}
\end{equation}
To see this, note that when $p_t>V_t(y_t)$, we can only observe the event $\one{V_t(y_t) < p_t}$ which occurs with probability $\bar{F} \paren{ V_{t}^{-1}(p_t) - \inner{ \bftheta  , \bfx_t }}$. 
When $p_t \leq V_t(y_t)$, we further observe the yield  $y_t$ of competitor's offer, which has density $f( y_t -  \inner{ \bftheta   , \bfx_t } ) $.
We let $\bar{y}$ be the upper bound of the yield of reasonably quoted prices. 
We define $u_F$ to be the maximum of the first order derivative of the likelihood $\ell_t$ under our range of consideration 
\begin{equation} \label{eq:def_u_F}
u_F \defeq \max_{ \abs{x} \leq  \bar{y} + W \xmax } \braces{  \min \braces{ - \log ' \paren{ \bar{F} \paren{x} } ,~  - \log' \paren{  f \paren{x} }  }  } ,
\end{equation}
where we recall that $ \bar{F}(x) = 1 - F(x)$. For a vector $\bfx \in \mathbb{R}^n$, we denote its projection to the Euclidean ball centered at the origin with radius $W$ by 
\begin{equation}  \label{eq:def_projection}
    \operatorname{Proj}_{ \mathcal{B}(W) } \paren{ \bfx } = \arg\min_{\bfv \in \mathbb{R}^n } \braces{ \|\bfx-\bfv\|_2:\, \|\bfv\|_2\le W }  . 
\end{equation}

The algorithm is fully detailed in Algorithm~\ref{algo:two_stage}.  Specifically, the algorithm runs in an episodic fashion, and the length of episodes grows exponentially. 
Such a design is common in dynamic pricing \citep{javanmard2019dynamic} and online learning literature \citep{even2006action, lattimore2020bandit}. 
In our case, it is critical to use only samples from the previous episode to make decisions during the current episode, in that it allows us to establish concentration inequalities for the maximum likelihood estimator (MLE).

In each episode, we run a two-stage estimation procedure:
\begin{itemize}
    \item In the first stage of episode $k$, observations of all securities are aggregated together to run an unregularized MLE to obtain $\bar{\bftheta}_{(k)}$ that estimates the common part $\bftheta_\star$. 
    \item In the second stage, we refine 
    the coefficient estimates  by conducting a separate regularized MLE for each individual security.
The regularization parameter $\lambda_{(k)}^{j}$ needs to be set properly  as
\begin{equation} \label{eq:lambda}
\lambda^j_{(k)}  =  \sqrt{\frac{8 u_F^2 \bar{x}^2 d \log \paren{2 d^2 M }}{ N_{(k)}^{j} }}  , 
\end{equation}
where $N_{(k)}^{j}$ is the number of observations of security $j$ in the $(k-1)$-th episode.\footnotemark{}\footnotetext{For the algorithm to be well-defined, we let $ \hat{\bftheta}^j_{(k)} = \bar{\bftheta}_{(k)}$ if $N_{(k)}^{j}=0$. } 
This tuning of regularization parameter ensures that the refining process can, on the one hand improve upon the pooling estimate $\bar{\bftheta}_{(k)}$ using the individual security data, and on the other hand, still inherit the accuracy from the multi-task learning.  
\end{itemize}


Unlike existing approaches in the literature, Algorithm~\ref{algo:two_stage} offers a distinct advantage as it runs {\it without} requiring prior knowledge of structural similarities and other instance-specific information. Specifically, the decision maker does not need to know the values of parameters such as  $\deltamax$, relatedness of bonds, or the arrival distribution. In contrast, \cite{chua2021fine}, for instance, necessitates an oracle with access to a predefined similarity level $\deltamax$. 

\begin{algorithm2e}[htbp]
\begin{small}
\caption{TSMT (Two-Stage Multi-Task) Pricing Policy }  \label{algo:two_stage}

\For{\text{each episode} $k=2,3\cdots $}{

Set the length of the $k$th episode $\tau_k \leftarrow 2^{k-1}$\;

Update the model parameter estimate $\braces{ \hat{\bftheta}_{(k)}^j }_{j=1}^{M}$ using the data in the previous episode\;

\textbf{Stage I: } aggregating data 
\begin{eqnarray}
    \bar{\bftheta}_{(k)}  &=&  \arg \min_{\bftheta \in \mathbb{R}^d }~ \bar{\calL}_{(k)} \paren{ \bftheta }  
        , \quad \text{with}~  \bar{\calL}_{(k)} \paren{ \bftheta }   \defeq 
        - \frac{1}{\tau_{k-1} } \sum_{t= \tau_{k-1} }^{\tau_k - 1}  \ell_t \paren{\bftheta}.  \nonumber 
\end{eqnarray}

\textbf{Stage II:  } refine the estimation  for every $j \in [M]$ 
\begin{eqnarray}
   \hat{\bftheta}^j_{(k)}  &=&  \arg \min_{ \bftheta^j \in \mathbb{R}^d   }   
   ~~ \calL^{j}_{(k)}  \paren{\bftheta^j} + \lambda^j_{(k)}  \normtwo{ \bftheta^j -    \bar{\bftheta}_{(k)}  }       , \nonumber  \\  
    & & \text{with} \quad \calL^{j}_{(k)}  \paren{\bftheta^j} \defeq 
    - \frac{1}{ N_{(k)}^{j}  }  \sum_{t= \tau_{k-1} }^{\tau_k - 1} \one{Z_t=j} \ell_t \paren{\bftheta^j}    , \nonumber \\
    & &  \text{and} \quad \lambda^j_{(k)} =  \sqrt{\frac{8 u_F^2 \bar{x}^2 d \log \paren{2 d^2 M }}{N_{(k)}^{j}}}  , ~ N_{(k)}^{j} \defeq  \sum_{t= \tau_{k-1} }^{\tau_k - 1} \one{Z_t=j}.  \nonumber 
\end{eqnarray}

For each time point $t$ in the $k$th episode, set $\hat{\bftheta}_t = \hat{\bftheta}^{Z_t}_{(k)}$ and quote according to 
\begin{equation} \label{eq:algo_pricing_formula} 
    p_{t} = p^{\star}_{t} \paren{ \inner{ \bfx_t,  \operatorname{Proj}_{\mathcal{B}(W)}  \paren{ \hat{\bftheta}_t } } }
\end{equation}
where $\operatorname{Proj}_{\mathcal{B}(W)}$ is defined in \eqref{eq:def_projection} and $p^\star_t$ is defined in \eqref{eq:p_star_as_b}. 
}
\end{small} 
\end{algorithm2e}



Before presenting our main result, we list the assumptions. 

First, we make the following assumption on $\bfSigma^j $, the covariance matrix of the context $\bfx_t$ given that it is security $j$.

\begin{assumption} \label{assumption:diverse_contexts}
    Assume that there exist $0 < \underline{\lambda}<\overline{\lambda}$ such that $ \underline{\lambda} <  \underset{j \in [M]}{\min} ~  \lambda_{\min} \paren{ \bfSigma^j } \leq \underset{j \in [M]}{\max} ~  \lambda_{\max} \paren{ \bfSigma^j } < \overline{\lambda} $. 
\end{assumption}

Assumption \ref{assumption:diverse_contexts} suggests that we see enough variation along all dimensions of the context vector, which is standard in contextual bandit literature; see, e.g., \cite{li2017provably}. 
A direct consequence of Assumption~\ref{assumption:diverse_contexts} is that 
\begin{eqnarray*}
   \lambda_{\min} \paren{ \bfSigma } &=& 
   \lambda_{\min} \paren{ \E{ \bfx_t \bfx_t^\top } } 
   = \lambda_{\min} \paren{  \sum_{j=1}^{M} \pi_j \E{ \bfx_t \bfx_t^\top | Z_t=j } }   \geq  \sum_{j=1}^{M} \pi_j \lambda_{\min} \paren{ \bfSigma^j } > \underline{\lambda} ,  
\end{eqnarray*}
where the first inequality follows since $\lambda_{\min} \paren{\cdot}$ is concave over positive definite matrices. 

Then, we make an assumption on the BCL yields. Recall the yield-to-price mapping defined in \eqref{eq:bond_yield-to-price} and  we define $A_t(r) \defeq 2 V_t'(r)  -  \frac{ V_t(r) V_t''(r) }{ V_t'(r) }$.
Let $r^{\star}_{t}(\cdot) = V_{t}^{-1} \paren{ p_{t}^{\star} (\cdot) } $.
\begin{assumption} \label{assumption:r_neighbourhood}
    There exists some $\bar{r} > 0$ such that for every $t\in[T]$, it holds that 
    \begin{equation*}
    A_t(r) \leq 0, \quad \forall ~ r < \bar{r} . 
    \end{equation*}
    In addition, for every $t \in [T]$, we assume that  
    \begin{equation*}
        r^{\star}_{t} ( \inner{\bfx_t, \bftheta_{\star}^{Z_t}}  ) \leq \bar{r} 
        \quad \text{and} \quad  r^{\star}_{t} (   \inner{\bfx_t, \hat{\bftheta}_{t} }    ) \leq \bar{r} . 
    \end{equation*}
\end{assumption}
Intuitively, this reflects the practical observation that BCL yields are typically confined within a reasonable range.
Thus, the assumption ensures that, 
both the optimal response given the true average yield  $\inner{\bfx_t, \bftheta_{\star}^{Z_t}}$ and the optimal response given the estimated average yield $\inner{\bfx_t, \hat{\bftheta}_{t} }  $ are upper bounded,  making the problem well-behaved.

Finally, we make the following assumption on the distribution of the noise.

\begin{assumption} \label{assumption:normal_noise}
We assume that the noise in \eqref{eq:linear_bcl} satisfies $\epsilon_t \overset{\text{i.i.d.}}{\sim} \mathcal{N}(0, \sigma^2)  $.
Let $\delta \in (0,1)$ be some fixed (small) constant. 
We also assume that the following holds almost surely,
    \begin{equation} \label{eq:normal_noise_sigma_assumption}
         \sqrt{\frac{1}{ 2 \pi}} \frac{1}{\sigma} e^{- \frac{1}{2} (\frac{\delta}{\sigma})^2 } 
             > \max_{ 1 \leq t \leq T } -   \frac{ V_t ' (0) }{ V_t (0)  } . 
    \end{equation}
\end{assumption}
A noise distribution with finite variance is crucial for ensuring that a reasonable size of changes in pricing lead to successful trades to compete with BCL. 
Moreover, \eqref{eq:normal_noise_sigma_assumption} requires that the true $\sigma$ in \eqref{eq:linear_bcl} lies in a confined range determined by bond primitives. 

The necessity of both Assumption~\ref{assumption:r_neighbourhood} and Assumption~\ref{assumption:normal_noise} arises from the unique pricing problem formulated in \eqref{eq:per_round_reward_expected}.  
    Specifically, these two assumptions ensure that a good estimation of the true parameter $\bftheta_{\star}^{Z_t}$ leads to a quoted price $p_t$ being close to the optimal price $p_t^\star$. 
    These conditions are \textit{independent} of the statistical learning aspect of the problem, namely the estimation error of $ \bftheta_{\star}^{j} $.  
    In Appendix~\ref{subsec:discussion_assumption_of_thm}, we provide a detailed discussion on how to relax these assumptions to the greatest extent possible, while minimizing any compromises in terms of the theoretical guarantee.

Now we are ready to state the main result. To ease the exposition, in the following theorem, we only report the dependence of the regret on $T, M, d$ and $\deltamax$.  The precise statement is deferred to the beginning of Appendix~\ref{app:omitted_proofs}.  

\begin{theorem} \label{thm:regret_simple_version}
Under Assumptions \ref{assumption:diverse_contexts}-\ref{assumption:normal_noise},
Algorithm~\ref{algo:two_stage} ensures that
\begin{eqnarray} 
\regret \paren{T}  &\lesssim&  \min \Big\{ 
\underbrace{ \sqrt{d \log \paren{M d }}  \sqrt{ T } \log \paren{T}  \cdot  \sum_{j=1}^{M} \sqrt{\pi_j} \cdot \deltamax   +
 d \log \paren{M d} \log \paren{T}   \sum_{j=1}^{M} \sqrt{\pi_j} }_{\mathrm{Term } ~ (\RN{1}) } , ~~ \nonumber \\
 & & \quad \underbrace{ M d \log \paren{M d} \log \paren{T}  }_{ \mathrm{Term } ~ (\RN{2}) } , ~~
 \underbrace{ \deltamax^2 T  \log \paren{T}  +  d \log(d) \log \paren{T}  }_{\mathrm{Term}~ (\RN{3} )}
    \Big \}  + M d  
\label{eq:main_theorem_simple_version} .     
\end{eqnarray}
\end{theorem}
The proof of Theorem \ref{thm:regret_simple_version} is deferred to Section \ref{sec:main_proof}. We highlight that the crux of the proof hinges on characterizing a bound on the expected estimation error for the Stage II estimators, as detailed in Lemma~\ref{lemma:expectation_bound_stage_2}. Notably, this bound is {\it new} in the statistical estimation literature and is crucial for deriving a sublinear regret bound within the context of dynamic pricing.


\begin{remark}[The additive $\order{Md}$ term] \label{rmk:additive_term}
The last additive term in \eqref{eq:main_theorem_simple_version} corresponds to the regret incurred due to the coarse estimation of $\bftheta_\star^j$, if the sample covariance matrix is rank deficient. This happens, for example, in early episodes that are of shorter length.
\end{remark}

Applying Cauchy's inequality on Term (\RN{1}) in \eqref{eq:main_theorem_simple_version}, the factor $\sum_{j=1}^{M} \sqrt{\pi_j} $ can be bounded by $\sqrt{M}$.  
It turns out that we can refine the estimate of $ \sum_{j=1}^{M} \sqrt{\pi_j} $  if there is a certain structure in the arrival distribution.
To ease the notation, we order the arrival distribution so that $\pi_1 \geq \pi_2 \geq \cdots \geq \pi_M$.
We define the {\it decay rate of arrival distribution} 
as how fast the sequence $\braces{\pi_j}_{j=1}^{M}$ decays. 
\begin{corollary} \label{cor:distribution_decay}
 We consider two cases of decay rate of arrival distribution. 
\begin{enumerate}
\item Exponential decay: Suppose that there are some constants $\beta , C > 0$ such that $ 
    \pi_j \leq C e^{- \beta j }$.
     Then under the assumptions as in Theorem~\ref{thm:regret_simple_version}, Algorithm~\ref{algo:two_stage} ensures that 
     \begin{eqnarray}
    \mathrm{Term }~ (\RN{1}) \lesssim \frac{\sqrt{C}}{\beta} e^{ - \frac{1}{2} \beta} \paren{ \sqrt{d \log \paren{M d }}  \sqrt{ T } \log \paren{T}   \deltamax  + d \log \paren{M d} \log \paren{T}  } . \nonumber
     \end{eqnarray}
\item Polynomial decay: Suppose that there are some constants $\alpha,  C> 0$ such that $\pi_j \leq C j^{-\alpha}$.
     Then under the same assumptions as in Theorem~\ref{thm:regret_simple_version}, Algorithm~\ref{algo:two_stage} ensures that 
    \begin{align}
    \mathrm{Term }~ (\RN{1})
    &\lesssim \sqrt{C}
    \left\{
    \begin{array}{ll}
    \log(M), & \alpha = 2, \\
    \dfrac{1-M^{1-\alpha/2}}{\alpha-2}, & \alpha \neq 2
    \end{array}
    \right. \nonumber \\
    &\quad\times
    \paren{ \sqrt{d \log \paren{M d }}\sqrt{T}\log\paren{T}\deltamax
    + d \log \paren{M d}\log\paren{T} } . \nonumber
    \end{align}
\end{enumerate}
\end{corollary} 
Intuitively, the faster the arrival distribution decays, the more benign the environment is. 
This is because, effectively, we will observe fewer securities during the same number of rounds.

Before concluding this section, we provide two sets of detailed discussions: one comparing multitask learning with benchmark methods, and the other addressing the technical challenges encountered in the proof.

\subsection{Discussion I: multitask versus benchmarks}

There are two extreme scenarios on the spectrum of utilizing data points of other securities to accelerate learning.
One is the \textit{individual learning strategy}, i.e., we run an MLE  for each bond separately in every episode.
The other is the \textit{pooling strategy}, in which we pool all the data together and use the estimator in Stage I for all securities. Intuitively, the individual learning is better when securities are indeed very different from each other, and hence utilizing data points of other securities may only contaminate the learning process.  On the other hand, the pooling strategy is better when all the securities are close to each other. 
A desirable policy shall {\it match up} the performance of these two extremes even {\it without} the knowledge of whether (or how) the securities are similar to each other. 
 
\begin{figure}[H]
\centering
	\subfigure[The blue (orange) curve corresponds to Term~\RN{2} (Term~\RN{3}) in \eqref{eq:main_theorem_simple_version}, which characterizes the worst-case regret upper bound when $T \geq \Theta \paren{ \frac{M d \log \paren{Md} }{\deltamax^2} }$ ($T \leq \Theta \paren{ \frac{M d \log \paren{Md} }{\deltamax^2} }$). ]   
 {\includegraphics[height=4cm]{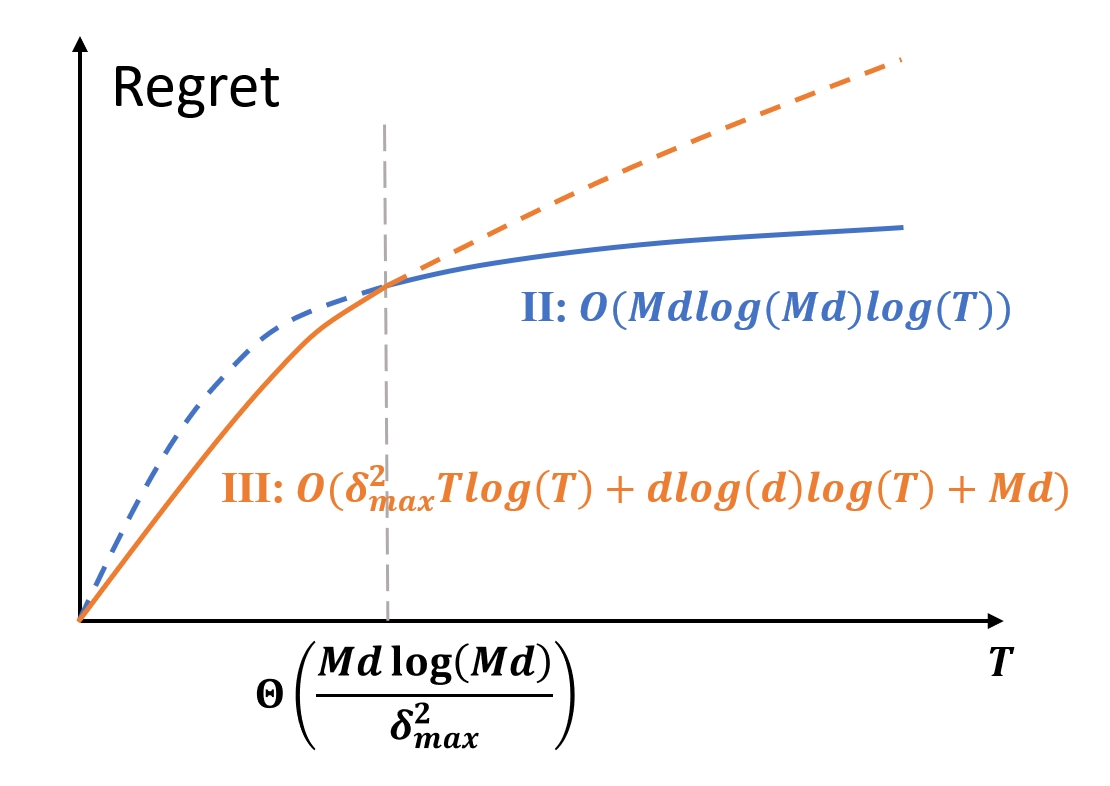}  \label{fig:regret_two_phases} 
 }%
 \hfill
	\subfigure[ 
 The two extreme cases when $\deltamax \rightarrow 0$ securities are essentially the same  and $\deltamax \rightarrow \infty$ securities are significantly different.
 ]   
 {\includegraphics[height=4cm]{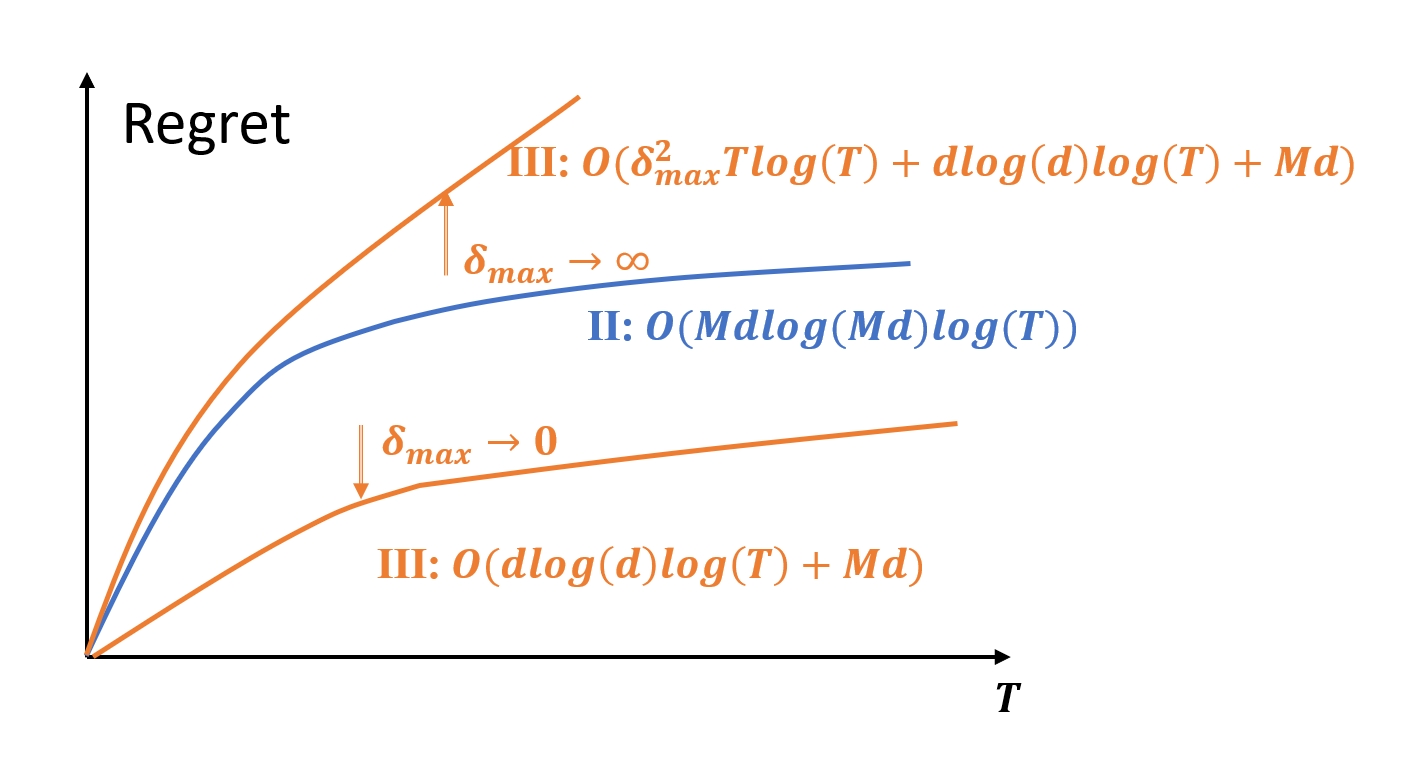} \label{fig:regret_two_extreme_cases}	}

\caption{ Algorithm~\ref{algo:two_stage} adaptively matches the performance of the pooling strategy and the individual learning strategy, without knowing $\deltamax$. 
 } 
\label{fig:uniform_group}
\end{figure}

Theorem~\ref{thm:regret_simple_version} shows that our design achieves a better regret of both extremes.
Indeed, Term (\RN{2}), which is linear in $M$, is comparable to the performance of individual learning \citep{javanmard2019dynamic}.  
Term (\RN{3}) is comparable to the performance of the pooling strategy.
We notice that one can give a coarse estimate of the factor $\sum_{j=1}^{M} \sqrt{\pi_j}$ by applying Cauchy's inequality, namely $\sum_{j=1}^{M} \sqrt{\pi_j}  \leq \sqrt{ M \sum_{j=1}^{M} \pi_j  } = \sqrt{M} $. Combining this observation with \eqref{eq:main_theorem_simple_version}, by straightforward calculation, we have 
\begingroup{}\setlength{\abovedisplayskip}{6pt}\setlength{\abovedisplayshortskip}{6pt}\setlength{\belowdisplayskip}{6pt}\setlength{\belowdisplayshortskip}{6pt}\begin{equation}  \label{eq:main_theorem_minimum_term}
     \regret \paren{T}  \lesssim 
\left\{
\begin{aligned}
&  \deltamax^2 T  \log \paren{T}  +  d \log(d) \log \paren{T}  
 + M d  
 & \text{if } T \leq \Theta \paren{ \frac{M d \log \paren{Md} }{\deltamax^2}  } \\
&  M d \log \paren{M d} \log \paren{T}  & \text{if } T \geq \Theta \paren{ \frac{M d \log \paren{Md}}{\deltamax^2} } 
\end{aligned}
\right. . 
\end{equation}
\endgroup{} Specifically, when $T \leq \Theta \paren{ \frac{M d \log \paren{Md} }{\deltamax^2} }$, Algorithm~\ref{algo:two_stage}  matches the performance of the pooling strategy, which we can think of as a fast learning period for warm-up. 
When $T$ gets larger, this advantage is diminishing. 
Conversely, when $T \geq \Theta \paren{ \frac{M d \log \paren{Md} }{\deltamax^2} }$, Algorithm~\ref{algo:two_stage}  aligns with the performance of the individual learning strategy, as shown in Figure~\ref{fig:regret_two_phases}.

We also note that, Term~(\RN{1}) never achieves the order-wise minimum among the three terms, due to the coarse estimation of the factor $\sum_{j=1}^{M} \sqrt{\pi_j}$.
In Corollary~\ref{cor:distribution_decay}, we further elaborate on this by identifying specific patterns in the arrival distribution that are either more benign or harder. 

The more similar the securities are to each other, the longer Term~(\RN{3}) will maintain an edge over Term~(\RN{2}).
Figure~\ref{fig:regret_two_extreme_cases} depicts 
two extreme cases when $\deltamax \rightarrow 0$ securities are essentially the same and when $\deltamax \rightarrow \infty$ securities are significantly different.
In the former case, Algorithm~\ref{algo:two_stage} is shown to enjoy the same worst-case performance as the pooling strategy; while in the latter case, our algorithm is as good as the individual strategy, which is desired.

\paragraph{Multi-task learning versus individual learning} 
As Term (\RN{2}) in \eqref{eq:main_theorem_simple_version} is comparable to the performance of individual learning, Theorem~\ref{thm:regret_simple_version} shows that our multi-task learning strategy automatically enjoys the performance guarantee of individual learning.
The power of multi-task learning boils down to a better estimation of coefficients. 
In episode $k$, let $N_{(k)}$ be the number of samples used for estimation of $\bar{\bftheta}_{(k)}$ and let $N^j_{(k)}$ be the number of samples used for estimation of $\hat{\bftheta}^{j}_{(k)}$.
We show in Lemma~\ref{lemma:stage_2} that, roughly speaking, the estimator $\hat{\bftheta}^{j}_{(k)}$ enjoys an estimation error bounded by 
\begin{equation*}
\min \braces{ \widetilde{\mathcal{O}} \paren{  \sqrt{\frac{d}{n^{j}_{(k)}}} \paren{ \frac{1}{N_{(k)}} \sum_{j=1}^{M} N_{(k)}^{j} \normtwo{\bfdelta_\star^j} + \sqrt{\frac{d}{N_{(k)}}}}   }  , ~  \widetilde{\mathcal{O}}  \paren{ \frac{d}{N^j_{(k)}} } }  , 
\end{equation*}
where the latter is also the estimation error that the individual learning estimator admits (based on $N^j_{(k)}$ number of samples of security $j$). 
Namely, when  securities are similar to each other and we do not have enough samples for security $j$, multi-task learning helps accelerate the learning of security $j$ compared to individual learning, by leveraging samples from other similar securities. 
As the precision of our estimator increases, there is also a diminishing benefit in leveraging the power of samples from other securities.

We delve deeper into the connection between the bound $\widetilde{\mathcal{O}} \paren{  \sqrt{\frac{d}{N^{j}_{(k)}}} \paren{ \deltamax + \sqrt{\frac{d}{N_{(k)}}}}   }$ and the two-stage estimation procedure in Algorithm~\ref{algo:two_stage}.  
Roughly speaking, the first stage produces an estimator for $\bftheta_\star$ with estimation error of order $\widetilde{\mathcal{O}} \paren{ \deltamax + \sqrt{\frac{d}{N_{(k)}}} } $.
The existence of the term $\deltamax$ is attributed by the heterogeneity among samples when we pool all securities together in the first stage. 
This estimation error will be mitigated by the refinement in the second stage. 



\paragraph{Multi-task learning versus pooling data}
 Term (\RN{3}) in \eqref{eq:main_theorem_simple_version} is comparable to the performance of the pooling strategy.
It suggests that the pooling strategy is expected to work well when securities are similar to each other ($\deltamax$ small) regardless of how many securities ($M$) there are. 
When all the securities are indeed the same, i.e. $\deltamax = 0$, such a strategy is natural and yields a desirable $\order{d \log(d) \log (T) } $ regret. 
Hence, Theorem~\ref{thm:regret_simple_version} shows that Algorithm~\ref{algo:two_stage} inherently aligns with the performance of the pooling strategy, even without the knowledge of $\deltamax$.  

\subsection{Discussion II: proof challenges}
The challenge of the proof of Theorem~\ref{thm:regret_simple_version} arises from several distinct perspectives. 
\begin{enumerate}
    \item From the modeling and optimization standpoint, we assume that the yield follows a linear contextual structure, and embed the yield-to-price function to address the pricing problem. While this modeling choice captures the essential structure of pricing credit market securities, it departs from the standard dynamic pricing framework, which typically assumes a linear relationship between price and context (see, e.g., \cite{javanmard2019dynamic, cohen2020feature, bu2022context}). As a result, our novel framework introduces a more complex optimization landscape, making it difficult to control the regret by the statistical error (i.e., Lemma~\ref{lemma:perround_regret_to_estimation_error}). 
    \item From the statistical perspective, we need to show that the TSMT {\it automatically outperforms} both the pooling strategy and the individual learning strategy in terms of the statistical error of the coefficients (i.e., Lemma~\ref{lemma:expectation_bound_stage_2}), even without knowing the similarity structure. 
    Hence, this bound is required to have three components that effectively reflect the comparable performances of both the pooling strategy and the individual learning strategy, in addition to what is unique to the multi-task learning strategy (i.e., Term (I) in \eqref{eq:main_theorem_simple_version}).
    \item Another challenge is inherent to the {\it online nature} of the problem, which introduces randomness in security arrivals and, consequently, makes estimation error harder to control. 
    To address this, we must carefully take into consideration these random events. Specifically, we demonstrate that:
    \begin{itemize}
        \item When the empirical frequency of the securities is sufficiently close to the nominal arrival probabilities $\pi_j$'s, the regret contributed by the estimation errors weighted by the arrival distribution scales sublinearly in $T$. 
        Securities that arrive more frequently have more refined estimations due to the law of large number. 
        Conversely, securities that arrive less frequently, despite having higher estimation errors, do not significantly contribute to the overall regret, as the chances of encountering them are relatively small.
        \item The events, when the empirical frequency of the securities is sufficiently far away from the nominal arrival probabilities $\pi_j$'s, are unlikely to happen.
        Even when summed over time, the probabilities of these events are small. 
    \end{itemize}
\end{enumerate}

\begin{remark}[Connections with literature] \label{rmk:technical_diff}
Viewing the dynamic pricing problem with linear structural similarity \eqref{eq:coefficient_structure} as an offline problem, modulo the yield-to-price structure, the feedback mechanism (censored versus uncensored) and dependency between the observations, one may apply algorithms developed in e.g. \cite{duan2022adaptive, tian2023learning}. 
However, our analysis is customized to overcome the challenges unique to the online learning framework. 
For example, should we directly apply the results from \cite{duan2022adaptive}, the estimation error of security $j$ in each episode would be of order $  \order{ \frac{d}{n} + k_w^2 \min \braces{ \delta_{\max}^2, \frac{d + \log M }{n_j} } } $, where $k_w = \frac{\max_{i \in [M]  } \sqrt{n_i} (\sum_{j=1}^{M} \sqrt{n_j}) }{n}$, and $n_j$ is the number of samples of security $j$, $n$ is the total samples in this episode. However, the order of $k_w$ depends on the security arrival distribution $\bfpi$, which for instance, can be easily as large as $\order{M^{1/2}}$. 
\cite{tian2023learning} consider equal sizes of samples for all the tasks, making it impractical for our settings nor for real-world applications.
Furthermore, a direct application of Theorem 1 therein yields worse dependence in the extreme case where $\deltamax=0$. 
\end{remark}



\input{sifin/numerical_experiments}\section{Proof of the Main Theorem 
}\label{sec:main_proof}
This section is devoted to the proof of Theorem~\ref{thm:regret_simple_version}. For some intermediate results, we further defer the proofs to Appendix~\ref{app:omitted_proofs}.\\

To set the stage for the analysis, we introduce some notations. 
Recall $f$ and $F$ are the $p.d.f.$ and $c.d.f.$ of the noise respectively. 
Given realizations of $\bfx_t, y_t, p_t$, 
we define $ \xi_t \paren{ \bftheta  } $ and $ \eta_t \paren{ \bftheta  } $ to be the following (deterministic) quantities related to gradient vector and  negative Hessian matrix of the likelihood function $\ell_t$ with respect to the variable $\bftheta$:
\begin{equation*}
\xi_t \paren{ \bftheta }  \defeq   \log ' \paren{ \bar{F} \paren{ V_{t}^{-1}(p_t) - \inner{ \bftheta  , \bfx_t }}}  \one{V_t(y_t) < p_t} 
+ \log ' \paren{ f( y_t -  \inner{ \bftheta   , \bfx_t } ) }  \one{ V_t(y_t) \geq p_t} , 
\end{equation*}
\begin{equation*}
 \eta_t \paren{ \bftheta  }  \defeq  
  - \log '' \paren{ \bar{F} \paren{ V_{t}^{-1}(p_t) - \inner{ \bftheta  , \bfx_t }}}  \one{V_t(y_t) < p_t} 
  - \log '' \paren{ f( y_t -  \inner{ \bftheta   , \bfx_t } ) }  \one{ V_t(y_t) \geq p_t}  . 
\end{equation*}
For any $\bftheta_1$ and $\bftheta_2$, there exists some $\tilde{\bftheta}$ which lies on the segment connecting both such that 
\begin{eqnarray}
   \abs{ \xi_t \paren{ \bftheta_1 } - \xi_t \paren{ \bftheta_2 } }  &\leq& 
   \normtwo{ \nabla_{\bftheta} \xi_t \paren{ \tilde{\bftheta}} } \normtwo{ \bftheta_1  - \bftheta_2 }   \nonumber  \\  
   &\leq& \braces{  \max_{ \abs{x} \leq \bar{y} + W  \xmax } \abs{  \log '' \paren{ \bar{F} \paren{ x} } }  +  \max_{ \abs{x} \leq \bar{y} + W  \xmax }  \abs{ \log '' \paren{ f( x ) }  } }  
   \xmax   \normtwo{ \bftheta_1  - \bftheta_2 } \nonumber   \\
   &=& L_F  \normtwo{ \bftheta_1  - \bftheta_2 }  , \label{eq:lip_xi}
\end{eqnarray}
for some absolute constant $L_F > 0$ due to log-concavity of the noise distribution $F$.
We recall that $\bar{y}$ and $\xmax$ are the upper bound of a reasonable yield and the upper bound of norm of the context, respectively. 
{ 
Recall \eqref{eq:def_u_F} that $u_F $ is the maximum possible value of $\xi_t \paren{ \bftheta  } $ under $\abs{x} \leq  \bar{y} + W \xmax$. 
}
Similarly, we define the minimum possible value of $\eta_t \paren{ \bftheta  } $ under our range of consideration as 
\begin{equation} \label{eq:def_ell_F}
\ell_F \defeq \min_{ \abs{x} \leq  \bar{y} + W \xmax } \braces{  \min \braces{ - \log '' \paren{ \bar{F} \paren{x} } ,~  - \log'' \paren{  f \paren{x} }  }  }    . 
\end{equation}
We recall that a log-concave density also implies a log-concave cumulative function. 
The strict log-concavity of $f$ and $F$ guarantees that $\ell_F>0$.

    


The core of the proof lies in the estimation error bound on our two-stage estimators. 
Hereafter, we present Lemmas~\ref{lemma:stage_1_common}--\ref{lemma:expectation_bound_stage_2}. The two of them are deterministic results for the Stage I estimator $\bar{\bftheta}_{(k)}$ and the Stage II estimators $\hat{\bftheta}_{(k)}^{j}$.
The other two of them translate the deterministic bounds to expectation bounds for the Stage I estimator and Stage II estimators, respectively.

Recall that $\epsilon_t$ denotes the noise in \eqref{eq:linear_bcl}. 
In our policy, the price $p_t$ is a function of the current context $\bfx_t$ and the samples observed in the \textit{previous} episode, not the current episode. 
To simplify the notation, we omit the episode index $k$ in the statements of the following four lemmas. When applying these lemmas to episode $k$, note that $t=1$ refers to the start index of the episode, and $t=n$ refers to the end index of the episode.
Furthermore, we denote $\hat{\bfSigma} \paren{n} = \frac{1}{n} \sum_{t=1}^{n}  \bfx_t \bfx_t^\top  $, 
$n^j = \sum_{t=1}^{n} \one{Z_t = j} $ and $\hat{\bfSigma}^{j} \paren{n^j} = \frac{1}{n^j} \sum_{t=1}^{n} \one{Z_t=j} \bfx_t \bfx_t^\top$. 


\begin{lemma}[Stage I Estimation Error]  \label{lemma:stage_1_common}
Let $\mathcal{H}_n = \braces{ Z_t, \bfx_t }_{t=1}^{n}$ and assume that $\lambda_{\min} \paren{\hat{\bfSigma}}  >0$.
Define $\bar{\calL} \paren{\bftheta} = - \frac{1}{n} \sum_{t=1}^{n}  \ell_t \paren{\bftheta; p_t, y_t, \bfx_t}$ and suppose $p_t$ is independent of $\{\epsilon_s\}_{s=1}^{n}$. 
Let $\bar{\bftheta}$ be the solution to the problem $\bar{\bftheta} = \argmin_{\bftheta \in \mathbb{R}^d} \bar{\calL} \paren{\bftheta}  $.  
Then, it holds almost surely that 
\begin{equation} \label{eq:theta_bar_deterministic_bound}
     \normtwo{ \bftheta_\star - \bar{\bftheta} } \leq 
     \frac{L_F \xmax}{\ell_F \lambda_{\min} \paren{\hat{\bfSigma}}} 
     \paren{ \frac{1}{n}  \sum_{j=1}^{M} n^j \normtwo{ \bfdelta_\star^j }  +   
     \normtwo{  \frac{1}{n}  \sum_{j=1}^{M} \sum_{t=1}^{n} \one{Z_t=j} \xi_t \paren{ \bftheta^j_\star } \bfx_t    } 
     }  . 
\end{equation} 

\end{lemma}
The proof of Lemma \ref{lemma:stage_1_common} is deferred to Appendix~\ref{appendix:statistical_perspective}.

\begin{lemma}[Stage I Expectation Bound] \label{lemma:expectation_bound_stage_1}
Under the assumptions of Lemma~\ref{lemma:stage_1_common},
we have 
\begin{equation}
   \E{  \normtwo{  \bftheta_\star -   \bar{\bftheta} } \Big|\, \mathcal{H}_n }  
\leq  \frac{L_F \xmax}{\ell_F \lambda_{\min} \paren{\hat{\bfSigma}}} 
     \paren{   \frac{1}{n}  \sum_{j=1}^{M} n^j \normtwo{ \bfdelta_\star^j }  
     + 3  \sqrt{\frac{8 u_F^2 \bar{x}^2 d \log \paren{ 2 d^2 }}{n}} } ,
 \nonumber  
\end{equation}
and 
\begin{equation} 
    \E{  \normtwo{  \bftheta_\star -  \bar{\bftheta} }^2  \Big|\, \mathcal{H}_n   }  
     \lesssim  \paren{ \frac{L_F \xmax}{\ell_F \lambda_{\min} \paren{\hat{\bfSigma}}  } }^2  
     \paren{   \paren{ \frac{1}{n}  \sum_{j=1}^{M} n^j \normtwo{ \bfdelta_\star^j } }^2   
     +  \frac{u_F^2 \bar{x}^2 d \log \paren{ d^2 }}{n}  }  \nonumber  . 
\end{equation}
\end{lemma}
The proof of Lemma~\ref{lemma:expectation_bound_stage_1} is standard and deferred to Appendix~\ref{appendix:statistical_perspective}.

\begin{lemma}[Stage II Estimation Error] \label{lemma:stage_2}
Let $\mathcal{H}_n=\braces{Z_t,\bfx_t}_{t=1}^{n}$. Assume that
\begin{equation*}
\lambda_{\min}\paren{\hat{\bfSigma}^{j}\paren{n^j}}>0.
\end{equation*} 
Let $\calL^j \paren{\bftheta} =  - \frac{1}{n^j}  \sum_{t= 1 }^{n} \one{Z_t=j} \ell_t \paren{\bftheta ; p_t, y_t, \bfx_t}$. 
Suppose $p_t$ is independent of $\braces{\epsilon_s}_{s=1}^{n}$. 
    Let $\hat{\bftheta}^j $ be the solution to the following regularized problem 
    \begin{equation} \label{eq:stage_2_problem}
       \hat{\bftheta}^j  = \argmin_{  \bftheta \in \mathbb{R}^d  } ~ \calL^j \paren{\bftheta}   + \lambda^j  \normtwo{ \bftheta - \bar{\bftheta}_{}  }  .  
    \end{equation}
We define
\begin{eqnarray*}
\textnormal{Term (I)} &=& \frac{\normtwo{\nabla_{\bftheta}\calL^{j}\paren{\bftheta^j_\star}}-\lambda^j}{\ell_F\lambda_{\min}\paren{\hat{\bfSigma}^{j}\paren{n^j}}}
+\frac{\sqrt{\lambda^j\paren{\normtwo{\bfdelta_\star^j}+\normtwo{\bftheta_\star-\bar{\bftheta}}}}}{\sqrt{\ell_F\lambda_{\min}\paren{\hat{\bfSigma}^{j}\paren{n^j}}}},\\
\textnormal{Term (II)} &=& \frac{\lambda^j+L_F\xmax\normtwo{\nabla\calL^j\paren{\bftheta_\star^j}}}{\ell_F\lambda_{\min}\paren{\hat{\bfSigma}^{j}\paren{n^j}}},\\
\textnormal{Term (III)} &=& \frac{\normtwo{\nabla\calL^j\paren{\bftheta_\star^j}}-\lambda^j+L_F\xmax\paren{\normtwo{\bfdelta_\star^j}+\normtwo{\bftheta_\star-\bar{\bftheta}}}}{\ell_F\lambda_{\min}\paren{\hat{\bfSigma}^{j}\paren{n^j}}}.
\end{eqnarray*}
It holds almost surely that 
\begin{equation}  \label{eq:lemma_stage_2}
    \normtwo{ \hat{\bftheta}^j - \bftheta_\star^j }  
    \lesssim  \braces{ \textnormal{Term (I)}, \textnormal{Term (II)} , \textnormal{Term (III)} }  
\end{equation}
for all $j \in [M]$.

\end{lemma}

The proof of Lemma \ref{lemma:stage_2} is deferred to Appendix~\ref{appendix:statistical_perspective}.

\begin{lemma}[Stage II Expectation Bound]\label{lemma:expectation_bound_stage_2}
Under the assumptions of Lemma~\ref{lemma:stage_1_common} and Lemma~\ref{lemma:stage_2}, by setting $\lambda^j = \sqrt{\frac{8 u_F^2 \bar{x}^2 d \log \paren{\frac{2d}{\delta}}}{n^j}}$, 
the output of \eqref{eq:stage_2_problem}  
satisfies 
\begin{equation}
     \E{ \normtwo{ \hat{\bftheta}^j - \bftheta_\star^j }^2 ~\big|~ \calH_n } \lesssim \min  \braces{  \textnormal{Term (I)},  \textnormal{Term (II)} ,  \textnormal{Term (III)} }
\end{equation}
for all $j \in [M]$, where
\begin{eqnarray*}
\textnormal{Term (I)} &=& \frac{(\lambda^j)^2\delta}{\ell_F^2\lambda_{\min}^2\paren{\hat{\bfSigma}^{j}\paren{n^j}}\log\paren{2d/\delta}}
+\frac{\lambda^j\paren{\normtwo{\bfdelta_\star^j}+\E{\normtwo{\bftheta_\star-\bar{\bftheta}}\mid\calH_n}}}{\ell_F\lambda_{\min}\paren{\hat{\bfSigma}^{j}\paren{n^j}}},\\
\textnormal{Term (II)} &=& \paren{\frac{L_F\xmax}{\ell_F\lambda_{\min}\paren{\hat{\bfSigma}^{j}\paren{n^j}}}}^2(\lambda^j)^2\paren{1+\delta+\frac{\delta}{\log\paren{2d/\delta}}},\\
\textnormal{Term (III)} &=& \paren{\frac{L_F\xmax}{\ell_F\lambda_{\min}\paren{\hat{\bfSigma}^{j}\paren{n^j}}}}^2
\paren{\frac{(\lambda^j)^2\delta}{\log\paren{2d/\delta}}+\normtwo{\bfdelta_\star^j}^2+\E{\normtwo{\bftheta_\star-\bar{\bftheta}}^2\mid\calH_n}}.
\end{eqnarray*}


\end{lemma}


Now we are well-prepared to prove Theorem~\ref{thm:regret_simple_version}. 

The length of the $k$th period is $\tau_k$. 
There are at most $\lceil \log_2 T \rceil$ episodes. 
We denote by $N_{(k)}^{j}$ the number of samples of security $j$ \textit{used} for the estimation in the $k$th episode. 
By design of the algorithm, the estimates are updated only at the beginning of each episode and only by using the samples from the previous episode. Therefore, the total number of samples used for estimates in episode $k$ is $ \sum_{j=1}^{M} N_{(k)}^{j} = \frac{1}{2} \tau_k$. 
We define $\mathcal{N}_k^j$ to be the event that security $j$ show up more frequently than half of the expected arrivals during the $k$th episode, namely 
\begin{equation*}
\mathcal{N}_k^j \defeq \braces{  N_{(k)}^{j} \geq \frac{1}{2} \cdot  \frac{1}{2 }\tau_k \cdot \pi_j}.
\end{equation*}
In addition, we let $\mathcal{E}_k^j$ be the event that the minimum eigenvalue of sample covariance matrices is larger than half of its expected value in the $k$th period, namely 
\begin{equation*}
\mathcal{E}_k^j \defeq  \braces{ \lambda_{\min} \paren{\hat{\bfSigma}^j \paren{ N_{(k)}^{j} }  }  \geq \frac{1}{2} \lambda_{\min} \paren{  \bfSigma^j }    }  . 
\end{equation*}
Likewise, we define $\calE_{k}^{\circ}$ to be the event that the minimum eigenvalue of the aggregate sample covariance matrix is larger than half of its expected value during the $k$th period, namely 
\begin{equation*}
\mathcal{E}^{\circ}_{k} \defeq \braces{ \lambda_{\min} \paren{ \hat{ \bfSigma } \paren{ \frac{1}{2} \tau_k  } } \geq \frac{1}{2} \lambda_{\min} \paren{\bfSigma} } .
\end{equation*}

Denote $ \regt \defeq \operatorname{reward}_t \paren{p_t^\star} - \operatorname{reward}_t \paren{p_t}$.
Let $\hat{\bftheta}_t$ denote the estimator used at time $t$. 
We proceed by breaking down the expected regret over the $k$th episode into various events:
\begin{eqnarray}
  \E{\regret \paren{ k \text{th episode}} } 
&=& \sum_{t = \tau_k }^{\tau_{k+1} - 1}  \E{  \regt }
=  \sum_{t = \tau_k }^{\tau_{k+1} - 1}  
\E{  \regt ; \calE_{k}^{\circ} } 
+ \sum_{t = \tau_k }^{\tau_{k+1} - 1} 
\E{  \regt ; \paren{  \calE_{k}^{\circ} }^\complement } \nonumber  \\ 
&\lesssim& \overline{\lambda}  \sum_{t = \tau_k }^{\tau_{k+1} - 1}  
  \E{ \normtwo{ \bftheta_\star^{Z_t} -  \hat{\bftheta}_t  }^2 ;  \calE_{k}^{\circ}  }
+ \overline{\lambda}  \sum_{t = \tau_k }^{\tau_{k+1} - 1} 
\E{  \normtwo{ \bftheta_\star^{Z_t} -  \hat{\bftheta}_t  }^2 ; \paren{   \calE_{k}^{\circ} }^\complement } .  \label{eq:bottleneck_step}
\end{eqnarray}
Recall $\overline{\lambda}$ is defined to be an upper bound of the the largest eigenvalue of the contexts' covariance matrix. 
The inequality follows from the pricing rule \eqref{eq:algo_pricing_formula} and 
Lemma~\ref{lemma:perround_regret_to_estimation_error}.
Both Assumption~\ref{assumption:r_neighbourhood} and Assumption~\ref{assumption:normal_noise} are imposed for deriving \eqref{eq:bottleneck_step}. 
We defer the proof to Appendix~\ref{appendix_sec:dynamic_pricing}. 

Next, we can further decompose the per-round estimation error  
\begin{eqnarray*}
& & \E{  \normtwo{ \bftheta_\star^{Z_t} -  \hat{\bftheta}_t  }^2 ; \calE_{k}^{\circ} }   
=
  \sum_{j=1}^{M} \pi_j \E{  \normtwo{ \bftheta_\star^{j} -  \hat{\bftheta}_t  }^2 ;    \calE_{k}^{\circ} ~\Bigg|~ Z_t=j }  \\ 
&=& \sum_{j=1}^{M} \pi_j \E{  \normtwo{ \bftheta_\star^{j} -  \hat{\bftheta}_t  }^2 ;    \calE_{k}^{\circ} \cap \calN_k^j  \cap \calE_k^j  } 
+  \sum_{j=1}^{M} \pi_j  \E{  \normtwo{ \bftheta_\star^{j} -  \hat{\bftheta}_t  }^2 ;  \calE_{k}^{\circ}  \cap \paren{ \calN_k^j  \cap \calE_k^j }^\complement }  . 
\end{eqnarray*}
The last step follows from the independence of the arrival of securities. 
Therefore, plugging this into \eqref{eq:bottleneck_step} yields that the total expected regret is bounded by 
\begin{eqnarray}
   & &  \sum_{k=1}^{\lceil \log_2 T \rceil} \E{\regret \paren{ k \text{th episode}} }  \nonumber \\ 
   &\lesssim&  \overline{\lambda}  \sum_{k=1}^{\lceil \log_2 T \rceil} \sum_{t = \tau_k }^{\tau_{k+1} - 1} 
   \sum_{j=1}^{M} \pi_j \E{  \normtwo{ \bftheta_\star^{j} -  \operatorname{Proj}_{\mathcal{B}(W)} \paren{ \hat{\bftheta}^{j}_{(k)} }  }^2  ~\big|~  \calE_{k}^{\circ} \cap \calN_k^j  \cap \calE_k^j  }   \nonumber \\
& & +   \overline{\lambda} W^2   
\sum_{k=1}^{\lceil \log_2 T \rceil} \sum_{t = \tau_k }^{\tau_{k+1} - 1}  \paren{ 
\pr{ \paren{ \calE_{k}^{\circ} }^\complement   }  
+ \sum_{j=1}^{M} \pi_j  
\pr{    \calE_{k}^{\circ} \cap \paren{ \calN_k^j  \cap \calE_k^j }^\complement }
}  \label{eq:phase_regret_decomposition}  . 
\end{eqnarray}
The inequality holds since $\hat{\bftheta}^{j}_{(k)}$ is projected back to $\mathcal{B}(W)$, hence bounded (c.f. \eqref{eq:algo_pricing_formula}) by the design of the algorithm.

\allowdisplaybreaks
In what follows, we study the two sums in \eqref{eq:phase_regret_decomposition} respectively.
\begin{enumerate}
    \item 
Recall  $\lambda^{j}_{(k)} = \sqrt{\frac{8 u_F^2 \bar{x}^2 d \log \paren{\frac{2d}{\delta}}}{N^{j}_{(k)}}}$ with $\delta = \frac{1}{Md}$. 
As for the first term in \eqref{eq:phase_regret_decomposition}, we make several observations in the sequel. 
{\setlength{\jot}{1pt}
\begin{align}
&\quad \overline{\lambda} \sum_{t = \tau_k }^{\tau_{k+1} - 1} \sum_{j=1}^{M} \pi_j
 \E{ \normtwo{ \bftheta_\star^{j} -      \operatorname{Proj}_{\mathcal{B}(W)} \paren{ \hat{\bftheta}^{j}_{(k)} }  }^2  ~\Bigg|~   \calE_{k}^{\circ} \cap \calN_k^j  \cap \calE_k^j   }   \nonumber \\ 
 &\stackrel{\rm (a)}{\leq} \overline{\lambda} \sum_{t = \tau_k }^{\tau_{k+1} - 1} \sum_{j=1}^{M} \pi_j
 \E{ \normtwo{ \bftheta_\star^j -   \hat{\bftheta}^{j}_{(k)}   }^2  ~\Bigg|~   \calE_{k}^{\circ} \cap \calN_k^j  \cap \calE_k^j   }   \nonumber  \\ 
&\stackrel{\rm (b)}{\lesssim}   
\overline{\lambda} \sum_{t = \tau_k }^{\tau_{k+1} - 1}   \sum_{j=1}^{M} \pi_j 
\min \Bigg \{ \frac{8 u_F^2 \bar{x}^2 d \log \paren{\frac{2d}{\delta}}}{ \frac{1}{4}  \pi_j \tau_k }  
 \frac{ 2 }{ \ell_F^2  \lambda^2_{\min} \paren{ \bfSigma^{j}  }}  
\delta \frac{1}{\log \paren{\frac{2 d}{\delta}}}  \nonumber \\ 
&\quad 
 + \frac{ 2 }{ \ell_F \lambda_{\min} \paren{ \bfSigma^j }}  \sqrt{\frac{8 u_F^2 \bar{x}^2 d \log \paren{\frac{2d}{\delta}}}{ \frac{1}{4}  \pi_j \tau_k }}  \paren{  \normtwo{ \bfdelta_\star^j } 
 + \E{ \normtwo{ \bftheta_\star - \bar{\bftheta}_{(k)} } ~\big |~  \calE_{k}^{\circ}   }  }   , \nonumber  \\
 &\quad \paren{ \frac{2 L_F \xmax}{\ell_F  \lambda_{\min} \paren{ \bfSigma^j } }  }^2 \frac{8 u_F^2 \bar{x}^2 d \log \paren{\frac{2d}{\delta}}}{ \frac{1}{4}  \pi_j \tau_k }  , \nonumber  \\  
 &\quad \paren{ \frac{2 L_F \xmax}{\ell_F  \lambda_{\min} \paren{ \bfSigma^j } }  }^2  \paren{ \frac{8 u_F^2 \bar{x}^2 d \log \paren{\frac{2d}{\delta}}}{ \frac{1}{4}  \pi_j \tau_k }  \delta \frac{1}{\log \paren{\frac{2 d}{\delta}}}  
 +  \normtwo{\bfdelta_\star^j}^2 + \E{ \normtwo{ \bftheta_\star - \bar{\bftheta}_{(k)} }^2  ~\big|~  \calE_{k}^{\circ} }   }  \Bigg \}  \nonumber  \\ 
 &\stackrel{\rm (c)}{\lesssim} 
 \overline{\lambda}  \sum_{j=1}^{M}  \pi_j \tau_k \cdot 
 \min \Bigg \{ \frac{ u_F^2 \bar{x}^2 d }{ \pi_j \tau_k } \frac{1}{\ell_F^2\underline{\lambda}^2}\frac{1}{dM} \nonumber \\
 &\quad + \frac{L_F\xmax}{\ell_F^2\underline{\lambda}^2}\sqrt{\frac{u_F^2\bar{x}^2d\log\paren{dM}}{\pi_j\tau_k}}
 \paren{\deltamax+\sqrt{\frac{u_F^2\bar{x}^2d\log\paren d}{\tau_k}}}, \nonumber \\
 &\quad \paren{  \frac{L_F \xmax}{\ell_F  \underline{\lambda}  }  }^2 \frac{ u_F^2 \bar{x}^2 d \log \paren{d M }}{  \pi_j \tau_k }  ,  \nonumber \\ 
 &\quad    \paren{  \frac{L_F \xmax}{\ell_F  \underline{\lambda}  }  }^2  
 \paren{ \frac{ u_F^2 \bar{x}^2 d }{   \pi_j \tau_k } \frac{1}{d M }    
 +  \delta_{\max}^2 + \paren{ \frac{L_F \xmax}{ \ell_F \underline{\lambda} }  }^2 \paren{ 
 \delta_{\max}^2 + \frac{u_F^2 \bar{x}^2 d \log(d) }{ \tau_k } }   }
     \Bigg \} \nonumber  \\
 &\lesssim \min \Bigg\{
 \frac{u_F^2\overline{\lambda}}{\ell_F^2\underline{\lambda}^2}
 +\frac{L_F\xmax u_F\overline{\lambda}}{\ell_F^2\underline{\lambda}^2}\sqrt{d\log\paren{dM}}\sqrt{\tau_k}\sum_{j=1}^{M}\sqrt{\pi_j}\deltamax, \nonumber\\
 &\quad \frac{L_F\xmax u_F^2\overline{\lambda}}{\ell_F^2\underline{\lambda}^2}d\log\paren{dM}\sum_{j=1}^{M}\sqrt{\pi_j}, \qquad
 \frac{L_F^2\xmax^2u_F^2\overline{\lambda}}{\ell_F^2\underline{\lambda}^2}Md\log\paren{dM}, \nonumber\\
 &\quad \frac{L_F^2\xmax^2u_F^2\overline{\lambda}}{\ell_F^2\underline{\lambda}^2}
 +\frac{L_F^4\xmax^4\overline{\lambda}}{\ell_F^4\underline{\lambda}^4}\tau_k\delta_{\max}^2
 +\frac{L_F^2\xmax^2\overline{\lambda}}{\ell_F^2\underline{\lambda}^2}\tau_k\delta_{\max}^2
 +\frac{L_F^4\xmax^4u_F^2\overline{\lambda}}{\ell_F^4\underline{\lambda}^4}d\log(d)\Bigg\}.
 \label{eq:phase_regret_main_term_bound}
\end{align}
}
Here, (a) holds since the projection to a convex set is a non-expansive mapping and securities arrive in an $i.i.d.$ fashion;
(b) follows from Lemma~\ref{lemma:expectation_bound_stage_2} by noting that $N_{(k)}^{j} \geq \frac{1}{2} \cdot  \frac{1}{2 }\tau_k \cdot \pi_j $ on event $\calN_{k}^{j}$;
(c) is due to Lemma~\ref{lemma:expectation_bound_stage_1}. 

\item 
The second sum in \eqref{eq:phase_regret_decomposition} corresponds to the regret incurred when we do not have precise estimates. 
In what follows, we show that the accumulated regret of this kind over the entire $T$ periods can be controlled by a quantity independent of $T$. 

We first note that 
\begin{eqnarray}
& & \sum_{k=1}^{\lceil \log_2 T \rceil} \sum_{t = \tau_k }^{\tau_{k+1} - 1}  
\paren{ 
\pr{ \paren{  \calE_{k}^{\circ} }^\complement   }  
+ \sum_{j=1}^{M} \pi_j  
\pr{     \calE_{k}^{\circ} \cap \paren{ \calN_k^j  \cap \calE_k^j }^\complement   }   } 
   \nonumber \\ 
&\leq&  \sum_{k=1}^{\lceil \log_2 T \rceil} \sum_{t = \tau_k }^{\tau_{k+1} - 1}  \paren{ 
\pr{  \paren{ \calE_{k}^{\circ} }^\complement } + \sum_{j=1}^{M} \pi_j \pr{ \paren{ \calN_k^j  \cap \calE_k^j }^\complement  } } 
.  \label{eq:bad_prob_main}
\end{eqnarray}
We study the above two terms respectively.

\begin{itemize}
    \item 
For the first term in \eqref{eq:bad_prob_main}, a direct application of Lemma~\ref{lemma:matrix_chernoff} yields that 
\begin{equation*}
\pr{ {\calE_{k}^{\circ}}^\complement } \leq d \cdot \paren{ \sqrt{\frac{e}{2}}  }^{ - \lambda_{\min}\paren{\bfSigma}\frac{    \tau_k  }{2\xmax^2}  }  . 
\end{equation*}

To proceed, we observe that summing probabilities exponentially small in the length of the current period yields a quantity that is independent of $T$. Let $ \rho = \frac{ \paren{\frac{1}{2}}^{\frac{1}{2}} }{ e^{-\frac{1}{2}}  } = \sqrt{\frac{e}{2}}$. \enlargethispage{6\baselineskip}
Namely, for $\alpha>0$,

\begin{eqnarray}
 & &    \sum_{k=1}^{\lceil \log_2 T \rceil} 
     \sum_{t = \tau_k }^{\tau_{k+1} - 1}  \rho^{ - \alpha \tau_k  }  
=  \sum_{k=1}^{\lceil \log_2 T \rceil} 
     \sum_{t = \tau_k }^{\tau_{k+1} - 1}  \rho^{ - \frac{1}{2}  \alpha 2 \tau_k  }  
\leq \sum_{k=1}^{\lceil \log_2 T \rceil} 
     \sum_{t = \tau_k }^{\tau_{k+1} - 1}  \rho^{ - \frac{1}{2}  \alpha t  }   \nonumber \\
&\leq& \int_{0}^{T} \rho^{  - \frac{1}{2}  \alpha t  } \dd t 
\leq \frac{2}{\alpha \log(\rho) }  .  \label{eq:sum_trick}
\end{eqnarray}
In the last inequality, we used the fact that 
$ \int_{ \tau }^{\infty} \rho^{ - \alpha t }  \dd t  = \frac{1}{\alpha \log(\rho) } \rho^{-  \alpha \tau  } $  for $\rho>1, \alpha > 0$.




Therefore, we conclude that 
\begin{equation}
     \sum_{k=1}^{\lceil \log_2 T \rceil} \sum_{t = \tau_k }^{\tau_{k+1} - 1}  \pr{ { \calE_{k}^{\circ} }^\complement }  
     \lesssim d \frac{\xmax^2}{ \underline{\lambda} }  . 
\end{equation}
\item We treat the second term in  \eqref{eq:bad_prob_main} in a slightly more complicated but similar fashion.  
By straightforward calculation and subadditivity of probability measure, we have  
\begin{eqnarray}
    && \sum_{j=1}^{M} \pi_j  \pr{ \paren{ \calN_k^j  \cap \calE_k^j }^\complement  } = \sum_{j=1}^{M} \pi_j  \pr{ \paren{ \calN_k^j }^\complement \cup \paren{ \calE_k^j }^\complement  }   \nonumber  \\
    &\leq& \sum_{j=1}^{M} \pi_j \paren{  \pr{ { \calN_k^j }^\complement } + \pr{ { \calE_k^j }^\complement  }  }  \nonumber  \\
    &=& \sum_{j=1}^{M} \pi_j \paren{  \pr{ { \calN_k^j }^\complement } 
    + \pr{ \calN_k^j \cap {\calE_k^j}^\complement   } 
    + \pr{ { \calN_k^j }^\complement \cap { \calE_k^j }^\complement   }  }  \nonumber \\ 
    &\leq& \sum_{j=1}^{M} \pi_j \paren{  2 \pr{ { \calN_k^j }^\complement } 
    + \pr{ \calN_k^j \cap {\calE_k^j}^\complement   }   
     }  .  \label{eq:residual_prob}
\end{eqnarray}
We bound the above two terms in \eqref{eq:residual_prob} in the sequel:
\begin{enumerate}
    \item  First, applying Lemma~\ref{lemma:multinomial_2} yields that 
\[
\begin{aligned}
\pr{ { \calN_k^j }^\complement }
&= \pr{ \frac{ N_{(k)}^{j} }{ \frac{1}{2} \tau_k } < \frac{1}{2} \pi_j } 
&\leq \exp \paren{ - \frac{1}{2} \frac{ \paren{ \frac{1}{2} \pi_j }^2 \frac{1}{2} \tau_k }{ \pi_j } } 
&= \exp \paren{ - \frac{1}{16} \pi_j \tau_k } .
\end{aligned}
\]
Also, we note that for $\alpha>0$, 
\begin{eqnarray}
  \sum_{k=1}^{\lceil \log_2 T \rceil} 
     \sum_{t = \tau_k }^{\tau_{k+1} - 1}  \exp \paren{ - \alpha \tau_k  }  
&= & \sum_{k=1}^{\lceil \log_2 T \rceil} 
     \sum_{t = \tau_k }^{\tau_{k+1} - 1}  \exp \paren{ - \frac{1}{2}  \alpha 2 \tau_k  }  \nonumber \\ 
&\leq& \sum_{k=1}^{\lceil \log_2 T \rceil} 
     \sum_{t = \tau_k }^{\tau_{k+1} - 1}  \exp \paren{ - \frac{1}{2}  \alpha t  }  \nonumber \\
&\leq& \int_{0}^{T} \exp \paren{ - \frac{1}{2}  \alpha t  } \dd t 
\leq \frac{2}{\alpha}  .  \nonumber 
\end{eqnarray}
where the first inequality holds by monotonicity.
Hence, we have   
\begin{equation*}
\sum_{j=1}^{M} \pi_j  \sum_{k=1}^{\lceil \log_2 T \rceil} \sum_{t = \tau_k }^{\tau_{k+1} - 1}  \pr{ { \calN_k^j }^\complement }   \lesssim \sum_{j=1}^{M} \pi_j  \frac{1}{\pi_j } = M. 
\end{equation*}

    \item  As for the second term in \eqref{eq:residual_prob}, we note that 
\begin{eqnarray}
    & & \pr{ \mathcal{N}_k^j  \cap {\mathcal{E}_k^j}^{\complement}  }  
    =  \pr{ {\mathcal{E}_k^j}^{\complement} \mid  \mathcal{N}_k^j    }  \pr{  \mathcal{N}_k  } \leq   \pr{  {\mathcal{E}_k^j}^{\complement} \mid  \mathcal{N}_k^j    }   \nonumber \\ 
    &=& \sum_{ n_{(k)}^j  } ~ \pr{ N_{(k)}^j = n_{(k)}^j \Big | N_{(k)}^{j} \geq \frac{1}{2} \pi_j \tau_k  }  
    \nonumber \\ && \quad {}\times \pr{ \lambda_{\min} \paren{ \hat{\bfSigma}^j \paren{ N_{(k)}^{j}   } } <   \frac{1}{2} \lambda_{\min} \paren{ \bfSigma^j  } \Big |  N_{(k)}^j = n_{(k)}^j  } \nonumber \\ 
    &\leq&  \sum_{ n_{(k)}^j  } ~ \pr{ N_{(k)}^j = n_{(k)}^j \Big | N_{(k)}^{j} \geq \frac{1}{2} \pi_j \tau_k  }  
    \times d \cdot 
   \rho^{ - \frac{ \lambda_{\min} \paren{\bfSigma} n_{ (k) }^{j}  }{\xmax^2}  }  
      \nonumber  \\ 
    &\leq& d \cdot 
    \rho^{ - \frac{ \lambda_{\min} \paren{\bfSigma} \frac{1}{2} \pi_j \tau_k  }{\xmax^2}  }     , \label{eq:invoking_smallest_eigen}\label{eq:second_term_bound} 
\end{eqnarray}
where we invoked Lemma~\ref{lemma:smallest_eigenvalue} to conclude \eqref{eq:invoking_smallest_eigen}.
Now, using \eqref{eq:sum_trick} again, we have 
\begin{eqnarray*}
 \sum_{k=1}^{\lceil \log_2 T \rceil} \sum_{t = \tau_k }^{\tau_{k+1} - 1}   \sum_{j=1}^{M} \pi_j \pr{ \mathcal{N}_k^j  \cap {\mathcal{E}_k^j}^{\complement}  }  
 \lesssim   \sum_{j=1}^{M} \pi_j d \frac{\xmax^2}{ \pi_j \underline{\lambda} } = M d  \frac{\xmax^2}{ \underline{\lambda} } .
\end{eqnarray*}
\end{enumerate}





\end{itemize}
\end{enumerate}
Now, the theorem is concluded by putting everything together.

\section{Concluding Remarks}

In this work, we study a contextual dynamic pricing framework for a large number of securities. 
Our approach introduces a multi-task learning strategy, capitalizing on the latent structural similarities among the securities. 
We provably show that the expected regret of the multi-task learning strategy performs better than the individual learning strategy and the pooling strategy. 
Moreover, the numerical experiments on real datasets support our theoretical findings, demonstrating superior performance of our proposed algorithm.



\section*{Acknowledgments and funding}
We would like to extend our gratitude to Peiqi Wang for the engaging discussions on real market challenges, which offered invaluable insights. 
Adel Javanmard is supported by the NSF Award DMS-2311024,
the Sloan fellowship in Mathematics, an Adobe Faculty Research Award and an
Amazon Faculty Research Award. Renyuan Xu is partially supported by the NSF
CAREER award DMS-2339240 and a JP Morgan Faculty Research Award.


\bibliographystyle{plain}
\bibliography{ref}

\appendix

\section{Omitted Proofs}\label{app:omitted_proofs}

We restate Theorem~\ref{thm:regret_simple_version} so that all constants in the model are included. 
\begin{theorem}[The Complete Statement of Theorem~\ref{thm:regret_simple_version}] \label{thm:regret_complete_version}
Under Assumption~\ref{assumption:diverse_contexts}, \ref{assumption:r_neighbourhood}, \ref{assumption:normal_noise}, 
Algorithm~\ref{algo:two_stage} ensures that 
\begin{eqnarray}
\regret \paren{T}  &\lesssim&  \min \Big\{ 
\frac{ u_F L_F \xmax  \overline{\lambda}}{ \ell_F^2 \underline{\lambda}^2} \sqrt{d \log \paren{dM}}  \sqrt{ T } \log \paren{T}    \sum_{j=1}^{M} \sqrt{\pi_j} \cdot \deltamax  \nonumber \\
& & +
 \frac{ u_F^2 L_F \xmax  \overline{\lambda}}{ \ell_F^2 \underline{\lambda}^2} d \log \paren{d M} \log \paren{T}   \sum_{j=1}^{M} \sqrt{\pi_j}  ,~~ \nonumber \\
 & & \quad \frac{ u_F^2 L_F^2 \xmax^2  \overline{\lambda} }{ \ell_F^2 \underline{\lambda}^2} M d \log \paren{d M} \log \paren{T}  , ~~  \nonumber \\
 & & 
\paren{ \frac{L_F^4 \xmax^4 }{ \ell_F^4 \underline{\lambda}^4 } + \frac{L_F^2 \xmax^2 }{ \ell_F^2 \underline{\lambda}^2 }  } \overline{\lambda}   \deltamax^2 T  \log \paren{T}  
 + \frac{L_F^4 \xmax^4 u_F^2 \overline{\lambda}}{ \ell_F^4 \underline{\lambda}^4 }  d \log(d) \log \paren{T}  
    \Big \}  \nonumber \\
& & + \frac{  W^2  \xmax^2  \overline{\lambda}}{\underline{\lambda}} M d 
\label{eq:main_theorem} .     
\end{eqnarray}
    
\end{theorem}

Below, we present the omitted proofs from Section~\ref{sec:main_proof}. 
In Appendix \ref{appendix_sec:dynamic_pricing}, we present the proof of a key lemma that provides an upper bound for \eqref{eq:bottleneck_step}. 
In Appendix \ref{subsec:discussion_assumption_of_thm}, we discuss how to relax Assumption~\ref{assumption:r_neighbourhood} and Assumption~\ref{assumption:normal_noise}. 
In Appendix  \ref{appendix:statistical_perspective}, we prove the lemmas to analyze the algorithm from a statistical learning perspective.

\subsection{Proofs regarding dynamic pricing challenges}
\label{appendix_sec:dynamic_pricing}

\begin{lemma} \label{lemma:perround_regret_to_estimation_error}
    Under Assumptions~\ref{assumption:r_neighbourhood} and~\ref{assumption:normal_noise}, we have 
    \begin{equation} \label{eq:perround_regret_to_estimation_error}
     R_t  \paren{ p_t^{\star} \paren{\inner{ \bftheta_\star^{Z_t}, \bfx_t }}   }
    - R_t \paren{ p_t^{\star} \paren{ \inner{ \hat{\bftheta}_t , \bfx_t } }   }
    \lesssim \normtwo{ \bftheta^{Z_t}_\star - \hat{\bftheta}_t }^2 \,
\end{equation}   
where the function $p_t^\star(\cdot)$ is defined in \eqref{eq:p_star_as_b}. Therefore, we have $ \E{ \operatorname{reward}_t ( p_t^\star ) - \operatorname{reward}_t ( p_t )  } 
 \lesssim  \overline{\lambda} \cdot   \E{ \normtwo{ \bftheta_\star^{Z_t} -  \hat{\bftheta}_t  }^2 }$.
\end{lemma}

\begin{myproof}
Without loss of generality, we consider $\gamma_t = 0$ in this proof. The case where $\gamma_t \neq 0$ can be similarly handled. 
Recall that given estimator $\hat{\bftheta}_t$, the pricing rule follows 
\begin{equation}   \label{eq:p_t_in_proof}
    p_t = \underset{ p \leq \bar{p} }{\argmax} ~ p \cdot F \paren{ V_{t}^{-1}(p) - \inner{\hat{\bftheta}_t, \bfx_t}  }.
\end{equation}

\noindent Connecting $R_t  \paren{ p^{\star} \paren{\inner{ \bftheta_\star^{Z_t}, \bfx_t }}   }
    - R_t \paren{ p^{\star} \paren{ \inner{ \hat{\bftheta}_t , \bfx_t } }   }$ and $ \abs{ p_t^\star - p_t }^2$:
Let $\lambda_t, \lambda_t^*$ be the Lagrangian multipliers of the inequality constraint associated with $p_t=p_{t}^{\star} \paren{ \inner{ \hat{\bftheta}_t , \bfx_t } }, ~ p_t^\star=p_{t}^{\star} \paren{ \inner{ \bftheta_\star^{Z_t}, \bfx_t }}$ respectively.  
By the smoothness of $F$ and $V_t^{-1}$, we know $R_t$ is twice continuously differentiable. 
We have 
\begin{eqnarray*}
  R_t ( p_t^\star ) - R_t ( p_t )  &=&  R_t ( p_t^\star ) + \lambda_{t}^{\star} \cdot \paren{ p_t^\star - \bar{p} }
       - R_t ( p_t ) - \lambda_t \cdot \paren{ p_t - \bar{p} }  \\
    &=& -   \underbrace{ \paren{ \frac{\dd R_t }{\dd p} (p_t^\star) + \lambda_{t}^{\star} } }_{=0} \paren{ p_t^\star - \bar{p} } 
    - \frac{1}{2} \frac{\dd^2 R_t }{\dd p^2} (p) \paren{ p_t^\star - p_t }^2, 
\end{eqnarray*}
for some $p$ between $p_t^\star$ and $p_t$. 
Here, the first equation follows from the complementary slackness, and the second equation is due to the Mean Value Theorem. Straightforward calculation yields that 
\begin{eqnarray*}
\frac{\dd^2 R_t }{\dd p^2} (p)
 &=& 
2 \frac{ f(r-b) }{ V_t'(r) }   + p \cdot\paren{  \frac{ f'(r-b) }{ V_t'(r)^2}  - \frac{f(r-b) V_t''(r-b) }{ V_t'(r)^3} }, 
\end{eqnarray*}
where $r = V^{-1}(p)$ and $b = \inner{ \bftheta_\star^{Z_t}, \bfx_t }$. 
We argue that $ \abs{ \frac{\dd^2 R_t }{\dd p^2} (p) } $ can be upper bounded by some universal constant. 
For ease of notation, we suppress the time index in expression of $V_t$. 
We recall that 
$
V_{t} (y) \defeq \sum_{ i=1}^{ n }  \frac{ c^{} P^{} }{ (1+y)^{\tau_i^{}} }  + \frac{P^{}}{(1+y)^{\tau_{n^{}}^{}}}  , 
$
and hence, 
$
V_t'(y) = -\sum_{i=1}^{n} cP\, \tau_i \,(1+y)^{-\tau_i-1} - P\, \tau_n \,(1+y)^{-\tau_n-1} . 
$
Since $\tau_i \leq \tau_n$, one can show that $ \frac{1}{\left|V^{\prime}(y)\right|} \leq \frac{(1+y)^{\tau_n+1}}{\sum_{i=1}^n c P \tau_i+P \tau_n} \leq \frac{ (1+y)^{\tau_n + 1} }{P \tau_n }$. 
Because the bond primitives lie in a compact set, $\tau_n$ is uniformly bounded above. 
Therefore, for $0 \leq y \leq \bar{r}$,  the term $\frac{ (1+y)^{\tau_n + 1} }{P \tau_n }$ can be uniformly upper bounded by a constant $\frac{1}{c_{V'}}$. In addition for the same reasoning, we note that $V^{\prime \prime}(y)=\sum_{i=1}^n \frac{c P \cdot \tau_i\left(\tau_i+1\right)}{(1+y)^{\tau_i+2}}+\frac{P \cdot \tau_n\left(\tau_n+1\right)}{(1+y)^{\tau_n+2}}  $, whose absolute value can be upper bounded by a universal constant $c_{V''}$ for $0 \leq y \leq \bar{r}$. 
Moreover, Assumption~\ref{assumption:normal_noise} implies that $|f(x)| \leq \frac{1}{\sqrt{2 \pi} \sigma}$ and $|f'(x)| \leq \frac{e^{-1/2}}{ \sqrt{2 \pi} \sigma^2}$ for any $x \in \mathbb{R}$. 
Therefore, to prove the lemma, it suffices to show that $ \abs{ p_t^\star - p_t } \lesssim   \abs{ \inner{\bfx_t, \bftheta_{\star}^{Z_t}  - \hat{\bftheta}_t   } } $.

\noindent\underline{Establishing $ \abs{ p_t^\star - p_t } \lesssim   \abs{ \inner{\bfx_t, \bftheta_{\star}^{Z_t}  - \hat{\bftheta}_t   } } $:}
Recall $r_t^\star(b) = V_t^{-1}(p_t^\star(b))$. 
Next, we proceed to invoke implicit function theorem to show that $r_t^\star$ is differentiable.

For notational convenience, in what follows, we suppress all the time index subscripts. 
For ease of discussion, we relax the inequality constraint $p \leq \bar{p}$. This relaxation does not compromise the correctness of the proof; in fact, the conclusion remains valid and holds even more strongly with the constraint in place.

We note that $p^\star$ is determined in the following way:
 given $b \in \mathbb{R}$, 
\begin{equation} \label{eq:the_p_problem}
  p^\star (b) =   \argmax_{p \leq \bar{p} } ~ p \cdot F \paren{V^{-1}(p) - b}  . 
\end{equation}
Recall that it follows from Assumption~\ref{assumption:normal_noise} that $f$ and $F$ are the $p.d.f.$ and $c.d.f.$ of a zero-mean normal distribution. 
Let $h$ denote the reversed hazard rate $h(x) = \frac{f(x)}{F(x)}$. 
Since $V$ is monotone, by change of variable $r = V^{-1}(p)$, the first-order-condition of \eqref{eq:the_p_problem} leads to the following equation: 
\begin{equation} \label{eq:foc_defining_trajectory}
    h(r-b) = - \frac{V'(r)}{V(r)} . 
\end{equation}
An equivalent way to express  \eqref{eq:foc_defining_trajectory} is 
\begin{equation} \label{eq:psi_function_defining_trajectory}
    \Psi(r,b) \defeq V'(r) F(r-b) + V(r) f(r-b) = 0 . 
\end{equation}
For points $(r,b)$ on the curve defined by \eqref{eq:foc_defining_trajectory}, 
\begin{eqnarray}
    \frac{\partial \Psi }{\partial r} (r,b) &=& F(r-b) + 2 V'(r) f(r-b) + V(r) f'(r-b)  \nonumber  \\
    &=& f(r-b)  \paren{ 2 V'(r)  -  \frac{ V(r) V''(r) }{ V'(r) } }  + V(r) f'(r-b)  \label{eq:psi_r_1} \\ 
    &=& f(r-b) \paren{ 2 V'(r)  -  \frac{ V(r) V''(r) }{ V'(r) } - \frac{r-b}{\sigma^2} } \label{eq:psi_r_2} . 
\end{eqnarray}
Equation~\eqref{eq:psi_r_1} follows due to \eqref{eq:psi_function_defining_trajectory}, and \eqref{eq:psi_r_2} follows by using Assumption~\ref{assumption:normal_noise}. 
To invoke implicit function theorem, it suffices to show $\frac{\partial \Psi}{\partial r} \neq 0$ for all the points on the curve defined by \eqref{eq:psi_function_defining_trajectory}.
By Assumption~\ref{assumption:r_neighbourhood}, we have $A(r) = 2 V'(r)  -  \frac{ V(r) V''(r) }{ V'(r) } \leq 0$. 
We proceed to show that under Assumption~\ref{assumption:normal_noise}, for points $(r,b)$ that satisfy \eqref{eq:psi_function_defining_trajectory}, it is guaranteed that $r - b \geq \delta$.  

Equation~\eqref{eq:foc_defining_trajectory} characterizes the set of points $(r,b)$ that lie on the corresponding trajectory. 
    First, we note that the modified duration on the R.H.S. of \eqref{eq:foc_defining_trajectory} is decreasing in $r$, as we recall  $V>0, V'<0, V''>0, V'''<0$. Its maximum over $[0, \infty)$ is reached at $r=0$:
    \begin{equation*}
    \max_{r \in [ 0 , \infty) } - \frac{V'(r)}{V(r)} = 
    -   \frac{V'(0)}{V(0)}  > 0  .  
    \end{equation*}

    Regarding the L.H.S. of \eqref{eq:foc_defining_trajectory}, we note that 
    \begin{eqnarray*}
       h( \delta )
       &=& \sqrt{\frac{2}{\pi}} \frac{1}{\sigma} e^{- \frac{1}{2} (\frac{\delta}{\sigma})^2 } 
       \frac{1}{\operatorname{erfc}(- \frac{\delta}{\sqrt{2} \sigma})}  \\ 
       &\geq& \sqrt{\frac{2}{\pi}} \frac{1}{\sigma} e^{- \frac{1}{2} (\frac{\delta}{\sigma})^2 } 
       \frac{1}{2}  \quad \text{ as $\operatorname{erfc}(x) \in [1,2] $ for $x \leq 0$ }  \\ 
       &>&  - \frac{V'(0)}{V(0)} \quad \text{ under Assumption~\ref{assumption:normal_noise}} . 
    \end{eqnarray*}
    As established in Lemma~\ref{claim:reverse_hazard_rate_normal}, the reversed hazard rate $h$ for a zero-mean normal distribution is decreasing. 
    Consequently, for any $(r,b)$ satisfying \eqref{eq:foc_defining_trajectory}, it must be that $r-b \geq \delta$.  
    
Hence, we conclude that  $\frac{\partial \Psi}{\partial r} \neq 0$ for all the points on the curve defined by \eqref{eq:psi_function_defining_trajectory}.
By implicit function theorem, 
there exists a unique function $r^\star: \mathbb{R} \rightarrow \mathbb{R}$ such that $ \Psi(r^\star(b), b) = 0$.
Moreover, $r^\star$ is continuously differentiable and its derivative with respect to $b$ is given by 
$
 \frac{\dd r^\star}{\dd b} = - \frac{\frac{\partial \Psi}{\partial b} }{ \frac{\partial \Psi}{\partial r} }  . 
$
Next, writing $V_{\max}\defeq \max_{r\geq 0}V(r)$, we realize that $| \frac{\dd r^\star}{\dd b} |$ is upper bounded by a universal constant:
\begin{eqnarray}
    \abs{  \frac{\dd r^\star}{\dd b} } &=& \abs{ \frac{V(r) \paren{ h(r-b) + \frac{r-b}{\sigma^2} } }{ \underbrace{A(r)}_{ \leq 0 } - \frac{r-b}{\sigma^2} } }  \leq   \frac{  V_{\max} \cdot  ( h(0) +  \frac{r-b}{\sigma^2})  }{  \frac{r-b}{\sigma^2}   }  
    \leq V_{\max} \cdot \left( h(0) \frac{\sigma^2}{\delta} + 1\right) , \label{eq:universal_constant_upper_bound_on_drdb}
\end{eqnarray}
where the first inequality invokes Lemma~\ref{claim:reverse_hazard_rate_normal} and the second inequality follows from the fact that $r-b \geq \delta$.

We recall the pricing rule, which states that
\begin{eqnarray*}
    \abs{ p_t^\star - p_t } &=& \abs{ p^{\star}_t( \inner{\bfx_t , \bftheta_{\star}^{Z_t}} ) - p^{\star}_t( \inner{\bfx_t , \hat{\bftheta}_t } )  }  \nonumber \\ 
    &=& \abs{ V_t \paren{  r_{t}^{\star}( \inner{\bfx_t , \bftheta_{\star}^{Z_t}} )   } 
    - V_t \paren{  r_{t}^{\star} ( \inner{\bfx_t , \hat{\bftheta}_t } )    }  }  .
\end{eqnarray*}
Combining with \eqref{eq:universal_constant_upper_bound_on_drdb}, we have 
\begin{eqnarray}
    \abs{ p_t^\star - p_t }
    &=& \abs{  V_t'( r^{\star} ( \xi ) ) \frac{\dd r^\star}{\dd b } (  \xi   )  \inner{\bfx_t, \bftheta_{\star}^{Z_t}  - \hat{\bftheta}_t   }       }   \\
    &\leq& \abs{ \sup_{r \geq 0} V_t '(r)  } 
    \abs{ \frac{\dd r^\star}{\dd b } (  \xi   ) }  
    \abs{ \inner{\bfx_t, \bftheta_{\star}^{Z_t}  - \hat{\bftheta}_t   } } \nonumber , 
\end{eqnarray}
for some $\xi$ that is a convex combination of $\inner{\bfx_t , \bftheta_{\star}^{Z_t}}$ and $\inner{\bfx_t , \hat{\bftheta}_t }$.
It is easy to see that $\abs{ \sup_{r \geq 0} V_t '(r) }  = \abs{ \lim_{r \rightarrow 0} V_t'(r) }
= \sum_{i=1}^{n} c P \tau_i + P \tau_n$,  which can be upper bounded by a universal constant since primitives belong to a compact set. 
The fact that $ \abs{ p_t^\star - p_t } \lesssim   \abs{ \inner{\bfx_t, \bftheta_{\star}^{Z_t}  - \hat{\bftheta}_t   } } $ is now established.

\noindent\underline{Upper bounding per-round regret:}

Finally, we put everything together,
\begin{eqnarray*}
    \E{ \operatorname{reward}_t ( p_t^\star ) - \operatorname{reward}_t ( p_t )  }   &=&  \E{ \E{ \operatorname{reward}_t ( p_t^\star ) - \operatorname{reward}_t ( p_t ) \mid \calF_{t} }   }  \\
&\lesssim&   \E{ \paren{ \inner{ \bftheta_\star^{Z_t} - \hat{\bftheta}_t , \bfx_t} }^2  }  \\ 
&=& \mathbb{E}_{} \left[  \sum_{j=1}^{M} \pi_j \E{  \paren{ \bftheta_\star^j -  \hat{\bftheta}_t }^\top \bfx_t \bfx_t^\top \paren{ \bftheta_\star^j -  \hat{\bftheta}_t }  \mid \calF_{t-1} \cup \braces{Z_t} }  \right] \\ 
&=& \mathbb{E}_{} \left[ \sum_{j=1}^{M} \pi_j  \paren{ \bftheta_\star^j -  \hat{\bftheta}_t }^\top  \E{  \bfx_t \bfx_t^\top  \mid \calF_{t-1} \cup \braces{Z_t} }  \paren{ \bftheta_\star^j -  \hat{\bftheta}_t } \right]  \\
&\leq&  \overline{\lambda} \cdot   \E{ \normtwo{ \bftheta_\star^{Z_t} -  \hat{\bftheta}_t  }^2 }  . 
\end{eqnarray*}
The first inequality holds since $ \abs{ p_t^\star - p_t } \lesssim   \abs{ \inner{\bfx_t, \bftheta_{\star}^{Z_t}  - \hat{\bftheta}_t   } } $, and the last inequality follows from Assumption~\ref{assumption:diverse_contexts} and the fact that $\sum_{j=1}^{M} \pi_j = 1$.


\end{myproof}

\subsection{Relaxation of Assumptions~\ref{assumption:r_neighbourhood} and~\ref{assumption:normal_noise} for Theorem~\ref{thm:regret_simple_version} }    
\label{subsec:discussion_assumption_of_thm}

In this subsection, we show how Assumption~\ref{assumption:r_neighbourhood} and Assumption~\ref{assumption:normal_noise} can be relaxed, which may potentially lead to a weaker result than Theorem~\ref{thm:regret_simple_version}. We emphasize  that the compromise introduced by this relaxation does not stem from the statistical properties of the learning algorithm, but rather from the inherent structure of the pricing problem.

Assumption~\ref{assumption:r_neighbourhood} and Assumption~\ref{assumption:normal_noise} are made to upper bound \eqref{eq:bottleneck_step}.
In the following lemma, we show that these assumptions can be dropped, at the cost of sacrificing the quadratic bound in \eqref{eq:perround_regret_to_estimation_error} in Lemma~\ref{lemma:perround_regret_to_estimation_error}
to a linear bound in the forthcoming Proposition~\ref{lemma:lip_regret_linear_order}. 
The key intuition is that when the true parameter $ \bftheta_{\star}^{Z_t}$ is well estimated, the per-round regret remains small even if the quoted price $ p_t $ may deviate {\it significantly} from the optimal price $ p_t^\star $, which is different from the argument in Theorem \ref{thm:regret_simple_version}. 
However, embedding the yield-to-price function does not lead to a favorable optimization landscape to establish concavity that permits a direct application of the standard Envelope Theorem.
To address this, we build our results based upon \cite{morand2018generalized}, which is a generalized Envelope Theorem.

We relax Assumption~\ref{assumption:r_neighbourhood} and Assumption~\ref{assumption:normal_noise} to the following assumption. 
\begin{assumption} \label{assumption:log_concave}
    Both the p.d.f. $f$ and c.d.f. $F$ of the noise distribution are continuously differentiable.
    The function $F(x)$ is strictly increasing. Furthermore, $F(x)$ and $1-F(x)$ are strictly log-concave in $x$. 
\end{assumption}
 Log-concavity is a commonly used assumption in auction design and dynamic pricing literature \citep{bagnoli2006log}.
Many common probability distributions are log-concave, such as normal, uniform, Gamma$(r,\lambda)$ for $r \geq 1$, Beta$(a,b)$ for $a,b \geq 1$, Subbotin$(r)$ with $r \geq 1$, and the truncated version of many other distributions.   
One can show that $\bar{f} = \sup_{ w \in \mathbb{R} } f( w ) < \infty $ for a log-concave distribution. 

\begin{proposition}  \label{lemma:lip_regret_linear_order}
Let $\hat{\bftheta}_t$ be the estimator used for pricing \eqref{eq:algo_pricing_formula} for round $t$.
Under Assumption~\ref{assumption:log_concave}, the following holds for all bonds with payment structure defined by \eqref{eq:bond_yield-to-price}: 
\begin{equation}
     R_t  \paren{ p^{\star}_{t} \paren{\inner{ \bftheta_\star, \bfx_t }}   }
    - R_t \paren{ p^{\star}_{t} \paren{ \inner{ \hat{\bftheta}_t , \bfx_t } }   }
    \lesssim \bar{p} \bar{f} \bar{x}  \normtwo{ \bftheta_\star - \hat{\bftheta}_t }   ,  
\end{equation}   
where { $\bar{f} = \sup_{ w \in \mathbb{R} } f( w ) $}, and recall that $\normtwo{\bfx_t}\leq \bar x$ (c.f.~\eqref{eq:x_iid}) and $p^{\star} \paren{ \inner{ \hat{\bftheta}, \bfx } } \leq \bar{p}$ (c.f.~\eqref{eq:p_star_as_b}).

Therefore, we have $ \E{ \operatorname{reward}_t ( p_t^\star ) - \operatorname{reward}_t ( p_t )  } 
 \lesssim  \overline{\lambda} \bar{p} \bar{f} \bar{x} \cdot   \E{ \normtwo{ \bftheta_\star^{Z_t} -  \hat{\bftheta}_t  } }$.
\end{proposition}

\begin{myproof}
First of all, we note that the log-concavity is to ensure that $\ell_F > 0$ in \eqref{eq:def_ell_F}. 
Moreover, one can show that under the log-concavity condition, $\bar{f} < \infty$.

For notational convenience, in this proof, we suppress all time index subscripts. 
With slight abuse of notation, we overload the notation $R$ by including one more argument 
\begin{equation*}
\begin{aligned}
R(p,\bftheta)&\defeq\paren{p-\gamma}\pr{p\leq V\paren{\inner{\bftheta,\bfx}+\epsilon}}\\
&=p\,\pr{V^{-1}(p)\geq\bftheta^\top\bfx+\epsilon}\\
&=p\,F\paren{V^{-1}(p)-\bftheta^\top\bfx}.
\end{aligned}
\end{equation*}
and $ \bar{R} ( \bftheta ) = \underset{p \leq \bar{p} }{\argmax}~ R(p, \bftheta) $.

By introducing intermediate terms, the per-round regret can be written as 
\begin{eqnarray}
& &   R \paren{ p^{\star} \paren{\inner{ \bftheta_\star, \bfx }}  , \bftheta_\star  }
    - R \paren{  
    p^{\star} \paren{ \inner{ \hat{\bftheta}, \bfx } }  ,   \bftheta_\star   } \nonumber   \\ 
&=&  
    R \paren{ p^{\star} \paren{\inner{ \bftheta_\star, \bfx }}  ,  \bftheta_\star  } 
    - 
    R \paren{  
    p^{\star} \paren{ \inner{ \hat{\bftheta}, \bfx } }  ,   \hat{\bftheta}  }  
    + 
    R \paren{ p^{\star} \paren{\inner{ \hat{\bftheta}, \bfx }}  ,  \hat{\bftheta} } 
    - R \paren{  
    p^{\star} \paren{ \inner{ \hat{\bftheta}, \bfx } }  ,  \bftheta_\star}  \nonumber  \\
&=&  
     \bar{R} \paren{\bftheta_\star}  - \bar{R} (\hat{\bftheta}) 
    + 
    R \paren{ p^{\star} \paren{\inner{ \hat{\bftheta}, \bfx }}  ,  \hat{\bftheta} } 
    - R \paren{  
    p^{\star} \paren{ \inner{ \hat{\bftheta}, \bfx } }  ,  \bftheta_\star}  \label{eq:two_diff_to_bound} . 
\end{eqnarray}
In what follows, we proceed to bound the two differences respectively. 
\begin{itemize}
    \item 
    For the latter difference in \eqref{eq:two_diff_to_bound}, by the Mean Value Theorem, we have 
    \begin{equation*}
    R \paren{ p^{\star} \paren{\inner{ \hat{\bftheta}, \bfx }}  ,  \hat{\bftheta} } 
    - R \paren{  p^{\star} \paren{ \inner{ \hat{\bftheta}, \bfx } }  ,  \bftheta_\star} 
    =  \frac{\partial R}{ \partial ( \bftheta^\top \bfx ) }  \paren{ p^\star \paren{ \inner{ \hat{\bftheta}, \bfx} } ,  \tilde{\bftheta} }  \inner{ \bftheta_\star - \hat{\bftheta} , \bfx} 
    \end{equation*}
    for some $ \tilde{\bftheta}$ on the line segment connecting $\hat{\bftheta}$ and $\bftheta_\star $.
    Hence, 
    \begin{align}
&\abs{R\paren{p^{\star}\paren{\inner{\hat{\bftheta},\bfx}},\hat{\bftheta}}
-R\paren{p^{\star}\paren{\inner{\hat{\bftheta},\bfx}},\bftheta_\star}} \nonumber\\
&=\abs{(-1)\,p^{\star}\paren{\inner{\hat{\bftheta},\bfx}}
 f\paren{V^{-1}\paren{p^{\star}\paren{\inner{\hat{\bftheta},\bfx}}}-\theta^\top\bfx}
 \inner{\bftheta_\star-\hat{\bftheta},\bfx}} \nonumber\\
&\leq \bar{p}\bar{f}\bar{x}\normtwo{\bftheta_\star-\hat{\bftheta}}.
\label{eq:two_diff_to_bound_1}
    \end{align}
where the inequality follows since $p^{\star} \paren{ \inner{ \hat{\bftheta}, \bfx } } \leq \bar{p}$, { $\bar{f} = \sup_{ w \in \mathbb{R} } f(w) $ } and $ \normtwo{\bfx_t} \leq \bar{x}$.   

    \item 
With slight abuse of notation, we overload the notation 
\begin{equation*}
R(p, b) = \paren{ p - \gamma } \pr{ p \leq V( b + \epsilon ) } 
= \paren{ p - \gamma } \pr{ V^{-1} ( p ) \geq  b + \epsilon  }
= \paren{ p - \gamma } F \paren{ V^{-1} ( p ) -  b   }
\end{equation*}
and 
\begin{equation*}
\bar{R} (b) = \underset{p \leq \bar{p}}{\max} ~ \paren{ p - \gamma } 
F \paren{ V^{-1} ( p ) -  b   } . 
\end{equation*}
Let $P^{\star}(b) = \underset{ p \in ( -\infty, \bar{p}] }{\argmax} ~ R(p,b)$.
The Gâteaux derivative of $R(p, b)$ at $b \in \mathbb{R}$ along direction $ x \in \mathbb{R}$ exists, and is equal to  
\begin{eqnarray} \label{eq:directional_derivative_R}
  \bar{R}' (b; x) = \max_{p^{\star}(b) \in P^{\star}(b)}  
  \braces{ (-1) \cdot \paren{ p^{\star}(b) - \gamma } \cdot f ( V^{-1}(p^{\star}(b) ) -b )  \cdot x }  . 
\end{eqnarray}
We prove \eqref{eq:directional_derivative_R} by invoking Proposition 3.3 in \cite{morand2018generalized}, which is a generalized version of the standard Envelope Theorem. 
In the following, we proceed to verify the conditions specified in the proposition.

Consider the problem  $\max_{p} R(p,b) ~\text{s.t.} ~ p - \bar{p} \leq 0.$
The Lagrangian function is 
\begin{equation}  \label{eq:lagrangian}
    \mathcal{L} (p, b, \lambda ) = (p - \gamma) \cdot F \paren{ V^{-1}(p) - b } 
    + \lambda \cdot (p - \bar{p})  . 
\end{equation}
We note that the prerequisites of Proposition 3.3 in \cite{morand2018generalized} are all satisfied: 
\begin{itemize}
    \item The strict Mangasarian–Fromovitz constraint qualification is trivially satisfied for all $p^*(b)$. As there is only one (inequality) constraint, and its gradient with respect to $p$ is $1$.
    \item The Clarke's hypothesis is satisfied as the inf-compactness condition is satisfied (c.f. the discussion after Theorem 2.1 in \cite{morand2018generalized}). 
    For every $b'$ in a neighborhood of $b$, the set 
    \begin{equation*}
\braces{ p : p \leq \bar{p} , ~   (p - \gamma) \cdot F \paren{ V^{-1}(p) - b' }  \geq 0   } 
    \end{equation*}
    is non-empty and contained in $[0, \bar{p}]$. 
    \item The function $R(p,b)$ is $\mathcal{C}^1$ in $b$. 
\end{itemize}
Now, the proof of \eqref{eq:directional_derivative_R} is complete. 

Next, we proceed to show that $\bar{R} (\cdot)$ is Lipschitz. 
For any $b_1, b_2$, define $\Psi: [0,1] \rightarrow \mathbb{R}$,  $\Psi(t) = \bar{R} \paren{ b_1 + t (b_2-b_1) } $.
By the Berge's Maximum Theorem (see e.g. Theorem 9.14 in \cite{sundaram1996first}),  
$\Psi$ is a continuous function on $[0,1]$. 
We note $\Psi$ is also differentiable, as for any $t \in (0,1)$, 
\begin{eqnarray} \label{eq:psi_prime_to_R_bar}
    \Psi'(t) &=& \lim_{\Delta t \rightarrow 0} \frac{ \bar{R} \paren{ b_1 + (t+\Delta t )(b_2-b_1) }  - \bar{R} \paren{ b_1 + t (b_2-b_1) }  }{\Delta t}  \nonumber \\
    &=& \bar{R} ' (  b_1 + t (b_2-b_1) ; b_2 - b_1 )     
\end{eqnarray}
is well defined due to \eqref{eq:directional_derivative_R}. 
The first and the second equation follows from the definition of derivative and Gâteaux directional derivative, respectively. 
Hence, by the Mean Value Theorem, we have 
\begin{align}
     \bar{R} (b_2) - \bar{R} (b_1) &=  \Psi(1) - \Psi(0) 
     =  \Psi'(t)  \quad \text{ for some } t \in (0,1) \nonumber \\ 
     &=  \bar{R} ' (  b_1 + t (b_2-b_1) ; b_2 - b_1 )  \label{eq:psi_prime_to_R_bar_2} \\ 
     &=  \max_{p^{\star}\in P^{\star}\paren{b_1+t(b_2-b_1)}}
     \left\{\begin{aligned}
     &(-1)\paren{p^{\star}-\gamma}
     f\paren{V^{-1}(p^{\star})-\paren{b_1+t(b_2-b_1)}}\\
     &\qquad\cdot(b_2-b_1)
     \end{aligned}\right\}
     \label{eq:plug_in_directional_derivative} \\
    &\leq  \bar{p} \bar{f} \abs{b_2-b_1} , 
\end{align}
where \eqref{eq:psi_prime_to_R_bar_2} follows from \eqref{eq:psi_prime_to_R_bar} and \eqref{eq:plug_in_directional_derivative} is due to \eqref{eq:directional_derivative_R}. 
By letting $b_1 = \inner{ \bftheta_\star, \bfx}, b_2 =  \inner{ \hat{\bftheta} , \bfx}$, it is now clear that the former difference in \eqref{eq:two_diff_to_bound} can be upper bounded by 
\begin{equation} \label{eq:two_diff_to_bound_2}
    \bar{R} \paren{\bftheta_\star}  - \bar{R} (\hat{\bftheta})  \leq \bar{p} \bar{f} \bar{x} \normtwo{ \bftheta_\star - \hat{\bftheta} } . 
\end{equation}

\end{itemize}

 By combining \eqref{eq:two_diff_to_bound}, \eqref{eq:two_diff_to_bound_1} and \eqref{eq:two_diff_to_bound_2}, the proof of the proposition is complete.

\end{myproof}


\subsection{Proofs regarding the statistical perspective }
\label{appendix:statistical_perspective}

\begin{lemma} \label{lemma:smallest_eigenvalue}
Given $\braces{Z_t}_{t=1}^{n}$, denote
\begin{equation*}
n^j=\sum_{t=1}^{n}\one{Z_t=j},\qquad
\hat{\bfSigma}^{j}\paren{n^j}=\frac{1}{n^j}\sum_{t=1}^{n}\one{Z_t=j}\bfx_t\bfx_t^\top.
\end{equation*}
Then, we have
\begin{equation}
    \pr{ \lambda_{\min} \paren{\hat{\bfSigma}^{j} \paren{n^j} } < \frac{1}{2} \lambda_{\min} \paren{  \bfSigma^j  }  }  
    \leq d \cdot 
    \paren{ \sqrt{\frac{1}{2} e }  }^{ - \frac{ \lambda_{\min} \paren{\bfSigma} \frac{1}{2} \tau_k  }{\xmax^2}  }  
    .  
\end{equation}
\end{lemma}

\begin{myproof}
This proof is standard and we include it merely for completeness. 
    The proof idea is to invoke Lemma~\ref{lemma:matrix_chernoff}.

We first verify that 
\begin{align*}
\lambda_{\max}\paren{\one{Z_t=j}\bfx_t\bfx_t^\top}
&\leq\lambda_{\max}\paren{\bfx_t\bfx_t^\top}
=\max_{\bfv\neq0}\frac{\normtwo{\bfx_t\bfx_t^\top\bfv}}{\normtwo{\bfv}}\\
&\leq\max_{\bfv\neq0}\frac{\normtwo{\bfx_t}\abs{\bfx_t^\top\bfv}}{\normtwo{\bfv}}
\leq\max_{\bfv\neq0}\frac{\normtwo{\bfx_t}^2\normtwo{\bfv}}{\normtwo{\bfv}}
\leq\bar{x}^2.
\end{align*}
{Therefore by Lemma~\ref{lemma:matrix_chernoff},} we obtain that provided that the sequence $\braces{Z_t}_{t=1}^{n} $ is known, 
\begin{align}
&\pr{\begin{aligned}
&\lambda_{\min}\paren{\sum_{t=1}^{n}\one{Z_t=j}\bfx_t\bfx_t^\top}<\tfrac12\lambda_{\min}\paren{\bfSigma^j}n^j,\\[-0.5ex]
&\lambda_{\min}\paren{\sum_{t=1}^{n}\mathbb{E}_{t-1}\brackets{\one{Z_t=j}\bfx_t\bfx_t^\top}}
\geq\lambda_{\min}\paren{\bfSigma^j}n^j
\end{aligned}} \nonumber\\
&\leq d\paren{\frac{e^{-1/2}}{(1/2)^{1/2}}}^{\lambda_{\min}\paren{\bfSigma^j}n^j/\bar{x}^2}
=d\paren{\sqrt{e/2}}^{-\lambda_{\min}\paren{\bfSigma^j}n^j/\bar{x}^2}.
\nonumber
\end{align}
where we note that $ \sqrt{\frac{1}{2} e }  > 1$. 
On the other hand, it is straightforward to notice that given $\braces{Z_t}_{t=1}^{n}$, 
\begin{eqnarray}
     \lambda_{\min}  \paren{  \sum_{t=1}^{n} \mathbb{E}_{t-1} \brackets{ \one{Z_t=j} \bfx_t \bfx_t^\top }  }  
     &=&    \lambda_{\min}  \paren{  \sum_{t=1}^{n}  \one{Z_t=j}  \mathbb{E}_{t-1} \brackets{\bfx_t \bfx_t^\top }  }     \nonumber \\ 
     &\geq& \sum_{t=1}^{n}  \one{Z_t=j} \lambda_{\min}  \paren{    \mathbb{E}_{t-1} \brackets{\bfx_t \bfx_t^\top }  }  \nonumber \\  
     &=& \sum_{t=1}^{n}  \one{Z_t=j} \lambda_{\min}  \paren{    \mathbb{E}\brackets{\bfx_t \bfx_t^\top }  }  \nonumber \
     \geq  \lambda_{\min} \paren{\bfSigma^j} n^j  \nonumber  ,
\end{eqnarray}
where the first inequality follows since the minimum eigenvalue is concave over positive semidefinite matrices; and the second equality follows because of the assumption that $\bfx_t$ is independent over time $t$ (c.f. \eqref{eq:x_iid}). 

\end{myproof}

\subsubsection{Proof of Lemma~\ref{lemma:stage_1_common}}

\begin{myproof}

Provided that $\lambda_{\min}\paren{\hat{\bfSigma}}>0$, the function $\bar{\calL}\paren{\bftheta}$ is strongly convex in $\bftheta$, since for any $\bfv\in\mathbb{R}^d$,
\begin{align*}
\bfv^\top\nabla^2\bar{\calL}\paren{\bftheta}\bfv
&=\bfv^\top\paren{\frac1n\sum_{j=1}^{M}\sum_{t=1}^{n}\one{Z_t=j}\eta_t\paren{\bftheta}\bfx_t\bfx_t^\top}\bfv\\
&\geq\ell_F\lambda_{\min}\paren{\hat{\bfSigma}}\bfv^\top I_d\bfv.
\end{align*}
In view of the optimality of $\bar{\bftheta}$ in the MLE procedure, we have that $\nabla \bar{\calL} \paren{ \bar{\bftheta} } = 0$.
{By the fundamental theorem of calculus,  we obtain    } 
\begin{eqnarray}   
  \normtwo{ \nabla \bar{\calL} \paren{ \bftheta_\star }   }   &=&  \normtwo{ \nabla \bar{\calL} \paren{ \bftheta_\star }  - \nabla \bar{\calL} \paren{ \bar{\bftheta} } } = \normtwo{ \int_{0}^{1}  \nabla^2 \bar{\calL} \paren{ \alpha \bftheta_\star + \paren{1 - \alpha }  \bar{\bftheta}  }  \dd \alpha \paren{ \bftheta_\star - \bar{\bftheta} }  } \nonumber\\
  &\geq& \ell_F \lambda_{\min} \paren{\hat{\bfSigma}}  \normtwo{ \bftheta_\star - \bar{\bftheta} } .   \label{eq:stage_1_size_gradient_1}
\end{eqnarray}
On the other hand, we note that 
\begin{eqnarray}
& &  \normtwo{ \nabla \bar{\calL} \paren{ \bftheta_\star } } \nonumber \\ 
 &=& 
 \normtwo{ \frac{1}{n}  \sum_{j=1}^{M} \sum_{t=1}^{n} \one{Z_t=j}  
 \paren{ \xi_t \paren{ \bftheta_\star }  -  \xi_t \paren{ \bftheta^j_\star }  } \bfx_t 
 + \frac{1}{n}  \sum_{j=1}^{M} \sum_{t=1}^{n} \one{Z_t=j} \xi_t \paren{ \bftheta^j_\star } \bfx_t  
 }  \nonumber \\ 
 &\leq&     \frac{1}{n}  \sum_{j=1}^{M} \sum_{t=1}^{n} \one{Z_t=j} \abs{\xi_t \paren{ \bftheta_\star }  -  \xi_t \paren{ \bftheta^j_\star } } \normtwo{\bfx_t} 
 + \normtwo{  \frac{1}{n}  \sum_{j=1}^{M} \sum_{t=1}^{n} \one{Z_t=j} \xi_t \paren{ \bftheta^j_\star } \bfx_t    }   \nonumber \\ 
 &\leq&   L_F \xmax \frac{1}{n}  \sum_{j=1}^{M} n^j \normtwo{ \bfdelta_\star^j } 
 + \normtwo{  \frac{1}{n}  \sum_{j=1}^{M} \sum_{t=1}^{n} \one{Z_t=j} \xi_t \paren{ \bftheta^j_\star } \bfx_t    }  ,  \label{eq:stage_1_size_gradient_2}
\end{eqnarray} 
where the inequalities follow from the triangular inequality. 
Combining \eqref{eq:stage_1_size_gradient_1} and \eqref{eq:stage_1_size_gradient_2} yields the result. 

\end{myproof}

\subsubsection{Proof of Lemma~\ref{lemma:expectation_bound_stage_1}}


\begin{myproof}

We denote
\begin{equation*}
G=\frac1n\sum_{j=1}^{M}\sum_{t=1}^{n}\one{Z_t=j}\xi_t\paren{\bftheta^j_\star}\bfx_t,
\qquad
\calG_{\varepsilon}\defeq\braces{\normtwo{G}\leq\varepsilon}.
\end{equation*}
We know
\begin{equation} \label{eq:prob_G_epsilon_c}
    \pr{ \calG_{\varepsilon}^{\complement} \Big|\, \mathcal{H}_n    } \leq 2 d \exp \paren{  - \frac{ \varepsilon^2 n }{ 8 u_F^2 \bar{x}^2 d} }  . 
\end{equation}

Indeed, to see this, we first recall that the expected gradient of the log-likelihood is zero, hence $\E{ \xi_t \paren{ \bftheta_\star^j } \bfx_t  \mid \mathcal{H}_n  } = 0$.
Noting the fact that $p_t$ is independent of $\braces{\epsilon_s}_{s=1}^{t}$ for all $t \in [n]$ and $\abs{\xi_t \paren{x}} \leq u_F$, 
therefore by Hoeffding's inequality, 
we obtain 
\begin{equation}
    \pr{ \frac{1}{n} \abs{  \sum_{j=1}^{M} \sum_{t=1}^{n} \one{Z_t = j} \xi_t \paren{ \bftheta_{\star}^j } \brackets{ \bfx_t }_{\ell}  } \leq \varepsilon ~ \Big | ~ \mathcal{H}_n  } \geq 1 - 2 \exp \paren{ - \frac{ n {\varepsilon}^2 }{ 8 u_F^2 \bar{x}^2  }}  \  , 
\end{equation}
where $[\bfx_t]_{\ell}$ denotes the $\ell$th coordinate of $\bfx_t$.
Then, by a union bound over $d$ coordinates, with probability at least $1-\delta$, 
\begin{align*}
&\normtwo{\frac{1}{n}\sum_{j=1}^{M}\sum_{t=1}^{n}
\one{Z_t=j}\xi_t\paren{\bftheta_{\star}^j}\bfx_t}\\
&\leq \sqrt{d}\,
\norminf{\frac{1}{n}\sum_{j=1}^{M}\sum_{t=1}^{n}
\one{Z_t=j}\xi_t\paren{\bftheta_{\star}^j}\bfx_t}\\
&\leq \sqrt{d}\sqrt{\frac{8u_F^2\bar{x}^2
\log\paren{2d/\delta}}{n}}.
\end{align*}

\textbf{\underline{Part (i) }} 
Let $\lambda =  \sqrt{\frac{8 u_F^2 \bar{x}^2 d \log \paren{ \frac{2d}{\delta} }}{n}}$ with $\delta = 1 / d$. 
Now, continuing from Lemma~\ref{lemma:stage_1_common} and taking expectation on both sides, we have  
\begin{eqnarray}
 &&\E{  \normtwo{  \bftheta_\star -   \bar{\bftheta} }  \mid \mathcal{H}_n  } \leq \E{ \frac{L_F \xmax}{\ell_F \lambda_{\min} \paren{\hat{\bfSigma}}}   \frac{1}{n}  \sum_{j=1}^{M} n^j \normtwo{ \bfdelta_\star^j }  
+  \frac{L_F \xmax}{\ell_F \lambda_{\min} \paren{\hat{\bfSigma}}}  \normtwo{  G  } ~\Big | ~\mathcal{H}_n   }    \nonumber \\
\quad &=&  \frac{L_F \xmax}{\ell_F \lambda_{\min} \paren{\hat{\bfSigma}}} \paren{  \frac{1}{n}  \sum_{j=1}^{M} n^j \normtwo{ \bfdelta_\star^j }  
+  \E{  \normtwo{  G} \one{ \calG_{\lambda }  } \mid \mathcal{H}_n 
 }  
+  \E{  \normtwo{ G }  \one{ \calG_{\lambda }^{\complement}   } \mid \mathcal{H}_n   } }.  \label{eq:stage_1_expectation_decomposition}  
\end{eqnarray}

To bound the second term in \eqref{eq:stage_1_expectation_decomposition}, we note that 
$  \E{  \normtwo{  G} \one{ \calG_{\lambda }  } \mid \mathcal{H}_n   }   \leq \lambda $.
For the last term in \eqref{eq:stage_1_expectation_decomposition}, we proceed by invoking the tail integral formula for expectation, 
\begin{eqnarray}
     &&\E{  \normtwo{ G }  \one{ \calG_{\lambda }^{\complement}   }  \mid \mathcal{H}_n  }  \nonumber 
     =   \int_{0}^{\infty}    \pr{ \normtwo{  G }  \one{ \calG_{\lambda }^{\complement}   } > \alpha }  \dd \alpha \nonumber  \\ 
    &=&  \int_{0}^{\lambda}    \pr{ \normtwo{  G }  \one{ \calG_{\lambda }^{\complement}   } > \alpha \mid \mathcal{H}_n  }  \dd \alpha
    +  \int_{\lambda}^{\infty}    \pr{ \normtwo{  G }  \one{ \calG_{\lambda }^{\complement}   } > \alpha \mid \mathcal{H}_n 
 }  \dd \alpha  \nonumber  \\ 
    &=&  \lambda \pr{ \calG_{\lambda }^{\complement} \mid \mathcal{H}_n   }  +  \lambda \int_{1}^{\infty}    \pr{ \normtwo{  G }  \one{ \calG_{\lambda }^{\complement}   } >  \lambda \alpha \mid \mathcal{H}_n  }  \dd \alpha  \label{eq:the_tricky_equation}  \\
    &\lesssim& \frac{\lambda}{d}   +   \lambda \int_{1}^{\infty}    \pr{  \calG_{ \lambda \alpha}^{\complement} \Big| \mathcal{H}_n 
 }  \dd \alpha  \label{eq:after_the_tricky_equation}   . 
\end{eqnarray}
In the last inequality, we used the fact that $\pr{ \calG_{\lambda }^{\complement} \mid \mathcal{H}_n   } \leq 2 d \exp \paren{  - \frac{ \lambda^2 n }{ 8 u_F^2 \bar{x}^2 d} } \lesssim \frac{1}{d} $. 
Equation~\eqref{eq:the_tricky_equation} holds as follows. For the first term $\int_{0}^{\lambda}    \pr{ \normtwo{  G }  \one{ \calG_{\lambda }^{\complement}   } > \alpha \mid \mathcal{H}_n  }  \dd \alpha$, when  $\alpha \in [0, \lambda]$, we have $ \pr{ \normtwo{  G }  \one{ \calG_{\lambda }^{\complement}   } > \alpha \mid \mathcal{H}_n  }  =  \pr{ \calG_{\lambda }^{\complement} \mid \mathcal{H}_n   }$. 
To see this, we note that on $\calG_{\lambda}$ we have $\normtwo{G} \leq \lambda$ hence the event $\braces{ \normtwo{  G }  \one{ \calG_{\lambda }^{\complement}   } > \alpha  }$ does not happen. 
On $\calG_{\lambda}^\complement$, we have $ \normtwo{G} \one{ \calG_{\lambda }^{\complement}   } = \normtwo{G} > \lambda \geq \alpha$.
For the second term in \eqref{eq:the_tricky_equation},  it holds by change of variable.

For the latter term in \eqref{eq:after_the_tricky_equation}, we observe that by \eqref{eq:prob_G_epsilon_c}
\begin{eqnarray*}
     \lambda \int_{1}^{\infty}    \pr{  \calG_{ \lambda \alpha}^{\complement} \mid \mathcal{H}_n  }  \dd \alpha &\leq& 
     \lambda   \int_{1}^{\infty}   2 d \exp \paren{  - \frac{ ( \lambda \alpha )^2 n }{ 8 u_F^2 \bar{x}^2 d} } \dd \alpha   \\
     &=& \lambda  \int_{1}^{\infty}   2 d \exp \paren{ - \alpha^2 \log \paren{2 d^2} } \dd \alpha  \\
     &=& 
     \lambda 2 d \int_{1}^{\infty}  \paren{ \sqrt{2} d }^{ -2 \alpha^2}  \dd \alpha
     \leq \lambda 2 d \int_{1}^{\infty}  \paren{ \sqrt{2} d }^{ -2 \alpha}  \dd \alpha
     \\
     &=&  \lambda 2 d  \frac{1}{ 4 d^2 \log \paren{ \sqrt{2} d } }, 
\end{eqnarray*}
where the first equality holds by the choice of $ \lambda$.

Putting everything together, we conclude that 
\begin{equation}
     \E{  \normtwo{  \bftheta_\star -  \bar{\bftheta} } \mid \mathcal{H}_n   }  
     \leq  \frac{L_F \xmax}{\ell_F \lambda_{\min} \paren{\hat{\bfSigma}}} 
     \paren{   \frac{1}{n}  \sum_{j=1}^{M} n^j \normtwo{ \bfdelta_\star^j }  
     + 3  \sqrt{\frac{8 u_F^2 \bar{x}^2 d \log \paren{ 2 d^2 }}{n}} } . 
\end{equation}

\textbf{\underline{Part (ii) }}  
In view of Lemma~\ref{lemma:stage_1_common}, squaring both sides of \eqref{eq:theta_bar_deterministic_bound} and taking expectation yield that
\begin{equation*}
 \E{  \normtwo{  \bftheta_\star -  \bar{\bftheta} }^2  \mid \mathcal{H}_n   }  
     \lesssim \paren{   \frac{L_F \xmax}{\ell_F \lambda_{\min}  \paren{\hat{\bfSigma}}}  }^2 
     \paren{   \paren{ \frac{1}{n}  \sum_{j=1}^{M} n^j \normtwo{ \bfdelta_\star^j } }^2   
     +  \E{ \normtwo{G}^2  \mid \mathcal{H}_n }  } . 
\end{equation*}
It suffices to upper bound $\E{ \normtwo{G}^2 \mid \mathcal{H}_n  }$.
Following the same reasoning as in \textbf{\underline{Part (i) }}, we have 
\begin{equation} \label{eq:break_G_to_two}
    \E{ \normtwo{G}^2 \mid \mathcal{H}_n  }  = 
\E{ \normtwo{G}^2 \one{ \calG_\lambda }\mid \mathcal{H}_n   } + \E{ \normtwo{G}^2 \one{ {\calG_\lambda}^\complement } \mid \mathcal{H}_n  } 
\leq  \lambda^2 + \E{ \normtwo{G}^2 \one{ {\calG_\lambda}^\complement } \mid \mathcal{H}_n  }  . 
\end{equation}
For the latter term in \eqref{eq:break_G_to_two}, 
\begin{eqnarray}
&& \E{  \normtwo{ G }^2  \one{  \paren{\calG^{}_{\lambda^{} }}^{\complement}   }  \mid \mathcal{H}_n  }  \nonumber  
     =   \int_{0}^{\infty}    \pr{ \normtwo{  G^{} }^2  \one{  \paren{\calG^{}_{\lambda^{} }}^{\complement}   } > \alpha \mid \mathcal{H}_n  }  \dd \alpha  \nonumber  \\ 
    &=&  \int_{0}^{ \lambda^2 }    \pr{ \normtwo{  G^{} }^2  \one{  \paren{\calG^{}_{\lambda^{} }}^{\complement}   } > \alpha \mid \mathcal{H}_n  }  \dd \alpha
    +  \int_{ \lambda^2 }^{\infty}    \pr{ \normtwo{  G^{} }^2  \one{  \paren{\calG^{}_{\lambda^{} }}^{\complement}   } > \alpha \mid \mathcal{H}_n  }  \dd \alpha  \nonumber   \\ 
    &=&  \lambda^2  \pr{  \paren{\calG^{}_{\lambda^{} }}^{\complement}  \mid \mathcal{H}_n  }  +  \lambda^2 \int_{1}^{\infty}    \pr{ \normtwo{  G^{} }^2  \one{  \paren{\calG^{}_{\lambda^{} }}^{\complement}   } >  (\lambda^{} )^2 \alpha \mid \mathcal{H}_n  }  \dd \alpha    \label{eq:tricky_eqn_2}  \\
    &\leq& \lambda^2 \pr{  \paren{\calG^{}_{\lambda^j }}^{\complement}  \mid \mathcal{H}_n  }    +   \lambda^2 \int_{1}^{\infty}    \pr{  \paren{\calG_{ \lambda^{}  \sqrt{\alpha} }^{}}^{\complement} \mid \mathcal{H}_n  }  \dd \alpha   .  \label{eq:gradient_sqr_expectation}
\end{eqnarray}
Here \eqref{eq:tricky_eqn_2} follows by the same reasoning as \eqref{eq:the_tricky_equation}. 
In the last inequality \eqref{eq:gradient_sqr_expectation}, we use the fact that $ \normtwo{ G^{}} \leq \lambda^{} \sqrt{\alpha} $ implies the event that $\normtwo{ G^{}}^2 \leq \lambda^2 \alpha $.
For the latter term in \eqref{eq:gradient_sqr_expectation}, following the same reasoning of \eqref{eq:after_the_tricky_equation}, we have that 
\begin{eqnarray*}
      \int_{1}^{\infty}   \pr{  \paren{\calG_{ \lambda^{}  \sqrt{\alpha} }^{}}^{\complement} \Big| \mathcal{H}_n  }    \dd \alpha 
     &\leq& 
         \int_{1}^{\infty}   2 d \exp \paren{  - \frac{ ( \lambda^{} \sqrt{\alpha} )^2 n^{} }{ 8 u_F^2 \bar{x}^2 d} } \dd \alpha  
     =   2 d  \int_{1}^{\infty}     \paren{ \frac{2 d }{\delta}  }^{-\alpha} \dd \alpha \\
     &=& 2 d \frac{1}{ \frac{2 d }{\delta}  \log \paren{ \frac{2 d }{\delta}  }} 
     =    \delta \frac{1}{\log \paren{2 d /  \delta}} . 
\end{eqnarray*}
Combining the above yields the result. 

\end{myproof}

\subsubsection{Proof of Lemma~\ref{lemma:stage_2}}

\begin{myproof}
\textbf{\underline{Part (i) }} 
By Taylor expansion of $\calL^j (\cdot)$ at $\bftheta_\star^j$, we have 
\begin{equation} \label{eq:stage_2_taylor}
    \calL^{j} \paren{\hat{\bftheta}^j} - \calL^{j} \paren{\bftheta^j_\star} = \paren{ \hat{\bftheta}^j - \bftheta_\star^j }^\top \nabla_{ }   \calL^{j} \paren{\bftheta^j_\star}  +  \frac{1}{2}  \paren{ \hat{\bftheta}^j - \bftheta_\star^j }^\top \nabla_{ }^{2}  \calL^{j} \paren{ \tilde{\bftheta}^j }   \paren{ \hat{\bftheta}^j - \bftheta_\star^j }  ,
\end{equation}
where  $\tilde{\bftheta}^j$ lies on the line segment connecting $\hat{\bftheta}^j$ and $\bftheta_\star^j$. 
The optimality of $\hat{\bftheta}^j $ to problem \eqref{eq:stage_2_problem} implies that  
\begin{equation}
    \calL^{j} \paren{ \hat{\bftheta}^j} + \lambda^j  \normtwo{ \hat{\bftheta}^j - \bar{ \bftheta } } \leq \calL^{j} \paren{\bftheta^j_\star} + \lambda^j  \normtwo{ \bftheta^j_\star - \bar{ \bftheta } }  \label{eq:stage_2_opt}  . 
\end{equation}
Noting that $ \nabla_{ }^{2}  \calL^{j} \paren{ \tilde{\bftheta}^j }  = -  \frac{1}{n^j} \sum_{t=1}^{n} \one{Z_t = j}  \nabla_{  }^{2}  \ell_t \paren{\tilde{\bftheta}^j }  =   \frac{1}{n^j} \sum_{t=1}^{n} \one{Z_t = j}  \eta_t \paren{ \tilde{\bftheta}^j }  \bfx_t \bfx_t^\top  $ and the definition of $\ell_F$ (c.f. \eqref{eq:def_ell_F}), we have 
\begin{eqnarray}
     \frac{1}{2}  \paren{ \hat{\bftheta}^j - \bftheta_\star^j }^\top \nabla_{ }^{2}  \calL^{j} \paren{ \tilde{\bftheta}^j }   \paren{ \hat{\bftheta}^j - \bftheta_\star} &\geq&  
     \frac{\ell_F}{2}   \paren{ \hat{\bftheta}^j - \bftheta_\star^j }^\top \frac{1}{n^j} \sum_{t=1}^{n} \one{Z_t = j}  \bfx_t \bfx_t   \paren{ \hat{\bftheta}^j - \bftheta_\star^j }  \nonumber  \\ 
    &\geq&   \frac{\ell_F}{2}  \lambda_{\min} \paren{ \hat{\bfSigma}^j \paren{n^j} }  \normtwo{ \hat{\bftheta}^j - \bftheta_\star^j }^2     , \label{eq:stage_2_theta_error_lower_bound}
\end{eqnarray}
provided that $\lambda_{\min} \paren{ \hat{\bfSigma}^{j} \paren{n^j} } > 0$.
By combining \eqref{eq:stage_2_taylor}, \eqref{eq:stage_2_opt} and \eqref{eq:stage_2_theta_error_lower_bound}, we obtain 
\begin{eqnarray}
& & \frac{\ell_F}{2}  \lambda_{\min} \paren{ \hat{\bfSigma}^{j} \paren{n^j} }  \normtwo{ \hat{\bftheta}^j - \bftheta_\star^j }^2   \nonumber \\ 
&\leq&   \lambda^j  \normtwo{ \bftheta^j_\star - \bar{ \bftheta } }  - \lambda^j  \normtwo{ \hat{\bftheta}^j - \bar{ \bftheta } }  - \paren{ \hat{\bftheta}^j - \bftheta_\star^j }^\top \nabla_{\bftheta}   \calL^{j} \paren{\bftheta^j_\star}  \nonumber  \\ 
&\leq&  \lambda^j  \normtwo{ \bftheta^j_\star - \bar{ \bftheta } }  - \lambda^j  \normtwo{ \hat{\bftheta}^j - \bar{ \bftheta } } +
    \normtwo{\hat{\bftheta}^j - \bftheta_\star^j } \normtwo{\nabla_{\bftheta}   \calL^{j} \paren{\bftheta^j_\star}}    
       \label{eq:apply_cauchy_stage_2} \\ 
    &\leq&  2 \lambda^j \paren{  \normtwo{ \bftheta_{\star}^{j} - \bftheta_{\star}^{ } } + \normtwo{ \bftheta_{\star}^{ } - \bar{\bftheta} } } 
    - \lambda^j \normtwo{\hat{\bftheta}^j - \bftheta_\star^j} 
    + \normtwo{\hat{\bftheta}^j - \bftheta_\star^j } \normtwo{\nabla_{\bftheta}   \calL^{j} \paren{\bftheta^j_\star}}   \label{eq:stage_2_upper_bound_part_1_1} , 
\end{eqnarray}
 where \eqref{eq:apply_cauchy_stage_2} follows from Cauchy's inequality, and \eqref{eq:stage_2_upper_bound_part_1_1} follows since (i) $\normtwo{ \bftheta^j_\star - \bar{ \bftheta } }  \leq \normtwo{ \bftheta^j_\star - \bftheta_{\star}^{ } } + \normtwo{ \bftheta_{\star}^{ } - \bar{ \bftheta }} $ and (ii) the following fact 
\begin{eqnarray}
  -  \normtwo{ \hat{\bftheta}^j - \bar{ \bftheta } } &=& -  \normtwo{ \hat{\bftheta}^j - \bftheta_{\star}^{j} + \bftheta_{\star}^{j} - \bar{ \bftheta } } \leq - \paren{ \normtwo{\hat{\bftheta}^j - \bftheta_{\star}^{j} } - \normtwo{\bftheta_{\star}^{j} - \bar{ \bftheta }} }   \nonumber 
\end{eqnarray}
where the inequality is due to the reverse triangular inequality.

In view of Lemma~\ref{lemma:basic_fact_qua_bound}, from \eqref{eq:stage_2_upper_bound_part_1_1}, we deduce that the following inequality holds almost surely 
\begin{eqnarray}
 \normtwo{ \hat{\bftheta}^j - \bftheta_\star^j }  &\leq&  \frac{2 }{ \ell_F \lambda_{\min} \paren{ \hat{\bfSigma}^{j} \paren{n^j} }}   
 \paren{ \normtwo{\nabla_{\bftheta}   \calL^{j} \paren{\bftheta^j_\star}} - \lambda^j } \nonumber \\
 & & + \sqrt{ \frac{ 4 }{ \ell_F \lambda_{\min} \paren{ \hat{\bfSigma}^{j} \paren{n^j} }}  }  \sqrt{ \lambda^j \paren{ \normtwo{ \bfdelta_\star^j }
 + \normtwo{ \bftheta_\star - \bar{\bftheta} }  } } \nonumber .  
\end{eqnarray}

\textbf{\underline{Part (ii) }} 
On the other hand, we show that the solution $\hat{\bftheta}^j$ to the regularized problem~\eqref{eq:stage_2_problem} is also not too far away from the solution to the unregularized problem 
\begin{equation}  \label{eq:eq:stage_2_unregulaized}
    \tilde{\bftheta}^j = \argmin_{\bftheta \in \mathbb{R}^d} ~  \calL^j  \paren{\bftheta}   . 
\end{equation}
First, we note that the estimation error of $\tilde{\bftheta}^j$ is bounded by the gradient of $\calL^j$ evaluated at the true coefficient $\bftheta_\star^j$. 
Indeed, following the analysis in Lemma~\ref{lemma:stage_1_common}, we know 
\begin{equation*}
\normtwo{  \tilde{\bftheta}^j - \bftheta_\star^j } \leq \frac{L_F \xmax}{ \ell_F \lambda_{\min} \paren{ \hat{\bfSigma}^{j} \paren{n^j} } }  
\normtwo{ \nabla \calL^j \paren{ \bftheta_\star^j } }
\end{equation*}
holds almost surely.

By optimality of $\hat{\bftheta}^j$, we have 
$
0 \in \nabla \calL^j \paren{ \hat{\bftheta}^j  } + \partial \paren{ \lambda^j \normtwo{  \hat{\bftheta}^j  - \bar{\bftheta}}}.$
In view of the fact that $\partial \normtwo{ \bfx} = \braces{ \frac{ \bfx }{ \normtwo{\bfx}} }$ for $\bfx \neq 0$ and $\partial \normtwo{ 0 } = \braces{ \bfx \mid \normtwo{\bfx} \leq 1 }  $,  we have 
\begin{equation} \label{eq:stage_2_gradient_subgradient}
\normtwo{ \nabla \calL^j \paren{ \hat{\bftheta}^j  } } = \lambda^j \partial \normtwo{ \hat{\bftheta}^j  - \bar{\bftheta} } \leq \lambda^j .       
\end{equation}
Similar to the reasoning of \eqref{eq:stage_1_size_gradient_1}, we deduce that 
\begin{equation} \label{eq:stage_2_norm_bdd_by}
\begin{aligned}
\normtwo{\hat{\bftheta}^j-\tilde{\bftheta}^j}
\leq\frac{\normtwo{\nabla\calL^j\paren{\hat{\bftheta}^j}-\nabla\calL^j\paren{\tilde{\bftheta}^j}}}{\ell_F\lambda_{\min}\paren{\hat{\bfSigma}^{j}\paren{n^j}}}=\frac{\normtwo{\nabla\calL^j\paren{\hat{\bftheta}^j}}}{\ell_F\lambda_{\min}\paren{\hat{\bfSigma}^{j}\paren{n^j}}}.
\end{aligned}
\end{equation}
Combining \eqref{eq:stage_2_gradient_subgradient} and \eqref{eq:stage_2_norm_bdd_by} yields that 
\begin{equation} \label{eq:distance_theta_hat_theta_tilde}
    \normtwo{  \hat{\bftheta}^j - \tilde{\bftheta}^j } \leq 
\frac{1}{\ell_F  \lambda_{\min} \paren{ \hat{\bfSigma}^{j} \paren{n^j} } }  \lambda^j   . 
\end{equation}

Putting the above together concludes that 
\begin{equation*}
\begin{aligned}
\normtwo{\hat{\bftheta}^j-\bftheta_\star^j}
&\leq\normtwo{\hat{\bftheta}^j-\tilde{\bftheta}^j}+\normtwo{\tilde{\bftheta}^j-\bftheta_\star^j}\\
&\leq\frac{\lambda^j}{\ell_F\lambda_{\min}\paren{\hat{\bfSigma}^{j}\paren{n^j}}}
+\frac{L_F\xmax\normtwo{\nabla\calL^j\paren{\bftheta_\star^j}}}{\ell_F\lambda_{\min}\paren{\hat{\bfSigma}^{j}\paren{n^j}}}.
\end{aligned}
\end{equation*}
as desired. 

\textbf{\underline{Part (iii) }} 
Following the consideration of \eqref{eq:stage_2_taylor}, \eqref{eq:stage_2_opt} and \eqref{eq:stage_2_theta_error_lower_bound}, by Taylor expansion of $\calL^j (\cdot)$ around  $\bar{\bftheta}$, it holds almost surely that 
\begin{equation*}
 \frac{\ell_F}{2}  \lambda_{\min} \paren{ \hat{\bfSigma}^{j} \paren{n^j} }  \normtwo{ \hat{\bftheta}^j - \bar{\bftheta} }^2 
 \leq  - \lambda^j \normtwo{ \hat{\bftheta}^j - \bar{\bftheta} } - \paren{ \hat{\bftheta}^j - \bar{\bftheta} }^{\top} \nabla \calL^j \paren{ \bar{\bftheta}}   . 
\end{equation*}
By applying Cauchy's inequality and the triangular inequality, we have 
\begin{eqnarray*}
& & - \lambda^j \normtwo{ \hat{\bftheta}^j - \bar{\bftheta} } - \paren{ \hat{\bftheta}^j - \bar{\bftheta} }^{\top} \nabla \calL^j \paren{ \bar{\bftheta}}   \\ 
    &\leq& - \lambda^j \normtwo{ \hat{\bftheta}^j - \bar{\bftheta} } 
    + \normtwo{\hat{\bftheta}^j -  \bar{\bftheta}} \normtwo{\nabla \calL^j \paren{ \bar{\bftheta}} } \\ 
    &\leq&  - \lambda^j \normtwo{ \hat{\bftheta}^j - \bar{\bftheta} }  
    +  \normtwo{\hat{\bftheta}^j -  \bar{\bftheta}} \normtwo{ \nabla \calL^j \paren{ \bftheta_\star^j }  }
    +  \normtwo{\hat{\bftheta}^j -  \bar{\bftheta}} \normtwo{ \nabla \calL^j \paren{ \bar{\bftheta}} - \nabla \calL^j \paren{ \bftheta_\star^j } }   \\
    &\leq& - \lambda^j \normtwo{ \hat{\bftheta}^j - \bar{\bftheta} } 
    +  \normtwo{\hat{\bftheta}^j -  \bar{\bftheta}} \normtwo{ \nabla \calL^j \paren{ \bftheta_\star^j }  }
    +  \normtwo{\hat{\bftheta}^j -  \bar{\bftheta}} L_F \xmax \normtwo{ \bar{\bftheta} - \bftheta_\star^j } ,
\end{eqnarray*}
where last inequality holds simply because 
\begin{equation*}
\normtwo{ \nabla \calL^j \paren{ \bar{\bftheta}} - \nabla \calL^j \paren{ \bftheta_\star^j } }  
= 
\normtwo{ \frac{1}{n^j} \sum_{t=1}^{n} \one{Z_t=j} \paren{ \xi_t \paren{\bar{\bftheta}} - \xi_t \paren{\bftheta_\star^j} } \bfx_t  } 
\leq 
L_F \xmax \normtwo{ \bar{\bftheta} - \bftheta_\star^j } . 
\end{equation*}
Therefore, applying Lemma~\ref{lemma:basic_fact_qua_bound} and rearranging the terms lead to 
\begin{equation*}
\normtwo{ \hat{\bftheta}^j - \bar{\bftheta} } \leq 
\frac{2  }{ \ell_F \lambda_{\min} \paren{ \hat{\bfSigma}^{j} \paren{n^j} }  }  
\paren{ \normtwo{  \nabla \calL^j \paren{ \bftheta_\star^j } }  - \lambda^j }
+ \frac{2 L_F \xmax  }{ \ell_F \lambda_{\min} \paren{ \hat{\bfSigma}^{j} \paren{n^j} }  }  \normtwo{ \bar{\bftheta} - \bftheta_\star^j }  . 
\end{equation*}

\end{myproof}

\subsubsection{Proof of Lemma~\ref{lemma:expectation_bound_stage_2}}

\begin{myproof}
Define the event 
\begin{equation*}
 \calG_{ \varepsilon }^{j} =  \braces{   \normtwo{\nabla_{}   \calL^{j} \paren{\bftheta^j_\star}}  \leq \varepsilon }.
\end{equation*}
 
{Following the same reasoning as \eqref{eq:prob_G_epsilon_c}}, we have $\pr{ {\calG_{\lambda^j }^{j}}^\complement \Big | \, \mathcal{H}_n  } \leq 2 d \exp \paren{  - \frac{ (\lambda^j)^2 n^j }{ 8 u_F^2 \bar{x}^2 d} } $. 
To set the stage, we first note that 
\begin{equation} \label{eq:diff_gradient_lambda_sqr}
    \E{  \normtwo{  \nabla_{}   \calL^{j} \paren{\bftheta^j_\star} }^2  \one{  \paren{\calG^{j}_{\lambda^j }}^{\complement}   } \Big|\, \mathcal{H}_n  } 
\leq (\lambda^j)^2  \paren{  \delta \frac{1}{\log \paren{\frac{2 d}{\delta}}} + \pr{  \paren{\calG^{j}_{\lambda^j }}^{\complement} \Big|\, \mathcal{H}_n  } }  . 
\end{equation}
The proof of this inequality follows the same argument for \eqref{eq:gradient_sqr_expectation} in \textbf{\underline{Part (ii)}} 
in the proof of Lemma~\ref{lemma:expectation_bound_stage_1}, hence omitted.

\textbf{\underline{Term (I) }} 
We start by recalling from Lemma~\ref{lemma:stage_2} that,  conditioned on $\calH_n$, it holds almost surely
\begin{eqnarray}
  \normtwo{ \hat{\bftheta}^j - \bftheta_\star^j }  &\lesssim&  
\frac{1 }{ \ell_F \lambda_{\min} \paren{ \hat{\bfSigma}^{j} \paren{n^j} }}   
 \paren{ \normtwo{\nabla_{\bftheta}   \calL^{j} \paren{\bftheta^j_\star}} - \lambda^j }  \nonumber  \\ 
    &&  + \sqrt{ \frac{ 1 }{ \ell_F \lambda_{\min} \paren{ \hat{\bfSigma}^{j} \paren{n^j} }}  }  \sqrt{ \lambda^j \paren{ \normtwo{ \bfdelta_\star^j }
 + \normtwo{ \bftheta_\star - \bar{\bftheta} }  } }. \label{eq:lemma_stage_2_1}
\end{eqnarray}
In what follows, we calculate $\E{   \normtwo{ \hat{\bftheta}^j - \bftheta_\star^j }^2  \Big|\, \mathcal{H}_n  } $ by decomposing it into two terms:  
\begin{equation}  \label{eq:theta_hat_second_stage_expectation_decomposition}
    \E{   \normtwo{ \hat{\bftheta}^j - \bftheta_\star^j }^2 \Big|\, \mathcal{H}_n  }  =
\E{   \normtwo{ \hat{\bftheta}^j - \bftheta_\star^j }^2 \one{ \calG_{ \lambda^j }^{j}  } \Big|\, \mathcal{H}_n  } 
+ \E{   \normtwo{ \hat{\bftheta}^j - \bftheta_\star^j }^2  \one{ {\calG_{ \lambda^j }^{j}}^{\complement}  } \Big|\, \mathcal{H}_n  }  . 
\end{equation}

\begin{enumerate}
    \item 
On event $\calG_{ \lambda^j }^{j} $, we conclude from \eqref{eq:lemma_stage_2_1} that 
\begin{eqnarray*}
  \normtwo{ \hat{\bftheta}^j - \bftheta_\star^j }  \one{ \calG_{ \lambda^j }^{j}  }   &\lesssim&  \sqrt{ \frac{ 1 }{ \ell_F \lambda_{\min} \paren{ \hat{\bfSigma}^{j} \paren{n^j} }}  }  \sqrt{ \lambda^j \paren{ \normtwo{ \bfdelta_\star^j }  + \normtwo{ \bftheta_\star - \bar{\bftheta} }  } }  \one{ \calG_{ \lambda^j }^{j}  }   .   
\end{eqnarray*}
Hence, squaring both sides and taking expectation, we have 
\begin{eqnarray}
   \E{ \normtwo{ \hat{\bftheta}^j - \bftheta_\star^j }^2  \one{ \calG_{ \lambda^j }^{j}  }  \Big|\, \mathcal{H}_n  }  &\lesssim& \frac{ 1 }{ \ell_F \lambda_{\min} \paren{ \hat{\bfSigma}^{j} \paren{n^j} }}  \lambda^j  \normtwo{ \bfdelta_\star^j }  \pr{ \calG_{ \lambda^j }^{j} \Big|\, \mathcal{H}_n  } \nonumber \\ 
    & &  + \frac{ 1 }{ \ell_F \lambda_{\min} \paren{ \hat{\bfSigma}^{j} \paren{n^j} }}  
    \lambda^j  \E{ \normtwo{ \bftheta_\star - \bar{\bftheta} } \one{  \calG_{ \lambda^j }^{j} } \Big|\, \mathcal{H}_n   }  .
\end{eqnarray}

\item   
On event ${\calG_{ \lambda^j }^{j}}^\complement $,  
by straightforward algebraic manipulation and the definitions of the event $\calG_{ \lambda^j }^{j} $, we have 
\begin{eqnarray*}
&&  \E{ \paren{\normtwo{\nabla_{\bftheta}   \calL^{j} \paren{\bftheta^j_\star}} - \lambda^j  }^2 \one{ {\calG_{ \lambda^j }^{j}}^{\complement}  }  \mid \mathcal{H}_n  } \\ 
    &=&    \E{ \paren{\normtwo{\nabla_{\bftheta}   \calL^{j} \paren{\bftheta^j_\star}}^2  + \paren{\lambda^j}^2 - 2 \lambda^j \normtwo{\nabla_{\bftheta}   \calL^{j} \paren{\bftheta^j_\star}} } \one{ {\calG_{ \lambda^j }^{j}}^{\complement}  }  \mid \mathcal{H}_n  }  \\ 
   &\leq&  \E{ \paren{\normtwo{\nabla_{\bftheta}   \calL^{j} \paren{\bftheta^j_\star}}^2  - \paren{\lambda^j}^2  } \one{ {\calG_{ \lambda^j }^{j}}^{\complement}  }  \mid \mathcal{H}_n  }   \\ 
   &\leq&  (\lambda^j)^2 
 \paren{ \pr{{\calG_{ \lambda^j }^{j}}^{\complement} \mid \mathcal{H}_n  }    
 + \delta \frac{1}{\log \paren{\frac{2 d}{\delta}}}   
 - \pr{{\calG_{ \lambda^j }^{j}}^{\complement} \mid \mathcal{H}_n  } }  \\ 
 &=&  \paren{\lambda^j}^2  \delta \frac{1}{\log \paren{\frac{2 d}{\delta}}}  . 
\end{eqnarray*}
The first inequality follows by noticing the definition of event $\one{ {\calG_{ \lambda^j }^{j}}^{\complement}  } $, and 
the second inequality follows from  \eqref{eq:diff_gradient_lambda_sqr}. Now,  proceeding from \eqref{eq:lemma_stage_2_1}, we have 
\begin{eqnarray}
    & &  \E{  \normtwo{ \hat{\bftheta}^j - \bftheta_\star^j }^2   \one{ {\calG_{ \lambda^j }^{j}}^{\complement}  } \Big|\,\mathcal{H}_n  }  \nonumber \\ 
    &\lesssim&  \frac{ 1 }{ \ell_F^2  \lambda^2_{\min} \paren{ \hat{\bfSigma}^{j} \paren{n^j} }}  
 (\lambda^j)^2  \delta \frac{1}{\log \paren{\frac{2 d}{\delta}}}  
    + \frac{ 1 }{ \ell_F \lambda_{\min} \paren{ \hat{\bfSigma}^{j} \paren{n^j} }}  \lambda^j  \normtwo{ \bfdelta_\star^j }  \pr{ {\calG_{ \lambda^j }^{j}}^{\complement} \Big|\, \mathcal{H}_n  }\nonumber\\
    &&+ \frac{ 1 }{ \ell_F \lambda_{\min} \paren{ \hat{\bfSigma}^{j} \paren{n^j} }}  
    \lambda^j  \E{ \normtwo{ \bftheta_\star - \bar{\bftheta} } \one{ { \calG_{ \lambda^j }^{j} }^{\complement}  } \Big|\, \mathcal{H}_n  }  . 
\end{eqnarray}

\end{enumerate}
Therefore, putting everything together, we have 
\begin{eqnarray*}
&& \E{ \normtwo{ \hat{\bftheta}^j - \bftheta_\star^j }^2 \mid \mathcal{H}_n  } \\ 
 &\lesssim&    
 \frac{ 1 }{ \ell_F^2  \lambda^2_{\min} \paren{ \hat{\bfSigma}^{j} \paren{n^j} }}    (\lambda^j)^2  \delta \frac{1}{\log \paren{\frac{2 d}{\delta}}}  
 + \frac{ 1 }{ \ell_F \lambda_{\min} \paren{ \hat{\bfSigma}^{j} \paren{n^j} }}  \lambda^j  \paren{  \normtwo{ \bfdelta_\star^j } 
 + \E{ \normtwo{ \bftheta_\star - \bar{\bftheta} } \mid \mathcal{H}_n   }  }   . 
\end{eqnarray*}


\textbf{\underline{ Term (II) }}  By noticing the second term in \eqref{eq:lemma_stage_2}, squaring the both sides and taking expectation, we have 
\begin{align*}
\E{\normtwo{\hat{\bftheta}^j-\bftheta_\star^j}^2\Big|\,\mathcal{H}_n}
&\lesssim
\paren{\frac{L_F\xmax}{\ell_F\lambda_{\min}\paren{\hat{\bfSigma}^{j}\paren{n^j}}}}^2
\left[\begin{aligned}
(\lambda^j)^2
&+\E{\normtwo{\nabla_{}\calL^{j}\paren{\bftheta^j_\star}}^2
\one{\calG^j_{\lambda^j}}\Big|\,\mathcal{H}_n}\\
&+\E{\normtwo{\nabla_{}\calL^{j}\paren{\bftheta^j_\star}}^2
\one{\paren{\calG^j_{\lambda^j}}^{\complement}}\Big|\,\mathcal{H}_n}
\end{aligned}\right]\\
&\lesssim
\paren{\frac{L_F\xmax}{\ell_F\lambda_{\min}\paren{\hat{\bfSigma}^{j}\paren{n^j}}}}^2
\left[\begin{aligned}
(\lambda^j)^2+(\lambda^j)^2
+(\lambda^j)^2\paren{1+\delta\frac{1}{\log\paren{2d/\delta}}}
\end{aligned}\right].
\end{align*}
The last inequality follows from the definition of the event $\calG^j_{\lambda^j}$ and \eqref{eq:diff_gradient_lambda_sqr}.

\textbf{\underline{Term (III) }} 
It follows from Lemma~\ref{lemma:stage_2} that 
\begin{equation*}
\begin{aligned}
\normtwo{\hat{\bftheta}^j-\bftheta_\star^j}
&\lesssim\frac{\normtwo{\nabla\calL^j\paren{\bftheta_\star^j}}-\lambda^j}{\ell_F\lambda_{\min}\paren{\hat{\bfSigma}^{j}\paren{n^j}}}+\frac{L_F\xmax\paren{\normtwo{\bfdelta_\star^j}+\normtwo{\bftheta_\star-\bar{\bftheta}}}}{\ell_F\lambda_{\min}\paren{\hat{\bfSigma}^{j}\paren{n^j}}}.
\end{aligned}
\end{equation*}
\begin{enumerate}
\item  On event $\calG_{ \lambda^j }^{j} $, it is easy to see that 
\begin{eqnarray*}
    & &  \E{ \normtwo{ \hat{\bftheta}^j - \bftheta_\star^j }^2  \one{\calG_{ \lambda^j }^{j}} \Big|\, \mathcal{H}_n  }  \\ 
    &\lesssim &  \paren{ \frac{ L_F \xmax  }{ \ell_F \lambda_{\min} \paren{ \hat{\bfSigma}^{j} \paren{n^j} }  } }^2 
\paren{ 
\normtwo{ \bfdelta_\star^j }^2 \pr{  \calG^{j}_{\lambda^j } \Big|\, \mathcal{H}_n   } 
+  \E{ \normtwo{ \bftheta_\star 
 - \bar{\bftheta}  }^2  \one{\calG_{ \lambda^j }^{j}} \Big|\, \mathcal{H}_n  }   }  . 
\end{eqnarray*}

\item On event ${\calG_{ \lambda^j }^{j}}^\complement $, following the same calculation of \textbf{\underline{Term (I) }}, we have 
\begin{eqnarray*}
    & & \E{ \normtwo{ \hat{\bftheta}^j - \bftheta_\star^j }^2  \one{ {\calG_{ \lambda^j }^{j}}^\complement } \Big|\, \mathcal{H}_n  } 
    \lesssim \paren{ \frac{ L_F \xmax  }{ \ell_F \lambda_{\min} \paren{ \hat{\bfSigma}^{j} \paren{n^j} }  } }^2  \\ 
& &  
\cdot \paren{ \paren{\lambda^j}^2  \delta \frac{1}{\log \paren{\frac{2 d}{\delta}}}   
+   \normtwo{ \bfdelta_\star^j }^2 \pr{ {\calG_{ \lambda^j }^{j}}^\complement \Big|\, \mathcal{H}_n  } + 
\E{ \normtwo{ \bftheta_\star 
 - \bar{\bftheta}  }^2  \one{ {\calG_{ \lambda^j }^{j} }^\complement } \Big|\, \mathcal{H}_n  } }  .
\end{eqnarray*}

\end{enumerate}
Therefore, we obtain that 
\begin{equation}
\begin{aligned}
\E{\normtwo{\hat{\bftheta}^j-\bftheta_\star^j}^2\mid\mathcal{H}_n}
&\lesssim\paren{\frac{L_F\xmax}{\ell_F\lambda_{\min}\paren{\hat{\bfSigma}^{j}\paren{n^j}}}}^2\paren{\frac{(\lambda^j)^2\delta}{\log\paren{2d/\delta}}+\normtwo{\bfdelta_\star^j}^2
+\E{\normtwo{\bftheta_\star-\bar{\bftheta}}^2\mid\mathcal{H}_n}}.
\nonumber
\end{aligned}
\end{equation}
\end{myproof}

\section{Additional Discussion}
\label{app:strategic_foundation}
In the main body, we treat the BCL yield $y_t$ in \eqref{eq:linear_bcl} as exogenous and therefore abstract from strategic interaction among dealers.
In this appendix, we provide an equilibrium justification for this modeling choice. Specifically, we \textit{approximate} dealers’ monetary profits to first order and show that the resulting game admits an exact symmetric equilibrium whose induced BCL yield has the linear form assumed in \eqref{eq:linear_bcl}.

We fix a round $t$ in which security $Z_t=j$ arrives and consider $N\geq 2$ dealers, indexed by $i\in\{1,\ldots,N\}$. All dealers observe the public context $\mathbf{x}_t$ and the yield-to-price mapping $V_t$. 
Let
\begin{equation*}
    m_t = \inner{\bfbeta^j,\bfx_t}
\end{equation*}
denote the public component of dealers' reservation yields. The coefficient $\bfbeta^j$ is common across dealers.
Dealer $i$ privately observes a shock $\xi_{i,t}$ and has reservation yield $m_t+\xi_{i,t}$, corresponding to the monetary cost $V_t(m_t+\xi_{i,t})$.
We further assume that the shocks are independent across dealers and rounds, independent of $(Z_t,\mathbf{x}_t)$, and drawn from a common distribution $G$ supported on $[-\Delta,\Delta]$. 
The distribution $G$ has a continuous positive density on the interior of its support.
Before quoting, each dealer observes its own shock but not the shocks of the other dealers.
Dealer $i$ chooses a yield markup $a_{i,t}\in[-\Delta,\Delta]$, quotes the yield $m_t+a_{i,t}$, and therefore submits the monetary price
$
    V_t\paren{m_t+a_{i,t}}.
$

Because $V_t$ is strictly decreasing, a higher yield markup corresponds to a lower monetary price. Thus, dealer $i$ wins the RFQ if and only if its markup is the highest among all dealers. Let
\begin{equation*}
    a_{-i,t}=\paren{a_{k,t}}_{k\neq i}
\end{equation*}
denote the vector of the other dealers' markups. The first argument below is dealer $i$'s own action, while the second argument collects its competitors' actions. Conditional on its private shock and the public information, dealer $i$'s exact monetary reward is
\begin{equation*}
    \operatorname{reward}_{i,t}\paren{a_{i,t};a_{-i,t}}
    =
    \brackets{
        V_t\paren{m_t+a_{i,t}}
        -
        V_t\paren{m_t+\xi_{i,t}}
    }
    \one{
        a_{i,t}\geq\max_{k\neq i}a_{k,t}
    }.
\end{equation*}
If dealer $i$ wins, the bracketed difference is its quoted price minus its monetary cost; otherwise, its reward is zero. 

\paragraph{The approximate reward}
We approximate dealers' monetary profits to first order around the public reservation yield $m_t$. For a dealer with private shock $\xi$ that chooses a yield markup $a$, the exact profit conditional on winning is
$
    V_t\paren{m_t+a}-V_t\paren{m_t+\xi}.
$
A first-order expansion around $m_t$ gives
\begin{equation*}
    V_t\paren{m_t+a}-V_t\paren{m_t+\xi}
    \approx
    -V_t'\paren{m_t}\paren{\xi-a}.
\end{equation*}
Accordingly, we define dealer $i$'s approximate reward by
\begin{equation}\label{eq:strategic_first_order_reward}
    \widetilde{\operatorname{reward}}_{i,t}
    \paren{a_{i,t};a_{-i,t}}
    =
    -V_t'\paren{m_t}
    \paren{\xi_{i,t}-a_{i,t}}
    \one{
        a_{i,t}\geq\max_{k\neq i}a_{k,t}
    }.
\end{equation}
From this point onward, we consider a game in which each dealer maximizes the expected approximate reward in~\eqref{eq:strategic_first_order_reward}, rather than its expected exact monetary reward. 

\paragraph{The best response}
A dealer's strategy maps each possible private shock into a yield markup. We seek a symmetric Bayes--Nash equilibrium in which every dealer uses the same mapping
\begin{equation*}
    b:[-\Delta,\Delta]\to[-\Delta,\Delta]
\end{equation*}
and dealer $i$ chooses
\begin{equation*}
    a_{i,t}=b\paren{\xi_{i,t}}.
\end{equation*}

The candidate equilibrium mapping is
\begin{equation*}
    b\paren{\xi}
    =
    \xi-
    \frac{
        \displaystyle\int_{-\Delta}^{\xi}G\paren{s}^{N-1}\,\mathrm{d}s
    }{
        G\paren{\xi}^{N-1}
    },
    \qquad \xi>-\Delta,
\end{equation*}
with $b\paren{-\Delta}=-\Delta$ by continuity. The function $b$ is increasing and satisfies
\begin{equation}\label{eq:equilibrium_bid_ode}
    b'\paren{\xi}
    =
    \paren{N-1}\frac{G'\paren{\xi}}{G\paren{\xi}}
    \paren{\xi-b\paren{\xi}}.
\end{equation}
To verify equilibrium, fix dealer $i$ with private shock $\xi$ and suppose that every other dealer uses $b$. Dealer $i$ chooses its own yield adjustment $a$. For $a$ in the range of $b$, the dealer wins if and only if every competing dealer has a shock at most $b^{-1}\paren{a}$. Independence of the $N-1$ competing shocks therefore gives the winning probability $G\paren{b^{-1}\paren{a}}^{N-1}$. Its expected approximate reward is
\begin{equation*}
    -V_t'\paren{m_t}\paren{\xi-a}
    G\paren{b^{-1}\paren{a}}^{N-1}.
\end{equation*}
Because $-V_t'\paren{m_t}>0$ does not depend on $a$, it does not affect either the maximizing action or the sign of the derivative. By the inverse-function rule,
\begin{equation*}
    \frac{\mathrm{d}}{\mathrm{d}a}b^{-1}\paren{a}
    =
    \frac{1}{b'\paren{b^{-1}\paren{a}}}.
\end{equation*}
After division by $-V_t'\paren{m_t}$, differentiating the expected reward with respect to $a$ gives
\begin{align}
    &-G\paren{b^{-1}\paren{a}}^{N-1}
    +
    \paren{\xi-a}\paren{N-1}
    G\paren{b^{-1}\paren{a}}^{N-2}
    \frac{
        G'\paren{b^{-1}\paren{a}}
    }{
        b'\paren{b^{-1}\paren{a}}
    }.
    \label{eq:approximate_reward_derivative}
\end{align}
At the proposed action $a=b\paren{\xi}$, we have $b^{-1}\paren{a}=\xi$, so setting~\eqref{eq:approximate_reward_derivative} equal to zero gives
\begin{equation*}
    -G\paren{\xi}^{N-1}
    +\paren{\xi-b\paren{\xi}}\paren{N-1}G\paren{\xi}^{N-2}
    \frac{G'\paren{\xi}}{b'\paren{\xi}}
    =0,
\end{equation*}
which is equivalent to~\eqref{eq:equilibrium_bid_ode}. This verifies the first-order condition, but we must still show that $b\paren{\xi}$ is a global best response.

Because $b$ is strictly increasing with $b\paren{-\Delta}=-\Delta$, first consider a markup $a\in\paren{-\Delta,b\paren{\Delta}}$, the interior of the range $[-\Delta,b\paren{\Delta}]$. 
The equilibrium ODE~\eqref{eq:equilibrium_bid_ode} holds for every private shock in $(-\Delta,\Delta)$. Substituting $b^{-1}\paren{a}$ for its argument and using $b\paren{b^{-1}\paren{a}}=a$ gives
\begin{align*}
    G\paren{b^{-1}\paren{a}}^{N-1}
    ={}&
    \paren{b^{-1}\paren{a}-a}\paren{N-1}
    G\paren{b^{-1}\paren{a}}^{N-2}
    \frac{
        G'\paren{b^{-1}\paren{a}}
    }{
        b'\paren{b^{-1}\paren{a}}
    }.
\end{align*}
Using this identity to replace the first term in~\eqref{eq:approximate_reward_derivative}, the derivative of dealer $i$'s conditional expected approximate reward with respect to its own markup $a$, after division by $-V_t'\paren{m_t}$, becomes
\begin{equation*}
    \paren{\xi-b^{-1}\paren{a}}\paren{N-1}
    G\paren{b^{-1}\paren{a}}^{N-2}
    \frac{
        G'\paren{b^{-1}\paren{a}}
    }{
        b'\paren{b^{-1}\paren{a}}
    }.
\end{equation*}
For $a\in\paren{-\Delta,b\paren{\Delta}}$, every factor except $\xi-b^{-1}\paren{a}$ is positive. Because $b$ is strictly increasing,
\begin{equation*}
    a<b\paren{\xi}
    \quad\Longleftrightarrow\quad
    b^{-1}\paren{a}<\xi.
\end{equation*}
Therefore, dealer $i$'s conditional expected approximate reward increases with its own markup when $a<b\paren{\xi}$ and decreases when $a>b\paren{\xi}$. For every $\xi\in(-\Delta,\Delta)$, the markup $b\paren{\xi}$ is therefore the unique maximizer over $[-\Delta,b\paren{\Delta}]$.

The endpoint types can be checked directly. If $\xi=-\Delta$, choosing $a=b\paren{-\Delta}=-\Delta$ gives zero expected approximate reward, whereas every $a>-\Delta$ in the range of $b$ gives negative profit whenever the dealer wins. If $\xi=\Delta$, the conditional expected approximate reward increases for every $a<b\paren{\Delta}$, so $b\paren{\Delta}$ is the maximizer over $[-\Delta,b\paren{\Delta}]$.

Finally, no feasible markup lies below $-\Delta$. If $a>b\paren{\Delta}$, dealer $i$ wins with probability one, just as it does at $a=b\paren{\Delta}$, but receives a strictly smaller profit conditional on winning. Therefore, every $a>b\paren{\Delta}$ is dominated by $b\paren{\Delta}$.

Consequently,
\begin{equation*}
    a_{i,t}=b\paren{\xi_{i,t}},
    \qquad i=1,\ldots,N,
\end{equation*}
is an exact symmetric Bayes--Nash equilibrium under the approximate objective.

\paragraph{The induced BCL yield}
Designate dealer $1$ as the focal dealer. Under the symmetric equilibrium, the BCL yield faced by this dealer is the highest yield quoted by dealers $2,\ldots,N$:
\begin{equation}\label{eq:equilibrium_bcl}
    y_t
    =
    \inner{\bfbeta^j,\bfx_t}
    +\max_{2\leq i\leq N}b\paren{\xi_{i,t}}.
\end{equation}
Noting that $\bfx_t$ contains an intercept, we can define $\bftheta_\star^j$ such that 
\begin{equation*}
    \inner{\bftheta_\star^j,\bfx_t}
    =
    \inner{\bfbeta^j,\bfx_t}
    +\E{\max_{2\leq i\leq N}b\paren{\xi_{i,t}}}.
\end{equation*}
Define the remaining centered fluctuation by
\begin{equation*}
    \epsilon_t
    =
    \max_{2\leq i\leq N}b\paren{\xi_{i,t}}
    -\E{\max_{2\leq i\leq N}b\paren{\xi_{i,t}}}.
\end{equation*}
Substituting these two definitions into~\eqref{eq:equilibrium_bcl} gives
\begin{equation}\label{eq:linear_bcl_from_equilibrium}
    y_t
    =
    \inner{\bftheta_\star^j,\bfx_t}
    +\epsilon_t.
\end{equation}
Thus, \eqref{eq:linear_bcl_from_equilibrium} has exactly the linear form in~\eqref{eq:linear_bcl}. By construction, $\E{\epsilon_t}=0$, and the maintained independence assumptions imply that $\braces{\epsilon_t}_{t\geq 1}$ is i.i.d. and independent of $(Z_t,\bfx_t)$. Hence, under the approximate reward, strategic competition indeed induces the BCL to have linear dependence on $\bfx_t$.

\section{Standard Results and Facts }
 This section states results from the literature that are used in our proofs. 
\begin{lemma}[Hoeffding's inequality]  \label{lemma:hoeffding}
Let $X_1, X_2, \ldots, X_n$ be independent random variables such that $a_i \leq x_i \leq b_i$ for each $i \in[n]$. Then for any $\epsilon>0$,
\begin{equation*}
\operatorname{Pr}\left[\left| \sum_{i=1}^n X_i-\mathbb{E}\left[\sum_{i=1}^n X_i\right]\right| \leq \epsilon\right] \geq 1-2 \exp \left(\frac{- \epsilon^2}{2 \sum_{i=1}^n\left(b_i-a_i\right)^2}\right)  . 
\end{equation*}
\end{lemma}

\begin{lemma}[Theorem 3.1 in \cite{tropp2011user}. Matrix Chernoff: Adapted Sequences]
\label{lemma:matrix_chernoff}
Consider a finite adapted sequence $\left\{\boldsymbol{X}_k\right\}$ of positive-semidefinite matrices with dimension $d$, and suppose that
\begin{equation*}
\lambda_{\max }\left(\boldsymbol{X}_t\right) \leq R \quad \text { almost surely. }
\end{equation*}
Define the finite series
\begin{equation*}
\boldsymbol{Y}:=\sum_t \boldsymbol{X}_t \quad \text { and } \quad \boldsymbol{W}:=\sum_t \mathbb{E}_{t-1} \boldsymbol{X}_t . 
\end{equation*}
For all $\mu \geq 0$,
\begin{equation*}
\pr{ 
\lambda_{\min }(\boldsymbol{Y}) \leq(1-\delta) \mu ~ \text { and } ~ \lambda_{\min }(\boldsymbol{W}) \geq \mu 
} 
\leq d \cdot\left[\frac{\mathrm{e}^{-\delta}}{(1-\delta)^{1-\delta}}\right]^{\mu / R} \quad \text { for } \delta \in[0,1) . 
\end{equation*}   
\end{lemma}

\begin{lemma} \label{claim:reverse_hazard_rate_normal}
Let $f, F$ be the p.d.f. and c.d.f. of a zero-mean normal distribution with standard deviation $\sigma$. 
    Its reversed hazard rate is $h(x) = \frac{f(x)}{F(x)} = \frac{\sqrt{\frac{2}{\pi }} e^{-\frac{x^2}{2 \sigma^2}}}{ \sigma \operatorname{erfc}\left(-\frac{x}{\sqrt{2} \sigma }\right)}$.
    Moreover, it is decreasing for all $x \in \mathbb{R}$. 
\end{lemma}

\begin{myproof}
 Let $\operatorname{erf}(z) \defeq \frac{2}{\sqrt{\pi}}\int_{0}^{z} e^{- t^2} \dd t$ and $\operatorname{erfc}(z) \defeq 1 - \operatorname{erf}(z) $. 
    One can show that
    \begin{equation*}
    h'(x) = -\frac{e^{-\frac{x^2}{ \sigma^2}} \left(\sqrt{2 \pi } x e^{\frac{x^2}{2  \sigma^2}} \left(\operatorname{erf}\left(\frac{x}{\sqrt{2}  \sigma}\right)+1\right)+2  \sigma\right)}{\pi   \sigma^3 \left(\operatorname{erf}\left(\frac{x}{\sqrt{2} \operatorname{\sigma}}\right)+1\right)^2}
    < 0 
\end{equation*}

    for all $x \in \mathbb{R}$. 
    Indeed, what remains to be checked is the sign of $\sqrt{2 \pi } x e^{\frac{x^2}{2  \sigma^2}} \left(\operatorname{erf}\left(\frac{x}{\sqrt{2}  \sigma}\right)+1\right)\allowbreak+2  \sigma$.
    The case when $x \geq 0$ is straightforward. 
    For the case when $x < 0$, we invoke Corollary 3 in \cite{chechile2011properties}. 
\end{myproof}

\begin{lemma} \label{lemma:basic_fact_qua_bound}
If $a x^2 - b x - c \leq 0 $, where $a, b, c >0$, then 
\begin{equation*}
\frac{b - \sqrt{b^2 + 4 ac}}{2a} \leq x \leq \frac{b + \sqrt{b^2 + 4 ac}}{2a}  \leq \frac{b}{a} + \sqrt{\frac{c}{a}}  . 
\end{equation*}
\end{lemma}

\begin{lemma}[Lemma 5 in \cite{kawaguchi2022robustness}] \label{lemma:multinomial_2}
If $X_1, X_2, \cdots, X_M$ are multinomially distributed with parameters $n$ and $\pi_1, \cdots, \pi_M$, then for any $t \geq 0 $
\begin{equation*}
\pr{ \pi_j - \frac{X_j}{n} > t } \leq \exp \paren{ - \frac{n t^2}{ 2 \pi_j}} .  
\end{equation*}
In particular, for any $\delta > 0$, with probability at least $1- \delta$, the following holds for all $i \in [M]$: 
\begin{equation}
    \pi_i - \frac{X_i}{n} \leq \sqrt{\frac{2 \pi_i \log \paren{M / \delta}}{n}}  . \nonumber 
\end{equation}
\end{lemma}




\end{document}

%% file: sifin/macros.tex
\usepackage{booktabs}

\usepackage{multirow} 

\usepackage[dvipsnames]{xcolor}

\newcommand{\RN}[1]{%
  \textup{\uppercase\expandafter{\romannumeral#1}}%
}

\usepackage[colorlinks=true, linkcolor=blue, citecolor=blue, urlcolor=blue, hypertexnames=false]{hyperref}

\usepackage[ruled, lined, linesnumbered, commentsnumbered, longend]{algorithm2e}
\SetKwInput{KwInput}{Input}
\SetKwInput{KwOutput}{Output}

\usepackage{endnotes}
\usepackage{bbm}
\usepackage{amsmath}
\usepackage{float}
\usepackage{epstopdf}
\usepackage{subfigure,caption,enumitem,thm-restate}
\usepackage{enumitem}
\usepackage{url}
\setlist{leftmargin=*}
\let\footnote=\endnote

\usepackage{natbib}
 \bibpunct[, ]{(}{)}{,}{a}{}{,}%
\usepackage{bm}
\usepackage{comment}

\newenvironment{myproof}{ \paragraph{Proof} } {\hfill$\square$}

\newcommand{\E}[1]{\mathbb{E} \left[ {#1} \right]}
\newcommand{\one}[1]{\mathbbm{1} \left[ {#1} \right]}
\newcommand{\inner}[1]{\langle {#1} \rangle }
\newcommand{\pr}[1]{\mathbf{Pr} \left[ {#1} \right]}

\definecolor{darkolivegreen}{rgb}{0.33, 0.42, 0.18}

\DeclareSymbolFont{extraup}{U}{zavm}{m}{n}
\DeclareMathSymbol{\varheart}{\mathalpha}{extraup}{86}
\DeclareMathSymbol{\vardiamond}{\mathalpha}{extraup}{87}

\usepackage[textsize=tiny,textwidth=2cm]{todonotes}
\reversemarginpar

\PassOptionsToPackage{usenames,dvipsnames}{xcolor} 

\newcommand{\bfx}{\mathbf{x}}

\newcommand{\bfv}{\mathbf{v}}
\newcommand{\bfV}{\mathbf{V}}

\newcommand{\bfbeta}{\boldsymbol{\beta}}

\newcommand{\bfdelta}{\boldsymbol{\delta}}

\newcommand{\bftheta}{\boldsymbol{\theta}}

\newcommand{\bfpi}{\boldsymbol{\pi}}

\newcommand{\bfXi}{\boldsymbol{\Xi}}

\newcommand{\bfSigma}{\boldsymbol{\Sigma}}

\newcommand{\xmax}{ \bar{x} }

\newcommand{\deltamax}{\delta_{\max}}

\newcommand{\dd}{\textsf{d}}

\newcommand{\defeq}{\stackrel{\rm def}{=}}

\newcommand{\regret}{\textsf{Regret}} 
\newcommand{\regt}{\textsf{reg}_t} 
\newcommand{\calL}{\mathcal{L}}
\newcommand{\calH}{\mathcal{H}}
\newcommand{\calF}{\mathcal{F}}
\newcommand{\calG}{\mathcal{G}}

\newcommand{\calD}{\mathcal{D}}
\newcommand{\calE}{\mathcal{E}}
\newcommand{\calN}{\mathcal{N}}

\newcommand{\order}[1]{\mathcal{O} \left({#1}\right)}

\newcommand{\paren}[1]{\left({#1}\right)}
\newcommand{\brackets}[1]{\left[{#1}\right]}
\newcommand{\braces}[1]{\left\{{#1}\right\}}
\newcommand{\abs}[1]{\left|{#1}\right|}
\newcommand{\normtwo}[1]{\left \| {#1} \right \|_{2}}
\newcommand{\norminf}[1]{\left \| {#1} \right \|_{\infty}}

\usepackage{changepage} 

%% file: sifin/numerical_experiments.tex
\section{Numerical Experiments} \label{sec:experiments}

In this section, we support our theoretical findings using a real-world dataset of U.S. corporate bonds.



\paragraph{Data sources} 

We construct the data set by combining transaction records from TRACE \citep{finra_trace}, daily bond information from OpenBond \citep{openbond}, contractual information from Mergent FISD \citep{mergent_fisd}, and interest-rate data from FRED \citep{fred}. 
TRACE reports the time, price, yield, quantity, and direction of each transaction. 
Following \cite{dick2014clean}, we remove duplicate reports and transactions that are subsequently cancelled, corrected, or reversed. We also exclude interdealer transactions, trades with special conditions, and observations with invalid prices, quantities, or timestamps. We focus on transactions in which a dealer sells a bond to a customer. Finally, we combine reports with the same bond identifier, execution time, and transaction price, since these reports may represent different pieces of the same transaction, and treat each resulting consolidated transaction as one RFQ.
OpenBond provides daily information about each bond, including its yield, price, credit spread, duration, convexity, remaining maturity, amount outstanding, and industry classification. For each TRACE RFQ, we use the latest OpenBond observation from a date strictly before the transaction date. We do not use the same-day observation because it may contain information from transactions occurring later that day. We retain an RFQ only when its OpenBond information is no more than 30 days old and the required bond information is available. We also obtain short-term interest rates and Treasury yields from FRED. These variables are lagged by two weekdays before being added to the data.

We use FISD to identify the contractual terms of each bond. We focus on fixed-rate, U.S.-dollar corporate bonds with standard principal and coupon structures. We exclude convertible, exchangeable, asset-backed, putable, preferred, perpetual, payment-in-kind, and other structured securities. We also exclude bonds with sinking funds or nonstandard redemption provisions. A callable bond is retained only when its call provision is a standard make-whole provision. 

For each RFQ at time $t$, we construct features describing the bond, its recent trading history, and prevailing market conditions. 
The bond features include its coupon, age, amount outstanding, remaining maturity, duration, convexity, and the most recently available OpenBond price, yield, and credit spread. 
Using earlier OpenBond observations, we calculate recent changes in the bond's price, yield, and credit spread, as well as its recent trading volume and yield volatility. Using previously reported TRACE transactions, we record the bond's latest transaction yield, the time elapsed since that transaction, and summaries of its recent transaction activity. 
We also calculate exponentially weighted average yields for other bonds in the same industry and for the corporate-bond market as a whole, using decay horizons of 30 minutes, two hours, one day, and seven days. The bond's own transactions are excluded when constructing these industry and market averages. 
Finally, we include lagged interest rates, Treasury yields, and the Treasury yield corresponding to the bond's remaining maturity. Thus, every feature used for an RFQ at time $t$ is constructed solely from information available before $t$. We use the same feature definitions for all learning algorithms.

\paragraph{Construction of an ``oracle'' benchmark} 
Because $\bftheta_j^\star$ is unknown in the real data, we construct a retrospective oracle by fitting a separate ridge regression estimator $\widehat{\bftheta}_j^{\,\mathrm{oracle}}$ for each bond $j$, using all of the bond's RFQs in the evaluation period. 
We assume that $\varepsilon_t\sim\mathcal{N}(0,\sigma^2)$, where $\sigma$ is constant across bonds and over time. We estimate this common value using RFQs from September through December 2023 and then hold it fixed for every bond, date, policy, and oracle during the 2024 evaluation.
We then compute
\[
p_t^{\mathrm{oracle}}
\in
\arg\max_p
(p-\gamma_t)
\Phi\left(
\frac{
V_t^{-1}(p)
-
\bfx_t^\top\widehat{\bftheta}_{Z_t}^{\,\mathrm{oracle}}
}{\sigma}
\right),
\]
where $\Phi$ is the standard normal distribution function and $\gamma_t=90$. After observing the TRACE transaction price $p_t^{\mathrm{TRACE}}$, we evaluate a quote $p$ using
\[
\widetilde r_t(p)
=
(p-\gamma_t)\mathbbm{1}\{p\leq p_t^{\mathrm{TRACE}}\}.
\]
The realized cumulative regret of policy $\pi$ is
\(
\widetilde{\operatorname{Reg}}_T(\pi)
=
\sum_{t=1}^{T}
\left[
\widetilde r_t(p_t^{\mathrm{oracle}})
-
\widetilde r_t(p_t^\pi)
\right].
\)
The full-period ridge estimates are used only to construct the evaluation oracle and are never available to the online policies. Because the oracle maximizes expected rather than realized reward, individual regret terms can be negative and cumulative regret can decrease.

\paragraph{Experiment setup and result}
We evaluate pooling, individual estimation, and TSMT on RFQs observed from January 2 through December 31, 2024. 
For each new evaluation month, all statistics are constructed in a rolling fashion. 
To price an RFQ in a given month, we use observations from the previous four calendar months together with observations from the current month that occurred earlier. 
Again, among these observations, we include only transactions whose TRACE reports were already available when the pricing decision was made.

\begin{figure}[htbp]
    \centering
    \subfigure[Industry composition.]{
        \includegraphics[width=0.42\linewidth]{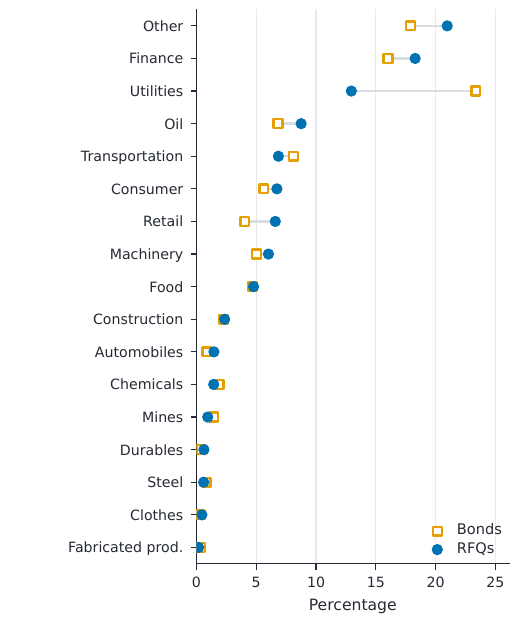}
        \label{fig:real_data_industry_composition}
    }%
    \hfill
    \subfigure[Trading intensity.]{
        \includegraphics[width=0.42\linewidth]{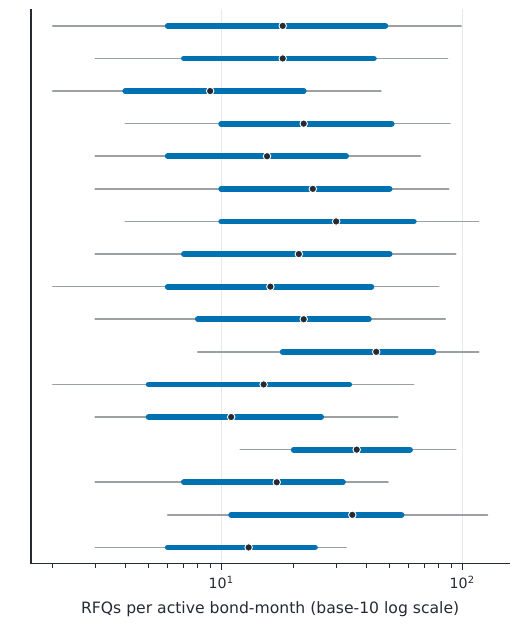}
        \label{fig:real_data_trading_intensity}
    }
    \caption{Industry composition and trading intensity in year of 2024. 
    Panel~(a) reports each Fama--French 17 industry's share of distinct bonds and RFQs. 
    Panel~(b) reports RFQs per active bond-month: 
    dots are medians, thick intervals contain the middle 50\% of observations, and thin intervals contain the middle 80\%. 
    Its horizontal axis uses a base-10 logarithmic scale. Industries are ordered by RFQ share. }
    \label{fig:real_data_sample_characteristics}
\end{figure}

Figure~\ref{fig:real_data_sample_characteristics} summarizes the composition of the dataset. Panel~\ref{fig:real_data_industry_composition} reports, for each industry, its percentage of the distinct bonds and its percentage of the RFQs among the entire dataset. 
Panel~\ref{fig:real_data_trading_intensity} shows the distribution of RFQs per active bond-month within each industry. We say that a bond is active in a given month if it receives at least one eligible RFQ during that month.
The dataset contains 1,232,665 RFQs and 3,532 bonds.

Pooling estimates one coefficient vector using all observations in the active window. The individual policy estimates a separate coefficient vector for each bond using only that bond's observations in the same window.\footnote{In the experiment, when implementing TSMT, in addition to the common part $\bftheta^\star$ shared by all bonds, we also specify a sector-specific common part $\bftheta^\star_s$ for each sector $s=1,2,\cdots,17$.}


For the reported backtest, all regularization parameters are held fixed across the twelve evaluation months. We set the ridge coefficient to $10^{-2}$ for every policy. 
TSMT uses $\lambda_t^j=0.1\sqrt{d/N_t^j}$, where $N_t^j$ is the number of observations of bond $j$ in the active window that are available before RFQ $t$.\footnote{When $N_t^j=0$, TSMT uses the corresponding sector-specific common part $\bftheta^\star_s$  directly.}

\begin{figure}[htbp]
    \centering
    \subfigure[Realized accumulated regret.]{
        \includegraphics[width=0.45\linewidth]{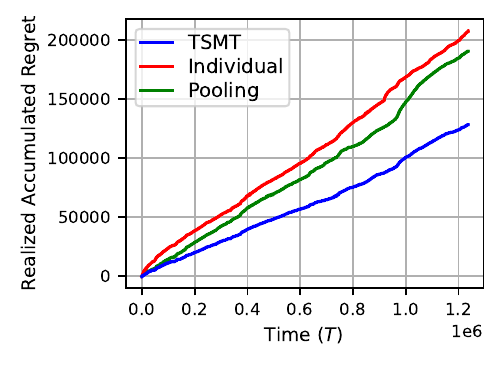}
        \label{fig:real_data_regret}
    }%
    \hfill
    \subfigure[Running yield-prediction MAE.]{
        \includegraphics[width=0.45\linewidth]{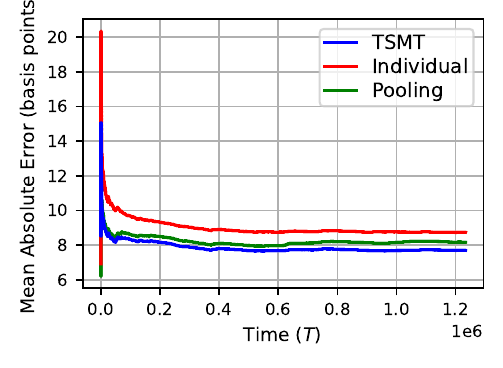}
        \label{fig:real_data_mae}
    }
    \caption{Regret and yield-prediction accuracy during the 2024 evaluation period. Each time step corresponds to one consolidated RFQ; lower values indicate better performance. TSMT is shown in blue, individual estimation in red, and pooling in green.}
    \label{fig:real_data_performance}
\end{figure}

Figure~\ref{fig:real_data_regret} reports the realized accumulated regret over the 2024 evaluation period.
It shows that TSMT has the lowest realized accumulated regret over the entire evaluation period.
It outperforms both the individual and pooling benchmarks consistently and by a large margin. 
TSMT reduces realized accumulated regret by 32.5\% relative to pooling and by 38.0\% relative to individual estimation.

To complement the regret measure, we also report the mean absolute error (MAE) of the prediction each policy yields. Whereas regret evaluates the resulting pricing decision, MAE measures the average absolute difference between the predicted yield and the realized transaction yield, in basis points. Figure~\ref{fig:real_data_mae} shows that TSMT has the lowest running MAE, ending at 7.724 basis points, compared with 8.195 for pooling and 8.771 for individual estimation.

\begin{remark}[Some limitations]
Note that we do not observe a clear sublinear behavior of the regret curves in Figure~\ref{fig:real_data_regret}.
One possible reason is that we cannot observe the true benchmark and must estimate it from noisy transactions, and our estimation suffers from some constant noises bounded away from zero.
Another possibility is that observations are discarded once they fall outside the four-month rolling window, so the amount of data available for learning does not keep growing over time.
We use such a rolling window because bond-market conditions change over time, so older observations may no longer be informative for current pricing decisions.
These limitations affect the shape of the curves, but not the main comparison: TSMT still has substantially lower regret than both benchmarks.  
\end{remark}

%% file: ref.bib
@article{bergault2021size,
  title={Size matters for OTC market makers: General results and dimensionality reduction techniques},
  author={Bergault, Philippe and Gu{\'e}ant, Olivier},
  journal={Mathematical Finance},
  volume={31},
  number={1},
  pages={279--322},
  year={2021},
  publisher={Wiley Online Library}
}

@article{cont2024dynamics,
  title={Dynamics of market making algorithms in dealer markets: Learning and tacit collusion},
  author={Cont, Rama and Xiong, Wei},
  journal={Mathematical Finance},
  volume={34},
  number={2},
  pages={467--521},
  year={2024},
  publisher={Wiley Online Library}
}

@article{cohen2024inventory,
  title={Inventory, market making, and liquidity in OTC markets},
  author={Cohen, Assa and Kargar, Mahyar and Lester, Benjamin and Weill, Pierre-Olivier},
  journal={Journal of Economic Theory},
  volume={222},
  pages={105917},
  year={2024},
  publisher={Elsevier}
}

@book{sundaram1996first,
  title={A first course in optimization theory},
  author={Sundaram, Rangarajan K},
  year={1996},
  publisher={Cambridge university press}
}

@article{morand2018generalized,
  title={Generalized envelope theorems: applications to dynamic programming},
  author={Morand, Olivier and Reffett, Kevin and Tarafdar, Suchismita},
  journal={Journal of Optimization Theory and Applications},
  volume={176},
  pages={650--687},
  year={2018},
  publisher={Springer}
}

@article{chechile2011properties,
  title={Properties of reverse hazard functions},
  author={Chechile, Richard A},
  journal={Journal of Mathematical Psychology},
  volume={55},
  number={3},
  pages={203--222},
  year={2011},
  publisher={Elsevier}
}

@article{atkins2024reinforcement,
  title={Reinforcement Learning for Corporate Bond Trading: A Sell Side Perspective},
  author={Atkins, Samuel and Fathi, Ali and Assefa, Sammy},
  journal={arXiv preprint arXiv:2406.12983},
  year={2024}
}

@article{cesa2024market,
  title={Market Making without Regret},
  author={Cesa-Bianchi, Nicol{\`o} and Cesari, Tommaso and Colomboni, Roberto and Foscari, Luigi and Pathak, Vinayak},
  journal={arXiv preprint arXiv:2411.13993},
  year={2024}
}

@inproceedings{xu2024dynamic,
  title={Dynamic Pricing in Securities Lending Market: Application in Revenue Optimization for an Agent Lender Portfolio},
  author={Xu, Jing and Hsu, Yung-Cheng and Biscarri, William},
  booktitle={Proceedings of the 5th ACM International Conference on AI in Finance},
  pages={513--520},
  year={2024}
}

@article{myerson1981optimal,
  title={Optimal auction design},
  author={Myerson, Roger B},
  journal={Mathematics of operations research},
  volume={6},
  number={1},
  pages={58--73},
  year={1981},
  publisher={INFORMS}
}

@inproceedings{kleinberg2003value,
  title={The value of knowing a demand curve: Bounds on regret for online posted-price auctions},
  author={Kleinberg, Robert and Leighton, Tom},
  booktitle={44th Annual IEEE Symposium on Foundations of Computer Science, 2003. Proceedings.},
  pages={594--605},
  year={2003},
  organization={IEEE}
}

@article{javanmard2019dynamic,
  title={Dynamic pricing in high-dimensions},
  author={Javanmard, Adel and Nazerzadeh, Hamid},
  journal={The Journal of Machine Learning Research},
  volume={20},
  number={1},
  pages={315--363},
  year={2019},
  publisher={JMLR. org}
}

@article{javanmard2017perishability,
  title={Perishability of data: dynamic pricing under varying-coefficient models},
  author={Javanmard, Adel},
  journal={Journal of Machine Learning Research},
  volume={18},
  number={53},
  pages={1--31},
  year={2017}
}

@article{cohen2020feature,
  title={Feature-based dynamic pricing},
  author={Cohen, Maxime C and Lobel, Ilan and Paes Leme, Renato},
  journal={Management Science},
  volume={66},
  number={11},
  pages={4921--4943},
  year={2020},
  publisher={INFORMS}
}

@article{keskin2014dynamic,
  title={Dynamic pricing with an unknown demand model: Asymptotically optimal semi-myopic policies},
  author={Keskin, N Bora and Zeevi, Assaf},
  journal={Operations research},
  volume={62},
  number={5},
  pages={1142--1167},
  year={2014},
  publisher={INFORMS}
}

@article{besbes2009dynamic,
  title={Dynamic pricing without knowing the demand function: Risk bounds and near-optimal algorithms},
  author={Besbes, Omar and Zeevi, Assaf},
  journal={Operations research},
  volume={57},
  number={6},
  pages={1407--1420},
  year={2009},
  publisher={INFORMS}
}

@article{broder2012dynamic,
  title={Dynamic pricing under a general parametric choice model},
  author={Broder, Josef and Rusmevichientong, Paat},
  journal={Operations Research},
  volume={60},
  number={4},
  pages={965--980},
  year={2012},
  publisher={INFORMS}
}

@article{qiang2016dynamic,
  title={Dynamic pricing with demand covariates},
  author={Qiang, Sheng and Bayati, Mohsen},
  journal={arXiv preprint arXiv:1604.07463},
  year={2016}
}

@article{bu2022context,
  title={Context-based dynamic pricing with partially linear demand model},
  author={Bu, Jinzhi and Simchi-Levi, David and Wang, Chonghuan},
  journal={Advances in Neural Information Processing Systems},
  volume={35},
  pages={23780--23791},
  year={2022}
}

@article{den2024pricing,
  title={Pricing and positioning of horizontally differentiated products with incomplete demand information},
  author={den Boer, Arnoud V and Chen, Boxiao and Wang, Yining},
  journal={Operations Research},
  volume={72},
  number={6},
  pages={2446--2466},
  year={2024},
  publisher={INFORMS}
}

@article{chen2023robust,
  title={Robust dynamic pricing with demand learning in the presence of outlier customers},
  author={Chen, Xi and Wang, Yining},
  journal={Operations Research},
  volume={71},
  number={4},
  pages={1362--1386},
  year={2023},
  publisher={INFORMS}
}

@inproceedings{li2017provably,
  title={Provably optimal algorithms for generalized linear contextual bandits},
  author={Li, Lihong and Lu, Yu and Zhou, Dengyong},
  booktitle={International Conference on Machine Learning},
  pages={2071--2080},
  year={2017},
  organization={PMLR}
}

@article{gorton2013fundamentals,
  title={The fundamentals of commodity futures returns},
  author={Gorton, Gary B and Hayashi, Fumio and Rouwenhorst, K Geert},
  journal={Review of Finance},
  volume={17},
  number={1},
  pages={35--105},
  year={2013},
  publisher={Oxford University Press}
}

@article{gorton2006facts,
  title={Facts and fantasies about commodity futures},
  author={Gorton, Gary and Rouwenhorst, K Geert},
  journal={Financial Analysts Journal},
  volume={62},
  number={2},
  pages={47--68},
  year={2006},
  publisher={Taylor \& Francis}
}

@article{szymanowska2014anatomy,
  title={An anatomy of commodity futures risk premia},
  author={Szymanowska, Marta and De Roon, Frans and Nijman, Theo and Van Den Goorbergh, Rob},
  journal={The Journal of Finance},
  volume={69},
  number={1},
  pages={453--482},
  year={2014},
  publisher={Wiley Online Library}
}

@article{ready2022order,
  title={Order flows and financial investor impacts in commodity futures markets},
  author={Ready, Mark J and Ready, Robert C},
  journal={The Review of Financial Studies},
  volume={35},
  number={10},
  pages={4712--4755},
  year={2022},
  publisher={Oxford University Press}
}

@article{lustig2011common,
  title={Common risk factors in currency markets},
  author={Lustig, Hanno and Roussanov, Nikolai and Verdelhan, Adrien},
  journal={The Review of Financial Studies},
  volume={24},
  number={11},
  pages={3731--3777},
  year={2011},
  publisher={Society for Financial Studies}
}

@article{andersen1998deutsche,
  title={Deutsche mark--dollar volatility: intraday activity patterns, macroeconomic announcements, and longer run dependencies},
  author={Andersen, Torben G and Bollerslev, Tim},
  journal={the Journal of Finance},
  volume={53},
  number={1},
  pages={219--265},
  year={1998},
  publisher={Wiley Online Library}
}

@article{evans2002order,
  title={Order flow and exchange rate dynamics},
  author={Evans, Martin DD and Lyons, Richard K},
  journal={Journal of political economy},
  volume={110},
  number={1},
  pages={170--180},
  year={2002},
  publisher={The University of Chicago Press}
}

@article{fama1970efficient,
  title={Efficient capital markets},
  author={Fama, Eugene F},
  journal={Journal of finance},
  volume={25},
  number={2},
  pages={383--417},
  year={1970}
}

@article{fama1993common,
  title={Common risk factors in the returns on stocks and bonds},
  author={Fama, Eugene F and French, Kenneth R},
  journal={Journal of financial economics},
  volume={33},
  number={1},
  pages={3--56},
  year={1993},
  publisher={Elsevier}
}

@article{aleti2025news,
  title={News and asset pricing: A high-frequency anatomy of the SDF},
  author={Aleti, Saketh and Bollerslev, Tim},
  journal={The Review of Financial Studies},
  volume={38},
  number={3},
  pages={712--759},
  year={2025},
  publisher={Oxford University Press}
}

@article{ait2025continuous,
  title={Continuous-time fama-macbeth regressions},
  author={A{\"\i}t-Sahalia, Yacine and Jacod, Jean and Xiu, Dacheng},
  journal={The Review of Financial Studies},
  pages={hhaf072},
  year={2025},
  publisher={Oxford University Press}
}

@article{gebhardt2005cross,
  title={The cross-section of expected corporate bond returns: Betas or characteristics?},
  author={Gebhardt, William R and Hvidkjaer, Soeren and Swaminathan, Bhaskaran},
  journal={Journal of financial economics},
  volume={75},
  number={1},
  pages={85--114},
  year={2005},
  publisher={Elsevier}
}

@article{chen2007corporate,
  title={Corporate yield spreads and bond liquidity},
  author={Chen, Long and Lesmond, David A and Wei, Jason},
  journal={The journal of finance},
  volume={62},
  number={1},
  pages={119--149},
  year={2007},
  publisher={Wiley Online Library}
}

@report{coalitionGreenwichReport,
author = {McPartland, Kevin and Kolchin,  Katie  },
title = {Understanding Fixed-Income
Markets in 2023},
institution = {Coalition Greenwich},
year = {2023}}

@article{gueant2019deep,
  title={Deep reinforcement learning for market making in corporate bonds: beating the curse of dimensionality},
  author={Gu{\'e}ant, Olivier and Manziuk, Iuliia},
  journal={Applied Mathematical Finance},
  volume={26},
  number={5},
  pages={387--452},
  year={2019},
  publisher={Taylor \& Francis}
}

@article{fermanian2016behavior,
  title={The behavior of dealers and clients on the European corporate bond market: the case of Multi-Dealer-to-Client platforms},
  author={Fermanian, Jean-David and Gu{\'e}ant, Olivier and Pu, Jiang},
  journal={Market microstructure and liquidity},
  volume={2},
  number={03n04},
  pages={1750004},
  year={2016},
  publisher={World Scientific}
}

@article{gabbi2005factors,
  title={Which factors affect corporate bonds pricing? Empirical evidence from eurobonds primary market spreads},
  author={Gabbi, Giampaolo and Sironi, Andrea},
  journal={The European Journal of Finance},
  volume={11},
  number={1},
  pages={59--74},
  year={2005},
  publisher={Taylor \& Francis}
}

@article{bongaerts2017asset,
  title={ },
  author={Bongaerts, Dion and De Jong, Frank and Driessen, Joost},
  journal={The Review of Financial Studies},
  volume={30},
  number={4},
  pages={1229--1269},
  year={2017},
  publisher={Oxford University Press}
}

@article{chen2024deep,
  title={Deep learning in asset pricing},
  author={Chen, Luyang and Pelger, Markus and Zhu, Jason},
  journal={Management Science},
  volume={70},
  number={2},
  pages={714--750},
  year={2024},
  publisher={INFORMS}
}

@article{black1972capital,
  title={The capital asset pricing model: Some empirical tests},
  author={Black, Fischer and Jensen, Michael C and Scholes, Myron and others},
  year={1972},
  publisher={Praeger New York}
}

@article{gu2020empirical,
  title={Empirical asset pricing via machine learning},
  author={Gu, Shihao and Kelly, Bryan and Xiu, Dacheng},
  journal={The Review of Financial Studies},
  volume={33},
  number={5},
  pages={2223--2273},
  year={2020},
  publisher={Oxford University Press}
}

@article{weigand2019machine,
  title={Machine learning in empirical asset pricing},
  author={Weigand, Alois},
  journal={Financial Markets and Portfolio Management},
  volume={33},
  pages={93--104},
  year={2019},
  publisher={Springer}
}

@article{kelly2023financial,
  title={Financial machine learning},
  author={Kelly, Bryan and Xiu, Dacheng and others},
  journal={Foundations and Trends{\textregistered} in Finance},
  volume={13},
  number={3-4},
  pages={205--363},
  year={2023},
  publisher={Now Publishers, Inc.}
}

@article{kelly2023modeling,
  title={Modeling corporate bond returns},
  author={Kelly, Bryan and Palhares, Diogo and Pruitt, Seth},
  journal={The Journal of Finance},
  volume={78},
  number={4},
  pages={1967--2008},
  year={2023},
  publisher={Wiley Online Library}
}

@article{kelly2019characteristics,
  title={Characteristics are covariances: A unified model of risk and return},
  author={Kelly, Bryan T and Pruitt, Seth and Su, Yinan},
  journal={Journal of Financial Economics},
  volume={134},
  number={3},
  pages={501--524},
  year={2019},
  publisher={Elsevier}
}

@article{giglio2022factor,
  title={Factor models, machine learning, and asset pricing},
  author={Giglio, Stefano and Kelly, Bryan and Xiu, Dacheng},
  journal={Annual Review of Financial Economics},
  volume={14},
  number={1},
  pages={337--368},
  year={2022},
  publisher={Annual Reviews}
}

@article{bianchi2021bond,
  title={Bond risk premiums with machine learning},
  author={Bianchi, Daniele and B{\"u}chner, Matthias and Tamoni, Andrea},
  journal={The Review of Financial Studies},
  volume={34},
  number={2},
  pages={1046--1089},
  year={2021},
  publisher={Oxford University Press}
}

@article{huang2023bond,
  title={Are bond returns predictable with real-time macro data?},
  author={Huang, Dashan and Jiang, Fuwei and Li, Kunpeng and Tong, Guoshi and Zhou, Guofu},
  journal={Journal of Econometrics},
  volume={237},
  number={2},
  pages={105438},
  year={2023},
  publisher={Elsevier}
}

@article{bryzgalova2019forest,
  title={Forest through the trees: Building cross-sections of stock returns},
  author={Bryzgalova, Svetlana and Pelger, Markus and Zhu, Jason},
  journal={Available at SSRN 3493458},
  year={2019}
}

@article{gu2021autoencoder,
  title={Autoencoder asset pricing models},
  author={Gu, Shihao and Kelly, Bryan and Xiu, Dacheng},
  journal={Journal of Econometrics},
  volume={222},
  number={1},
  pages={429--450},
  year={2021},
  publisher={Elsevier}
}

@article{caruana1997multitask,
  title={Multitask learning},
  author={Caruana, Rich},
  journal={Machine learning},
  volume={28},
  pages={41--75},
  year={1997},
  publisher={Springer}
}

@article{breiman1997predicting,
  title={Predicting multivariate responses in multiple linear regression},
  author={Breiman, Leo and Friedman, Jerome H},
  journal={Journal of the Royal Statistical Society Series B: Statistical Methodology},
  volume={59},
  number={1},
  pages={3--54},
  year={1997},
  publisher={Oxford University Press}
}

@inproceedings{romera2013multilinear,
  title={Multilinear multitask learning},
  author={Romera-Paredes, Bernardino and Aung, Hane and Bianchi-Berthouze, Nadia and Pontil, Massimiliano},
  booktitle={International Conference on Machine Learning},
  pages={1444--1452},
  year={2013},
  organization={PMLR}
}

@article{zhang2018overview,
  title={An overview of multi-task learning},
  author={Zhang, Yu and Yang, Qiang},
  journal={National Science Review},
  volume={5},
  number={1},
  pages={30--43},
  year={2018},
  publisher={Oxford University Press}
}

@article{yu2020gradient,
  title={Gradient surgery for multi-task learning},
  author={Yu, Tianhe and Kumar, Saurabh and Gupta, Abhishek and Levine, Sergey and Hausman, Karol and Finn, Chelsea},
  journal={Advances in Neural Information Processing Systems},
  volume={33},
  pages={5824--5836},
  year={2020}
}

@article{chua2021fine,
  title={How fine-tuning allows for effective meta-learning},
  author={Chua, Kurtland and Lei, Qi and Lee, Jason D},
  journal={Advances in Neural Information Processing Systems},
  volume={34},
  pages={8871--8884},
  year={2021}
}

@article{bastani2021predicting,
  title={Predicting with proxies: Transfer learning in high dimension},
  author={Bastani, Hamsa},
  journal={Management Science},
  volume={67},
  number={5},
  pages={2964--2984},
  year={2021},
  publisher={INFORMS}
}

@article{gu2022robust,
  title={Robust angle-based transfer learning in high dimensions},
  author={Gu, Tian and Han, Yi and Duan, Rui},
  journal={arXiv preprint arXiv:2210.12759},
  year={2022}
}

@article{li2022transfer,
  title={Transfer learning for high-dimensional linear regression: Prediction, estimation and minimax optimality},
  author={Li, Sai and Cai, T Tony and Li, Hongzhe},
  journal={Journal of the Royal Statistical Society Series B: Statistical Methodology},
  volume={84},
  number={1},
  pages={149--173},
  year={2022},
  publisher={Oxford University Press}
}

@article{duan2022adaptive,
  title={Adaptive and robust multi-task learning},
  author={Duan, Yaqi and Wang, Kaizheng},
  journal={The Annals of Statistics},
  volume={51},
  number={5},
  pages={2015--2039},
  year={2023},
  publisher={Institute of Mathematical Statistics}
}

@article{huang2024mean,
  title={Mean--Variance Portfolio Selection by Continuous-Time Reinforcement Learning: Algorithms, Regret Analysis, and Empirical Study},
  author={Huang, Yilie and Jia, Yanwei and Zhou, Xun Yu},
  journal={arXiv preprint arXiv:2412.16175},
  year={2024}
}

@article{huang2025sublinear,
  title={Sublinear regret for a class of continuous-time linear-quadratic reinforcement learning problems},
  author={Huang, Yilie and Jia, Yanwei and Zhou, Xun Yu},
  journal={SIAM Journal on Control and Optimization},
  volume={63},
  number={5},
  pages={3452--3474},
  year={2025},
  publisher={SIAM}
}

@article{jia2022policy,
  title={Policy evaluation and temporal-difference learning in continuous time and space: A martingale approach},
  author={Jia, Yanwei and Zhou, Xun Yu},
  journal={Journal of Machine Learning Research},
  volume={23},
  number={154},
  pages={1--55},
  year={2022}
}

@article{dai2024data,
  title={Data-driven option pricing},
  author={Dai, Min and Jin, Hanqing and Yang, Xi},
  journal={arXiv preprint arXiv:2401.11158},
  year={2024}
}

@article{cohen2021generalised,
  title={Generalised correlated batched bandits via the arc algorithm with application to dynamic pricing},
  author={Cohen, Samuel and Treetanthiploet, Tanut},
  journal={arXiv preprint arXiv:2102.04263},
  year={2021}
}

@article{tian2023learning,
  title={Learning from similar linear representations: Adaptivity, minimaxity, and robustness},
  author={Tian, Ye and Gu, Yuqi and Feng, Yang},
  journal={Journal of Machine Learning Research},
  volume={26},
  number={187},
  pages={1--125},
  year={2025}
}

@inproceedings{finn2017model,
  title={Model-agnostic meta-learning for fast adaptation of deep networks},
  author={Finn, Chelsea and Abbeel, Pieter and Levine, Sergey},
  booktitle={International conference on machine learning},
  pages={1126--1135},
  year={2017},
  organization={PMLR}
}

@article{xu2021learning,
  title={Learning across bandits in high dimension via robust statistics},
  author={Xu, Kan and Bastani, Hamsa},
  journal={arXiv preprint arXiv:2112.14233},
  year={2021}
}

@incollection{friedman2018double,
  title={The double auction market institution: A survey},
  author={Friedman, Daniel},
  booktitle={The double auction market},
  pages={3--26},
  year={2018},
  publisher={Routledge}
}

@article{cavallanti2010linear,
  title={Linear algorithms for online multitask classification},
  author={Cavallanti, Giovanni and Cesa-Bianchi, Nicolo and Gentile, Claudio},
  journal={The Journal of Machine Learning Research},
  volume={11},
  pages={2901--2934},
  year={2010},
  publisher={JMLR. org}
}

@article{fan2016multitask,
  title={Multitask quantile regression under the transnormal model},
  author={Fan, Jianqing and Xue, Lingzhou and Zou, Hui},
  journal={Journal of the American Statistical Association},
  volume={111},
  number={516},
  pages={1726--1735},
  year={2016},
  publisher={Taylor \& Francis}
}

@inproceedings{finn2019online,
  title={Online meta-learning},
  author={Finn, Chelsea and Rajeswaran, Aravind and Kakade, Sham and Levine, Sergey},
  booktitle={International conference on machine learning},
  pages={1920--1930},
  year={2019},
  organization={PMLR}
}

@article{hospedales2021meta,
  title={Meta-learning in neural networks: A survey},
  author={Hospedales, Timothy and Antoniou, Antreas and Micaelli, Paul and Storkey, Amos},
  journal={IEEE transactions on pattern analysis and machine intelligence},
  volume={44},
  number={9},
  pages={5149--5169},
  year={2021},
  publisher={IEEE}
}

@inproceedings{kveton2021meta,
  title={Meta-thompson sampling},
  author={Kveton, Branislav and Konobeev, Mikhail and Zaheer, Manzil and Hsu, Chih-wei and Mladenov, Martin and Boutilier, Craig and Szepesvari, Csaba},
  booktitle={International Conference on Machine Learning},
  pages={5884--5893},
  year={2021},
  organization={PMLR}
}

@article{bastani2022meta,
  title={Meta dynamic pricing: Transfer learning across experiments},
  author={Bastani, Hamsa and Simchi-Levi, David and Zhu, Ruihao},
  journal={Management Science},
  volume={68},
  number={3},
  pages={1865--1881},
  year={2022},
  publisher={INFORMS}
}

@article{taylor2009transfer,
  title={Transfer learning for reinforcement learning domains: A survey.},
  author={Taylor, Matthew E and Stone, Peter},
  journal={Journal of Machine Learning Research},
  volume={10},
  number={7},
  year={2009}
}

@article{zhuang2020comprehensive,
  title={A comprehensive survey on transfer learning},
  author={Zhuang, Fuzhen and Qi, Zhiyuan and Duan, Keyu and Xi, Dongbo and Zhu, Yongchun and Zhu, Hengshu and Xiong, Hui and He, Qing},
  journal={Proceedings of the IEEE},
  volume={109},
  number={1},
  pages={43--76},
  year={2020},
  publisher={IEEE}
}

@book{lattimore2020bandit,
  title={Bandit algorithms},
  author={Lattimore, Tor and Szepesv{\'a}ri, Csaba},
  year={2020},
  publisher={Cambridge University Press}
}

@article{aqsha2024strategic,
  title={Strategic Learning and Trading in Broker-Mediated Markets},
  author={Aqsha, Alif and Drissi, Fay{\c{c}}al and S{\'a}nchez-Betancourt, Leandro},
  journal={arXiv preprint arXiv:2412.20847},
  year={2024}
}

@article{boyce2024market,
  title={Market making with exogenous competition},
  author={Boyce, Robert and Herdegen, Martin and S{\'a}nchez-Betancourt, Leandro},
  journal={arXiv preprint arXiv:2407.17393},
  year={2024}
}

@article{wu2024broker,
  title={Broker-Trader Partial Information Nash Equilibria},
  author={Wu, Xuchen and Jaimungal, Sebastian},
  journal={arXiv preprint arXiv:2412.17712},
  year={2024}
}

@article{even2006action,
  title={Action elimination and stopping conditions for the multi-armed bandit and reinforcement learning problems.},
  author={Even-Dar, Eyal and Mannor, Shie and Mansour, Yishay and Mahadevan, Sridhar},
  journal={Journal of machine learning research},
  volume={7},
  number={6},
  year={2006}
}

@article{tropp2011user,
  title={User-friendly tail bounds for matrix martingales},
  author={Tropp, Joel A},
  year={2011},
  publisher={California Institute of Technology}
}

@inproceedings{bagnoli2006log,
  title={Log-concave probability and its applications},
  author={Bagnoli, Mark and Bergstrom, Ted},
  booktitle={Rationality and Equilibrium: A Symposium in Honor of Marcel K. Richter},
  pages={217--241},
  year={2006},
  organization={Springer}
}

@inproceedings{kawaguchi2022robustness,
  title={Robustness implies generalization via data-dependent generalization bounds},
  author={Kawaguchi, Kenji and Deng, Zhun and Luh, Kyle and Huang, Jiaoyang},
  booktitle={International Conference on Machine Learning},
  pages={10866--10894},
  year={2022},
  organization={PMLR}
}

@article{dick2014clean,
  title={How to clean enhanced TRACE data},
  author={Dick-Nielsen, Jens},
  journal={Available at SSRN 2337908},
  year={2014}
}

@online{finra_trace,
  author       = {{Financial Industry Regulatory Authority}},
  title        = {{TRACE: The Source for Real-Time Bond Market Transaction Data}},
  year         = {2024},
  url          = {https://www.finra.org/rules-guidance/key-topics/trace},
  note         = {Accessed: 2024-06-06}
}

@online{openbond,
  author       = {{Open Source Bond Asset Pricing}},
  title        = {{Open Source Bond Asset Pricing: Replication Data and Code}},
  year         = {2026},
  url          = {https://openbondassetpricing.com/},
  note         = {Accessed: 2026-09-08}
}

@online{mergent_fisd,
  author       = {{LSEG Mergent}},
  title        = {{Mergent Fixed Income Securities Database (FISD)}},
  year         = {2026},
  url          = {https://wrds-www.wharton.upenn.edu/pages/about/data-vendors/lseg-mergent/},
  note         = {Accessed through Wharton Research Data Services: 2026-09-08}
}

@online{fred,
  author       = {{Federal Reserve Bank of St. Louis}},
  title        = {{Federal Reserve Economic Data (FRED)}},
  year         = {2026},
  url          = {https://fred.stlouisfed.org/},
  note         = {Accessed: 2026-09-08}
}

@article{nasdaq_data,
      title  = "NASDAQ ITCH Data",
      author = "{{NASDAQ ITCH Data}}",
      note   = "\url{https://emi.nasdaq.com/ITCH/}",
      year   = "2022",
    }
